\documentclass[aps,prd,nopacs,nofootinbib,notitlepage]{revtex4-1}

\usepackage{tikz}
\usetikzlibrary{calc}
 \usetikzlibrary{arrows.meta}

\usepackage{amsfonts}
\usepackage{amsmath}
\usepackage{xcolor}
\usepackage{bm}
\usepackage{fancybox}
\usepackage{comment}
\usepackage{multirow}
\usepackage{tipa}
\usepackage{longtable}
\definecolor{refs}{RGB}{245,156,74}
\usepackage[colorlinks=true,hyperfootnotes=true,citecolor=cyan]{hyperref}

\usepackage{enumitem}
\newlist{postulates}{enumerate}{1}
\setlist[postulates]{
  label=\textbf{P\arabic*.},
  ref=P\arabic*,
  leftmargin=2.5em,
  labelsep=0.7em
}

\newcommand{\be}{\begin{equation}}
\newcommand{\ee}{\end{equation}}
\newcommand{\ba}{\begin{eqnarray}}
\newcommand{\ea}{\end{eqnarray}}
\newcommand{\bs}{\begin{subequations}}
\newcommand{\es}{\end{subequations}}

\newcommand{\bbe}{\boldsymbol{\mathrm{e}}}
\newcommand{\e}{\mathrm{e}}
\newcommand{\ie}{\text{\textschwa}}
\newcommand{\bbie}{\boldsymbol{\textbf{\textschwa}}}

\newcommand{\bdelta}{\boldsymbol{\delta}}

\newcommand{\blambda}{\boldsymbol{\lambda}}

\newcommand{\bomega}{\boldsymbol{\omega}}
\newcommand{\bpsi}{\boldsymbol{\psi}}
\newcommand{\bTheta}{\boldsymbol{\Theta}}

\newcommand{\bbell}{\boldsymbol{\ell}}

\newcommand{\bPsi}{\boldsymbol{\psi}}
\newcommand{\bA}{\boldsymbol{A}}
\newcommand{\bba}{\boldsymbol{a}}

\newcommand{\bE}{\boldsymbol{E}}
\newcommand{\bF}{\boldsymbol{F}}
\newcommand{\bG}{\boldsymbol{G}}

\newcommand{\bJ}{\boldsymbol{J}}
\newcommand{\bj}{\boldsymbol{j}}
\newcommand{\bK}{\boldsymbol{K}}
\newcommand{\bL}{\boldsymbol{L}}

\newcommand{\bQ}{\boldsymbol{Q}}
\newcommand{\bq}{\boldsymbol{q}}
\newcommand{\br}{\boldsymbol{r}}
\newcommand{\bR}{\boldsymbol{R}}
\newcommand{\bS}{\boldsymbol{S}}
\newcommand{\bbs}{\boldsymbol{s}}
\newcommand{\bT}{\boldsymbol{T}}
\newcommand{\bt}{\boldsymbol{t}}

\newcommand{\bX}{\boldsymbol{X}}

\newcommand{\bY}{\boldsymbol{Y}}

\newcommand{\bPi}{\boldsymbol{\Pi}}

\newcommand{\mH}{\mathcal{H}}
\newcommand{\mI}{\mathcal{I}}
\newcommand{\mS}{\mathcal{S}}

\newcommand{\aL}{\mathrm{\Lambda}}
\newcommand{\iL}{\mathrm{V}}
\newcommand{\V}{X}
\newcommand{\bmE}{\boldsymbol{\mathcal{E}}}

\newcommand{\Diff}{\textrm{D}}
\newcommand{\bdiff}{\boldsymbol{\mathrm{d}}}
\newcommand{\bDiff}{\boldsymbol{\mathrm{D}}}

\newcommand{\lp}{\left(}
\newcommand{\rp}{\right)}
\newcommand{\lb}{\left[}
\newcommand{\rb}{\right]}

\newcommand{\nn}{\nonumber}

\newcommand{\1}{1$^\text{st}$}
\newcommand{\2}{2$^\text{nd}$}
\newcommand{\3}{3$^\text{rd}$}
\newcommand{\4}{4$^\text{th}$}

\newcommand{\area}{\boldsymbol{B}}
\newcommand{\termopot}{\Omega}

\hypersetup{hidelinks}

\begin{document}

\title{Hamiltonian surface charges in General Parallel Relativity}

\author{Tomi S. Koivisto}
\address{Laboratory of Theoretical Physics, Institute of Physics, 
University of Tartu, W. Ostwaldi 1, 50411 Tartu, Estonia}
\address{National Institute of Chemical Physics and Biophysics, R\"avala 
pst. 10, 10143 Tallinn, Estonia}
\email{tomi.koivisto@ut.ee}

\begin{abstract}

General Parallel Relativity is a dynamically equivalent formulation of Einstein gravity in which the parallel frame supplies explicit boundary reference data.
In this formulation, finite-region energy--momentum and horizon entropy arise as distinct Hamiltonian surface charges. Diffeomorphisms along canonical frame directions define a quasi-local energy--momentum including a matter surface term, while normal-plane boosts at a smooth horizon give entropy. These covariant formulas are applied to the Reissner--Nordström black hole and to a dilatonic charged black hole. In both examples, the matter contribution and its Legendre transform complete the   
Misner–Sharp mass into the Hamiltonian energy appropriate to the chosen electromagnetic boundary datum.
The resulting horizon charges satisfy the thermodynamic relation obtained independently from the Euclidean boundary action.

\end{abstract}  

\maketitle

\tableofcontents

\section{Introduction}

Energy and entropy in general relativity are boundary observables. 
The energy of a gravitational system may not be represented by a covariant local bulk density, but by a quasi-local surface charge associated with the boundary of the region \cite{Szabados:2009eka}.
Black-hole entropy can likewise be represented by a Noether charge evaluated on a horizon cross-section \cite{Iyer:1994ys}.

The physical interpretation of a surface charge depends on the transformation it generates, as well as on the boundary terms and boundary conditions that enter the construction of the symplectic structure. Different approaches provide this reference structure in different ways.
At spatial infinity, asymptotic fall-off conditions select the Poincar{\'e} generators whose surface charges define total energy--momentum \cite{REGGE1974286}. In the Brown--York construction, quasi-local energy is associated with unit proper-time evolution on a prescribed timelike boundary \cite{Brown:1992br}. Covariant phase-space methods can determine whether a surface charge associated with a given transformation and the allowed boundary variations is a Hamiltonian generator \cite{Harlow:2019yfa,Chandrasekaran:2021vyu}. 
Edge-mode and corner-symmetry constructions further enlarge the regional phase space so that transformations acting at the corner and the associated boundary reference data are represented within the subsystem description \cite{Donnelly:2016auv,Speranza:2017gxd,Freidel:2020xyx}.

Parallel geometry offers a way to make the reference structure explicit. M{\o}ller used absolute parallelism to construct gravitational energy--momentum expressions that are tensorial under coordinate transformations \cite{Moller1961Conservation}. The formulation makes the reference frame part of the geometrical description, avoiding the coordinate-dependence of Einstein's original pseudotensor prescription\footnote{In modern Hamiltonian language, pseudotensors can be understood as quasi-local expressions tied to particular boundary terms, reference choices and displacement vectors \cite{Chang:1998wj}.} \cite{Einstein1916Hamilton}. The dynamics of Einstein's theory are not modified, but a parallel frame is introduced as an intrinsic reference structure for gravitational observables. 
However, the physical choice of the frame remains undetermined. 
The ambiguity persists, shifted from the spacetime coordinates to the tangent-space frame. Although M{\o}ller's scheme has been repeatedly revisited in the teleparallel literature \cite{deAndrade:2000kr,Maluf:2002zc,Nashed:2005kn,Lucas:2009nq,Krssak:2018ywd,Maluf:2023rwe,Bahamonde:2021gfp}, its core ambiguity and relation to quasi-local observables have remained open.  

The ambiguity admits a canonical resolution in General Parallel Relativity (G$_\parallel$R) \cite{Gomes:2023hyk,Koivisto:2026yxt}. 
The role of the reference frame is fixed by two conditions. First, variations of the parallel frame affect the gravitational action only through a boundary term, so that the frame does not contribute to the bulk dynamics \cite{Koivisto:2018aip}. Second, it is required that the reference frame itself carry no local spacetime energy--momentum \cite{BeltranJimenez:2019bnx}. The former principle confines the frame-dependence to the boundary, and the latter principle selects the canonical representative. 
Thus, the teleparallel sector of geometry (nonmetricity and torsion) is canonically determined by the 
bare metric sector of geometry (curvature). 
G$\parallel$R is not a replacement for the Riemannian description of Einstein
gravity \cite{BeltranJimenez:2019esp}, but can be viewed as its canonical completion in an extended geometry that reflects the
boundary-sensitive parallel reference structure. 

We shall compare two formulations of teleparallel gravity. In the
metric-affine formulation \cite{HEHL19951,JimenezCano:2021rlu}, flatness is imposed by a Lagrange multiplier \cite{Obukhov:2002tm}. In
absolute parallelism, the connection is generated by a frame and is flat by
construction \cite{Golovnev:2017dox}. The two formulations encode the same classical bulk dynamics,
but they organise off-shell variations and boundary terms differently. This
difference is reflected in the corresponding covariant surface charges.

The main result concerns diffeomorphism (Diff) charges. In a matter-coupled
teleparallel theory the physical charge is not, in general, exhausted by the
gravitational superpotential. A direct contribution is supplied by the matter
symplectic potential, in accordance with the covariant phase space method \cite{Frodden:2017suc}. 
In the applications
below this contribution is essential: for Einstein--Maxwell fields it turns
the gravitational Misner--Sharp part into the finite-radius Hamiltonian
energy with the electromagnetic correction consistently adapted to the 
polarisation. In the fixed-potential Dirichlet polarisation, the matter surface term reverses the 
finite-radius electromagnetic correction relative to the gravitational Misner--Sharp contribution.

The second result concerns General Linear (GL) symmetry. In the metric-affine
formulation the curvature momentum appears as the Noether potential of a
tangent-space GL transformation. In the Einstein-equivalent constitutive law
of G$_\parallel$R, its normal-plane boost projection at a horizon reduces to the Iyer--Wald area term \cite{Iyer:1994ys}. 
Thus the Hamiltonian energy and the entropy arise from distinct symmetry structures: the energy is the
matter-completed diffeomorphism charge, while the entropy is obtained from the
normal-plane GL charge.

The general formulae are tested in two charged black-hole geometries. The
Reissner--Nordström solution provides the simplest calibration, while a
dilatonic charged black hole probes a nontrivial canonical time normalisation. 
Special attention is paid to the boundary conditions of the electromagnetic sector
and the corresponding interpretation of the Hamiltonian energy.  
The distinction between fixed-potential and fixed-charge boundary conditions is well established in Hamiltonian black-hole thermodynamics \cite{Braden:1990hw,Louko:1996dw,Hawking:1995ap}. 
To our knowledge, however, the corresponding polarisation dependence of the finite-radius Hamiltonian energy has not been made explicit.
The consistency of the horizon thermodynamics is verified, using the free energy potential obtained independently in the Euclidean boundary 
framework \cite{BeltranJimenez:2024ufa}. 

We review and analyse generic teleparallel theories, although the applications are restricted to the G$_\parallel$R framework.
Thus, we begin in Section \ref{gravitytheory} with a systematic and general formulation of a teleparallel theory.
The Diff and GL surface charges are presented in Section \ref{surfacecharges}, and Section \ref{applications} applies
them to charged black holes. Section \ref{conclusions} concludes.

\section{Gravity theory} 
\label{gravitytheory}

We define teleparallelism by $\bDiff^2=0$, where $\bDiff$ is the exterior covariant derivative associated with GL($n$) connection on an $n$-dimensional spacetime. 
Throughout this and the next section the dimension $n$ is arbitrary, and tangent-space indices run as \(a,b=0,\ldots,n-1\).
A local transformation \(\delta_\epsilon\) is a {\it gauge symmetry} when $\delta_\epsilon \bL=\bdiff\bbell_\epsilon$,
for some \((n-1)\)-form \(\bbell_\epsilon\). The symmetry is exact when
\(\bbell_\epsilon=0\). The local GL transformations considered below
are exact symmetries, whereas a diffeomorphism satisfies $\delta_\xi \bL
    =\bdiff(\xi\lrcorner \bL)$. 
    Transformations
whose parameters vanish at the boundary are redundancies, while those
acting nontrivially at the boundary may carry surface charges.

A frame transformation has a different role \cite{Koivisto:2022oyt}. It changes the flat
parallel structure relative to the metric and coframe, while leaving
the Riemannian geometry unchanged. In
G$_\parallel$R, such a relative frame change  
modifies the Lagrangian only by an exact form. 
The {\it frame symmetry} is not a pure redundancy of
the boundary phase space, since it changes the boundary reference data,
the symplectic potential and, consequently, the quasi-local charges.

Section \ref{mag} introduces the notation, Section \ref{tele} reviews the relevant teleparallel formalism
and Section \ref{GPR} then states the postulates \ref{postulate1} and \ref{postulate2} that dictate the constitutive law for G$_\parallel$R. 

\subsection{Gravitational fields}
\label{mag}

\begin{center}
\begin{table}[h]
\begin{tabular}{  | l c | c || l   c  |}
\hline
  metric  0-form & $g^{ab}$ &  $n(n+1)/2$ & metrical energy $n$-form  &  $-\frac{1}{2}\bG_{ab}   =   
\frac{\partial\bL_G}{\partial g^{ab}}  + \bQ_a{}^c\wedge\bq_{cb}$ \\  \hline
 coframe 1-form  & $\bbe^a$ & $n^2$ & canonical energy $(n-1)$-form & $ {\bt}_a   =  - 
\frac{\partial\bL_G}{\partial \bbe^{a}}$  \\  \hline
 connection  1-form & $\bA_a{}^b$  & $n^3$ & conjugate potential $(n-1)$-form & $\bmE^a{}_b=2\bq^a{}_b -\bbe^a\wedge\bbs_b$ \\  \hline \hline
 frame matrix 0-form & $\Lambda^a{}_b$  & $n^2$ & connection EoM $n$-form & $\iL_c^a\bDiff\bmE^c{}_b$ \\  \hline
\end{tabular}

\caption{The gravitational fields. The second column indicates the number of independent components in $n$ dimensions, and the third column indicates the variational conjugate determined from the Lagrangian density $n$-form $\bL_G$. The first three rows are the fields of a metric-affine theory. In absolute parallelism the connection is replaced by the frame matrix as a fundamental field. 
\label{fields}}
\end{table}
\end{center}

\begin{center}
\begin{table}[h]
\begin{tabular}{| l  c l l  | l c    |    l     |}
\hline
  nonmetricity 1-form & 
$\bQ^{ab}$ & $ = -\bDiff g^{ab}$ & $= -\bdiff g^{ab} + 2\bA^{ab}$ &  Tolman $(n-1)$-form & $\bq_{ab}   =  
\frac{\partial\bL}{\partial\bQ^{ab}}$   & $\bDiff\bQ^{ab}  =  2\bR^{(ab)}$ \\  \hline
 torsion 2-form & $\bS^a$ & 
$= \bDiff \bbe^a$ & $= \bdiff \bbe^a - \bA_b{}^a\wedge\bbe^b$ & Freud $(n-2)$-form & $\bbs_a = \frac{\partial\bL}{\partial\bS^{a}}$ & $\bDiff\bS^a  =  -\bR_b{}^a\wedge\bbe^b$ \\  \hline
  curvature 2-form & 
$\bR_a{}^b$ & $ = (\bDiff \bA){}_a{}^b$ & $= \bdiff \bA_a{}^b + 
\bA_a{}^c\wedge\bA_c{}^b$ & Lagrange $(n-2)$-form & $\br^a{}_b = \frac{\partial\bL}{\partial\bR_a{}^b}$ & $\bDiff\bR_a{}^b =  0$ \\  \hline \hline
  connection 1-form & 
$\bA_a{}^b$ & $ = \iL_a^c\bdiff\aL^b_c$ & $=-\aL^b{}_c\bdiff\iL^c{}_a$ & conj. pot. $(n-1)$-form & $\bmE^a{}_b=2\bq^a{}_b -\bbe^a\wedge\bbs_b$ & $\bDiff\bmE^a{}_b =  0$ \\  \hline
\end{tabular}
\caption{The geometric forms derived from the fields in Table \ref{fields}.  The conjugate forms are defined as variations of the Lagrangian density and thus depend on the theory. The Bianchi identities in the last column follow from the definitions of the respective $p$-forms in the first three rows. 
In absolute parallelism, \(\bDiff\mathcal \bmE^{a}{}_{b}\approx 0\), whereas in G$_{\parallel}$R, Eq.~(\ref{important})
promotes it to an off-shell identity. \label{fields2}}
\end{table}
\end{center}

Our notation for the fundamental fields is summarised in Table \ref{fields}, and the
corresponding field strengths and conjugate forms are listed in Table \ref{fields2}.

\subsection{Parallel gravity}
\label{tele}


The analysis below is not specific to G$_\parallel$R but concerns the structure of teleparallel gravity in general.

\subsubsection{Action variation}

With the conventions from \ref{mag}, the action for a generic theory of teleparallel gravity, coupled to matter $\bL_M$, reads
\be \label{action}
I = \int \bL_G(g^{ab},\bbe^a,\bQ^{ab},\bS^a) + \int\br^a{}_b\wedge\bR_a{}^b + \int \bL_M(g^{ab},\bbe^a,\bpsi,\bdiff\bpsi)\,. 
\ee
Its variation takes the form
\be
\label{variations}
\delta\bL = \delta g^{ab}\bE_{ab} + \delta\bbe^a\wedge\bE_a + \delta\bA_a{}^b\wedge\hat{\bE}^a{}_b +  
\delta\br^a{}_b\wedge\bR_a{}^b + \delta\bpsi\wedge\bE_\psi+ \bdiff\bTheta\,,
\ee
and gives the equations of motion (EoMs),
\bs
\label{EoM}
\ba
\bE_{ab} & = & g_{ac}\bDiff\bq^c{}_b - \frac{1}{2}\bG_{ab} - \frac{1}{2}\bT_{ab}\,, \label{EoMg} \\
\bE_a & = & \bDiff\bbs_a - {\bt}_a  - \bT_a\,, \label{EoMe}\\
\hat{\bE}^a{}_b & = &  2\bq^a{}_b - \bbe^a\wedge\bbs_b  + \bDiff \br^a{}_b\,, \label{EoMc} \\
\bE_\psi & = & \partial\bL_M/\partial\bpsi - (-1)^p\bdiff \bPi\,, 
\ea
\es
and the (pre)symplectic potential 
\be
\bTheta = -\delta g^{ab}\bq_{ab} + \delta\bbe^a\wedge\bbs_a + \delta 
\bA_a{}^b\wedge\br^a{}_b + \delta\bPsi\wedge\bPi\,. \label{symplectic}
\ee
We have taken the matter fields $\bPsi$ to be $p$-forms and denoted the variations of $\bL_M$ as
\bs
\ba
\bPi & = & \frac{\partial\bL_M}{\partial \bdiff\bpsi}\,, \\
\bT_{ab}  & = &  -2\frac{\partial\bL_M}{\partial g^{ab}} = -\bbe_a\wedge\bT_b\,, \label{canonicalmatter2} \\
\bT_a & = & -\frac{\partial\bL_M}{\partial \bbe^{a}} = -\bbie_b\lrcorner\bT^b{}_a\,. \label{canonicalmatter}
\ea
\es
Relations (\ref{canonicalmatter2}), (\ref{canonicalmatter}) between the canonical matter current $\bT_a$ and the metrical matter current $\bT_{ab}$ are due to our initial restriction in (\ref{action}) to
matter fields $\bpsi$ which do not couple directly to the gravitational connections\footnote{In the generalised case, there would be two conventional options. In the terminology of 
Ref.~\cite{BeltranJimenez:2020sih},  the {\it minimal} coupling principle would result in $\bdiff\bpsi\rightarrow\bDiff\bpsi$, whereas  the {\it metrical} coupling principle would imply $\bdiff\bpsi\rightarrow\mathring{\bDiff}\bpsi$. The minimal coupling is motivated in a Yang--Mills-type gauge theory. However, the parallel-geometric formulations reviewed here do not support the standard gauge-theoretic interpretation for the connection, which may rather be regarded as a composite field.}, but may couple to the fields $(g^{ab},\bbe^a)$ through e.g. the Hodge operation and tangent space projections. 

For the gravitational fields, the analogous simple relations would equate $\bt_a$ and $\bbie_a\lrcorner \bG^b{}_a$. However, in the presence of torsion, the metrical and the canonical currents are not directly proportional since
\be \label{Gt}
\bG^a{}_b = \bS^a\wedge\bbs_b  -\bbe^a\wedge{\bt}_b\,.
\ee
This follows from the definitions in Table \ref{fields2}. Further, one can derive the explicit forms
\bs
\ba
\bG^a{}_b & = & \delta^a_b\bL_G - \lp Q_{bcd}\ast q^{acd} - S^{acd}\ast s_{bcd} + 2S_{cdb}\ast s^{cda}\rp\ast \boldsymbol{1}\,, \\
\bt_a & = & -\bbie_a\lrcorner\bL_G - \lp Q_b{}^{cd}\ast q_{acd} - 2S^{cd}{}_b\ast s_{cda}\rp\ast \bbe^b\,.
\ea
\es
The Lagrange multiplier $\br^a{}_b$ sets the curvature to vanish, and thus imposes the restriction to parallel geometry. 

\subsubsection{Transformations}

The action (\ref{action}) is quasi-invariant under an infinitesimal Diff $\delta_\xi$ with some vector field $\xi$, and exactly invariant under GL transformations given by some matrix $M_a{}^b$. These act upon the fields as
\bs
\begin{itemize}
\label{gauge}
\item Diff: 
\be \delta_\xi g^{ab} = \xi\lrcorner\bdiff g^{ab}\,, \quad
\delta_\xi \bbe^a = \{\bdiff,\xi\lrcorner\}\bbe^a\,, \quad
\delta_\xi \bA_a{}^b  = \bDiff\lp\xi\lrcorner\bA_a{}^b\rp + \xi\lrcorner\bR_a{}^b\,, \quad
\delta_\xi \bPsi = \{\bdiff,\xi\lrcorner\}\bPsi\,.\label{gaugeDiff} \ee 
\item GL: \be\delta_M g^{ab} = 2M^{(ab)}\,, \quad
\delta_M \bbe^a = M_b{}^a\bbe^b\,, \quad
\delta_M \bA_a{}^b = \bDiff M_a{}^b\,, \quad \delta_M \br^{a}{}_{b}
=
M^{a}{}_{c}\br^{c}{}_{b}
-
M^{c}{}_{b}\br^{a}{}_{c}\,, \label{gaugeGL} \ee
\end{itemize}
where the GL-covariant derivative $\bDiff$ treats the 0-form $\xi\lrcorner\bA_a{}^b$ like a GL-tensor. Under Diffs, the $\br^a{}_b$ transforms by the Lie derivative like a matter field $\bpsi$. 
The gauge symmetry of the action is thus GL$\times$Diff. We can also Lie-drag the fields in a covariant fashion by forming a combination of the above transformations as
\begin{itemize}
\item CDiff: 
\be \delta_\Xi g^{ab} = -\Xi\lrcorner\bQ^{ab}\,, \quad
\delta_\Xi \bbe^a = \bDiff\lp\Xi\lrcorner\bbe^a\rp + \Xi\lrcorner\bS^a\,, \quad
\delta_\Xi \bA_a{}^b  = \Xi\lrcorner\bR_a{}^b\,, \quad
\delta_\Xi \bPsi = \{\bDiff,\Xi\lrcorner\}\bPsi\,.\label{gaugeCDiff} \ee 
\end{itemize}
\es
In addition, the action may feature a frame symmetry. The existence and the form of the frame symmetry depend on the constitutive law of the theory. For the G$_\parallel$R theory, the frame symmetry is the full $\underline{\text{GL}}$, but for generic modified models, the frame symmetry is broken to a subset (not necessarily a subgroup) of the group of
$\underline{\text{GL}}$ transformations\footnote{
If the surviving frame transformations do not close into a subgroup,
they cannot be regarded as a residual gauge group with a fixed orbit
structure. The distinction between gauge and dynamical frame components
may then depend on the background, and the corresponding degree-of-freedom
count is typically not uniform. This is the setting in which extra,
strongly coupled, or ghost-like frame modes can appear in modified
teleparallel theories \cite{Li:2010cg,Gomes:2023tur}. Transverse--Diff-invariant ($\underline{\text{TDiff}}$) models in
symmetric-parallel geometry provide an example where the residual
subgroup structure has been used to construct ghost-free classes of
theories \cite{Bello-Morales:2024vqk}.} 
The frame transformation can be realised in two ways, parameterised by a matrix $m_a{}^b \in \mathfrak{gl}(n)$ , as
\bs
\begin{itemize}
\label{frameGL}
\item \underline{\text{GL}}: 
\be 
\delta_m \bA_a{}^b  = \bDiff m_a{}^b\,.\label{frameGL1} \ee 
\item alternative realisation: \be\delta_m g^{ab} = 2m^{(ab)}\,, \quad
\delta_m \bbe^a = m_b{}^a\bbe^b\,, \quad \delta_m \br^{a}{}_{b}
=
m^{a}{}_{c}\br^{c}{}_{b}
-
m^{c}{}_{b}\br^{a}{}_{c}\,. \label{frameGL2} \ee
\end{itemize}
\es
Thus, the frame transformation is simply a change of the parallel connection. The same effect is achieved by keeping the connection invariant and transforming instead all the other fields according to (\ref{gaugeGL}) where $M_a{}^b \rightarrow -m_a{}^b$.
As will be clarified below, the frame symmetry is a crucial determinant of a theory's structure. However, it is not associated with Noether charges in the same way that a gauge symmetry is.
  
 \subsubsection{Special cases}

There are two interesting special cases of parallel geometry, which can be
obtained by imposing additional constraints with Lagrange multipliers. These restrictions {\it partially} eliminate the GL gauge symmetry. 
\begin{itemize}
\item Symmetric parallelism: $\bS^a = 0$. The remaining GL freedom is reduced to the integrable transformations
GL$_{\parallel}$, i.e. to the Jacobian realisation of diffeomorphisms
on holonomic tangent frames\footnote{
Although a GL$_{\parallel}$ acts on tangent indices, its integrability renders it 
the Jacobian image of a Diff in a holonomic frame. Hence, locally,
\be
\text{GL}_{\parallel}\simeq \mathrm{Diff}/\mathrm{Trans}_{\rm global},
\qquad
\text{GL}_{\parallel}/\text{GL}_{\rm global}
\simeq
\mathrm{Diff}/\mathrm{Aff}_{\rm global}\,. \nn 
\ee}.
\item Metric parallelism: $\bQ^{ab} = 0$. This is parallelism in the orthonormal $g_{ab}=\eta_{ab}$ geometry, wherein the GL symmetry is reduced to the Lorentz transformations SO.
\end{itemize} 
Metric-parallel geometry is the most conventional teleparallel framework \cite{Aldrovandi:2013wha,Maluf:2023rwe,Krssak:2018ywd,Bahamonde:2021gfp}. Interest in symmetric-parallelism has emerged more recently \cite{Nester:1998mp,Adak:2005cd,Adak:2008gd,Jimenez:2017tkx,Jarv:2018bgs,Heisenberg:2023lru}. Yet only a handful of studies have explored general-parallel geometries
\cite{Hohmann:2021ast,Hohmann:2023xxx,Heisenberg:2022xxx,Heisenberg:2023tho,Adak:2023xxx,Bajardi:2024xxx}.  

There are two conventional gauge fixings that {\it completely} eliminate the GL redundancy and lead to alternative formulations of the theory.
\begin{itemize}
\item The Palatini gauge: $\bbe^a = \bdelta^a$. The variables are then the spacetime metric and the spacetime affine connection, which in this gauge can be identified with $(g^{ab},\bA_a{}^b)$, respectively. 
\item The Weitzenb{\"o}ck gauge: $\bA_a{}^b = 0$. The variables are then the metric $g^{ab}$ and the coframe field $\bbe^a$. The coframe then generates the teleparallel spacetime affine connection. 
\end{itemize} 
In the presence of the $\underline{\text{GL}}$ symmetry (or in the geometrically restricted settings, its relevant subgroup), both conditions can be imposed simultaneously. 
\begin{itemize}
\item The coincident ``gauge'' is defined by the vanishing spacetime affine connection \cite{BeltranJimenez:2022azb}.   
\end{itemize}  
More properly, the zero-affine connection configuration is a choice of frame. The configuration cannot be Diff-invariant because the affine connection is subject to  inhomogeneous coordinate transformations.

\subsubsection{Absolute parallelism}

The EoM obtained by varying the Lagrange multiplier imposes
\(\bR^a{}_b\approx 0\). Locally, the flat GL connection can therefore be written as
\be
\bA_a{}^b \approx \iL_a^c\bdiff\aL^b_c\,, \quad \text{where} \quad \iL^a_c \aL^c_b = \iL^c_b\aL^a_c = \delta^a_b\,. \label{Rsolution}
\ee
We regard $\aL^a_b \in$ GL as a parameterisation of a group element, and $\iL^a_b = (\aL^{-1})^a{}_b$. Under infinitesimal GL transformations, the matrix elements transform as $\delta_M \aL^a_b = M_c{}^a\aL^c_b$ and $\delta_M \iL^a_b = -M_b{}^c\iL^a_c$, which consistently results in
$\delta_M \bA_a{}^b = \bDiff M_a{}^b$.
Plugging the solution (\ref{Rsolution}) back into the action (\ref{action}), we obtain the action in terms of reduced variables, 
\be \label{Aaction}
I = \int\bL_G[g^{ab},\bbe^a,\bQ^{ab},\bS^a] + \int\bL_M[g^{ab},\bbe^a,\bpsi,\bdiff\bpsi]\,, 
\ee
where it is now understood that $\bQ^{ab}=\bQ^{ab}[g,\aL]$, $\bS^a = \bS^a[\bbe,\aL]$ are functionals not of an independent connection but of the frame matrix. 
This is the action principle for absolute parallelism. 
The connection is now flat by construction, as it is no longer a variational degree of freedom, but a composite field.

\subsection{G$_\parallel$R}
\label{GPR}

The preceding subsections described the general teleparallel setting. We now specialise to a quadratic theory
\be \label{GPRaction}
\bL_G = \frac{1}{2}\bQ^{ab}\wedge\bq_{ab} + \frac{1}{2}\bS^a\wedge\bbs_a\,. 
\ee
Further, we will specialise to the Einstein-equivalent constitutive law that defines G$_\parallel$R.
This specialisation avoids new propagating frame degrees
of freedom, but allows us to regard the frame field $\bbie_a$ as a boundary-sensitive reference
structure for quasi-local observables. 
The role of the reference frame $\bbie_a$ as a tool of description excludes its active participation in the physics. One may impose this with the following two postulates.
\begin{postulates}
\item \label{postulate1}
The frame should not contribute to the dynamics. Thus: $\bdiff (\delta\bL_G/\delta\bbie_a) = 0\,.$ 
\item \label{postulate2}
The frame should not contribute to the observables. Thus, in a canonical frame: $\partial \bL_G/\partial\bbie_a =0\,.$
\end{postulates}
Postulate \ref{postulate1} implies that the variations of the frame field induce only a total derivative into the action. Nevertheless, the induced total derivative term may give a useful handle on the physical boundary conditions when fixed by the postulate \ref{postulate2}.  

A priori there is no guarantee for the existence of an action principle consistent with these postulates. It can be shown, however, that the  postulate \ref{postulate1} selects the Lagrangian of a theory essentially uniquely \cite{BeltranJimenez:2019odq}. The overall normalisation is left free, and we denote the gravitational coupling parameter in $n$ dimensions by $\mu_n$, such that for $n=4$ we recover $\mu_4 = m_P^2$, where $m_P$ is the Planck mass.   
The Lagrangian $\bL_G$ for G$_\parallel$R is specified by the constitutive relation
\bs
\label{potentials}
\ba
\bbs_a  & = & \frac{\mu_n}{2}\lp Q^{[bc]}{}_a + \delta^{[b}_a \V^{c]}  + \frac{1}{2}S_a{}^{bc} - S^{[bc]}{}_a \rp \ast\lp \bbe_b\wedge\bbe_c\rp\,, \label{spotential}  \\
\bq_{ab} & = & \frac{\mu_n}{4}\lb Q^c{}_{ab} - 2Q_{(ab)}{}^c + \delta^c_{(a}\lp Q_{b)} + 2S_{b)}\rp - g_{ab}\V^c
- 2S_{(ab)}{}^c  \rb \ast\bbe_c\,, \label{qpotential}
\ea
\es
where we have introduced the traces of the tensor components as
\be \label{vectors}
Q^a = Q^{ab}{}_b\,, \quad \tilde{Q}^a = Q^{ba}{}_b\,, \quad S_a = S^b{}_{ab}\,, \quad \V^a = Q^a - \tilde{Q}^a + 2S^a\,.  
\ee
By construction, the Lagrangian (\ref{GPRaction}) is now quasi-invariant under both groups of gauge transformations (\ref{gauge}), and also under the $\underline{\text{GL}}$ frame transformations. An important consequence of the latter symmetry is that 
\be \label{important}
\bS^a\wedge\bbs_b - \bbe^a\wedge\bDiff\bbs_b = 2\bDiff\bq^a{}_b \quad \Leftrightarrow \quad \bDiff\bmE^a{}_b=0\,,
\ee
which can be verified explicitly from (\ref{potentials}). 
Thus, the action (\ref{GPRaction}) with the constitutive relation (\ref{potentials}) possesses the extended symmetry $\underline{\text{GL}}\times$GL$\times$Diff. 

Let us now show the dynamical equivalence of G$_\parallel$R with standard general relativity. 
Defining the distortion 1-form $\bX_a{}^b = \bA_a{}^b - \mathring{\bA}_a{}^b = X^a{}_{cb}\bbe^c$, we obtain explicitly 
\be
X^a{}_{bc} = \frac{1}{2}Q^a{}_{bc} - Q_{(bc)}{}^a  + \frac{1}{2}S^a{}_{bc} -  S_{(bc)}{}^a\,. 
\ee   
We then express the curvature as $\bR_a{}^b = \mathring{\bR}_a{}^b + \mathring{\bDiff}\bX_a{}^b + \bX_a{}^c\wedge\bX_c{}^b$. Contracting this 2-form expression into a scalar, and taking into account the parallelism $\bR_a{}^b = 0$, we obtain 
\be
\mathring{R} = \bbie_b\lrcorner\bbie_a\lrcorner\mathring{\bR}^{ab} = -\bbie_b\lrcorner\bbie_a\lrcorner\lp  \mathring{\bDiff}\bX^{ab} + \bX^{ac}\wedge\bX_c{}^b\rp
=  -2X^a{}_{[a\lvert b \rvert}X^b{}_{c]}{}^c - \mathring{\nabla}_a \V^a\,. \label{formula}
\ee
We note that the vector $X^a$ introduced at (\ref{vectors}) is equivalently the trace difference $X^a = X^a{}_{d}{}^{d}-X^d{}_{d}{}^{a} = 2X^{[a}{}_b{}^{b]}$.  
A direct algebraic calculation gives
\be
-2\mu_n X^a{}_{[a\lvert b \rvert}X^b{}_{c]}{}^c\ast 1 =  \bQ^{ab}\wedge\bq_{ab} + \bS^a\wedge\bbs_a\,. \label{X2}
\ee 
Combining (\ref{formula}) and (\ref{X2}), we see that the G$_\parallel$R Lagrangian (\ref{GPRaction}) is equivalent to the Einstein-Hilbert term, up to a total derivative,  $\bL_G = \ast \mu_n(\mathring{R} + \mathring{\nabla}_a \V^a)/2$. 

The construction becomes physically relevant at the level of observables. The
postulate \ref{postulate2} selects canonical reference frames which themselves
do not carry local energy-momentum, $\bt_a=0$. In such frames the conserved currents and
surface charges acquire a definite quasi-local interpretation. We shall turn to
their Noether definition after a brief summary of parallel-gravity formulations. 

\subsection{Summary}

The table below compares the three principal formulations and their symmetric and metric restrictions. The displayed frame symmetry assumes \ref{postulate1} and is generally reduced in modified models.
\[
\boxed{
\begin{array}{c|c|c|c}
  \text{symmetries}
 & \textbf{Metric-affine parallelism}
 & \textbf{Absolute parallelism}
 & \textbf{Minim. cov. parall.}
\\[2mm]
\hline
\begin{array}{l}
\text{Gauge: GL$\times$Diff}\\
\text{Frame: \underline{GL} if \ref{postulate1}}
\end{array}
&
\begin{array}{c}
(g^{ab},\bbe^a,\bA_a{}^b,\br^a{}_b)\quad \text{\cite{Hohmann:2021ast}} \\
\text{flatness imposed by EoM: } \\
\delta \bL/\delta\br^a{}_b \sim \bR_a{}^b\approx 0
\end{array}
&
\begin{array}{c}
(g^{ab},\bbe^a,\Lambda^a{}_b)  \\
\text{flat by construction:} \\
\bR_a{}^b = 0\,\, \Leftarrow\,\, \bA_a{}^b=\iL_a{}^c\bdiff\aL^b{}_c 
\end{array}
&
\begin{array}{c}
(g^{ab},\bbe^a) \quad \text{\cite{BeltranJimenez:2019odq}} \\
\text{Weitzenb\"ock gauge-fixed.} \\
\text{G$_\parallel$R}
\end{array}
\\[6mm]
\hline
\begin{array}{l}
\text{Gauge: GL$_\parallel\times$Diff}\\
\text{Frame: \underline{GL$_\parallel$} if \ref{postulate1}}
\end{array}
&
\begin{array}{c}
(g^{ab},\bbe^a,\bA_a{}^b,\br^a{}_b,\blambda_a) \quad \text{\cite{Adak:2005cd}} \\
\text{symmetry imposed by EoM:} \\
\delta\bL/\delta\blambda_a \sim \bS^a \approx 0
\end{array}
&
\begin{array}{c}
(g^{ab},\bbe^a, \xi^a)\\
\text{symmetric by construction}: \\
\bS^a=0\,\, \Leftarrow\,\, \Lambda^a{}_b=\bbie_b\lrcorner\bdiff\xi^a
\end{array}
&
\begin{array}{c}
(g^{ab},\xi^a) \quad \text{\cite{Koivisto:2018aip}} \\
\text{Diffs Kretschmannized.} \\
\text{$\xi^a=x^a \Rightarrow$ 1916 GR}
\end{array}
\\[6mm]
\hline
\begin{array}{l}
\text{Gauge: SO$\times$Diff}\\
\text{Frame: \underline{SO} if \ref{postulate1}}
\end{array}
&
\begin{array}{c}
(g^{ab},\bbe^a,\bA^a{}_b,\br^a{}_b, \blambda_{ab}) \quad \text{\cite{Obukhov:2002tm}} \\
\text{metricity imposed by EoM:} \\
\delta\bL/\delta\blambda_{ab} \sim \bQ^{ab} \approx 0 
\end{array}
&
\begin{array}{c}
(g^{ab},\bbe^a, L^a{}_b) \\
\text{metric by construction:} \\
\bQ^{ab} = 0\,\, \Leftarrow\,\, g_{ab}=L^c{}_a L^d{}_b\eta_{cd} 
\end{array}
&
\begin{array}{c}
(\bbe^a,L^a{}_b) \quad \text{\cite{Golovnev:2017dox}} \\
\text{Lorentz St\"uckelberged.} \\
L^a{}_b=\delta^a_b \Rightarrow \text{1928 GR}
\end{array}
\\[6mm]
\end{array}
}
\]

Alternative perspectives on parallel gravity are presented in the reviews \cite{Maluf:2023rwe,Krssak:2018ywd,Bahamonde:2021gfp}. 
In particular, Maluf {\it et al.} \cite{Maluf:2023rwe} have established an extensive Hamiltonian programme of their own, addressing gravitational energy–momentum, angular momentum and their reference-frame interpretation in the original parallel formulation.
Kr\v{s}\v{s}\'ak {\it et al.} \cite{Krssak:2018ywd} emphasise the role of the inertial spin connection in separating inertial effects and maintaining local Lorentz covariance.
Bahamonde {\it et al.} \cite{Bahamonde:2021gfp} provide a broad review, including extensive accounts of popular modified teleparallel gravity models with main motivation in cosmological applications.
They survey the general, symmetric and metric sectors of metric-affine parallelism and display their GR-equivalent actions. For metric parallelism, they identify an integrated flat-connection parametrisation $L^a{}_b$, although the corresponding absolute variational principle $I[\bbe,L]$ is not developed. The tetrad formulations considered by Kr\v{s}\v{s}\'ak {\it et al.} and Maluf {\it et al.} are restricted to the metric-parallel sector from the outset. Neither engages with the frame-variational problem: Kr\v{s}\v{s}\'ak {\it et al.} use a locally Lorentz-covariant bulk formulation $I[\bbe,\bomega]$ which does not retain the inertial connection as variational data, whereas Maluf {\it et al.} employ the pure-tetrad formulation $I[\bbe]$, which locally corresponds, in the variables used here, to the gauge choice $L^a{}_b=\delta^a_b$ and recovers the original formulation of Ref.~\cite{Einstein:1928tp}. Collectively, all these works discuss energy, frame dependence, and Noether and Hamiltonian constructions, but address different reference-structure questions from the frame-variational one considered here. Thus, these reviews are complementary to the present article. They do not mention the canonical frame in parallel gravity. 

\section{Surface charges}
\label{surfacecharges}

We state the result (Sec. \ref{mainresult}) and discuss its general (Sec. \ref{expressions}) and G$_\parallel$R-specific (Sec. \ref{implications}) implications. The derivation is in Appendix \ref{derivations}. 

\subsection{The main result}
\label{mainresult}

The charge $P$ associated with a vector field $\xi$ is $P[\xi]
=
-\oint
\left[
(\xi\lrcorner \bbe^a)\bbs_a
+
(\xi\lrcorner \bpsi)\wedge\bPi
\right]$. 
Equivalently, for the canonical frame direction $\bbie_a$,
\begin{equation} \label{charge}
\boxed{
P_a
=
-\oint \bj_a\,,
\qquad
\bj_a
=
\bbs_a
+
(\bbie_a\lrcorner\bpsi)\wedge\bPi\,.
}
\end{equation}
The first term is the gravitational contribution, while the second is the direct matter contribution coming from the symplectic potential $\bTheta_M$. The integration $\oint$ is over a closed $(n-2)$-surface $\mS$\footnote{We leave this implicit in $\oint = \oint_{\mS}$ and $P[\xi]=P[\xi, \mS]$, to ease notation.}. 

The GL Noether charge, associated with the symmetry under transformation parameterised by a GL matrix $M_a{}^b$, has the simple form
\be
\boxed{
P[M] = -\oint M_a{}^b\br^a{}_b\,. 
} \label{GLcharge}
\ee
Alternatively, $P^a{}_b = -\oint \br^a{}_b$. This charge also has a physical interpretation, as we shall clarify in the following.

\subsubsection{Boundary data and Hamiltonian integrability}

The integrals in Eqs.~(\ref{charge}) and (\ref{GLcharge}) are Noether charges defined on an arbitrary
closed codimension-two surface \(\mathcal S\). 
They acquire a
Hamiltonian interpretation once \(\mS\) is regarded as the corner of
a finite subsystem and boundary conditions are imposed on the
worldtube \(\Gamma_{\mS}\) swept out by it.

This follows directly from the variation (\ref{variations}). On shell, the variation
of the action reduces to the boundary term determined by the
presymplectic potential (\ref{symplectic}), $\delta I \approx \int_{\Gamma_{\mS}}\iota^*_{\Gamma_{\mS}}\bTheta$. 
We therefore require the pullback of
\(\bTheta\) to vanish for all allowed variations at
\(\Gamma_{\mS}\). This is the condition for a well-posed
spatial-boundary variational problem in the sense of
Ref.~\cite{Harlow:2019yfa}. We may choose the Dirichlet polarisation in
which the pullbacks of the geometric fields, the parallel reference
structure and the matter potential are fixed at
\(\Gamma_{\mathcal S}\). Sufficient choices in the two formulations, respectively, are
\bs
\label{dirichlet}
\ba 
\text{Metric-affine parallelism}: &\quad&
\left.
\delta(g^{ab},\bbe^a,\bA_a{}^b,\bpsi)
\right|_{\Gamma_{\mathcal S}}=0\,, \\
\text{Absolute parallelism}: &\quad&
\left.
\delta(g^{ab},\bbe^a,\aL_a{}^b,\bpsi)
\right|_{\Gamma_{\mathcal S}}=0\,.
\ea
\es
For gauge matter $\bpsi$, this
polarisation includes the boundary value of the gauge potential, as
will be seen explicitly in the charged black-hole examples of Section \ref{applications}. The symmetry transformation must also preserve the chosen boundary conditions. For a CDiff this means, in
particular, that the pullbacks of
\be
\delta_\Xi g^{ab}=-\Xi\lrcorner \bQ^{ab}\,,\qquad
\delta_\Xi \bbe^a=\bDiff(\Xi\lrcorner \bbe^a)+\Xi\lrcorner \bS^a\,,
\ee
and of the corresponding matter variation preserve the fixed boundary
data on \(\Gamma_S\).

A symmetry transformation $\xi$
which preserves these boundary data and is allowed to act non-trivially at
the corner then has a Hamiltonian generator $H_\xi$, provided the corresponding
surface variation is integrable on the chosen phase space. The variation of $H_\xi$ is given 
by the presymplectic form $\Omega$ by \cite{Lee:1990nz,Iyer:1994ys,Harlow:2019yfa}
\be
\delta H_\xi = \Omega(\delta,\delta_\xi) = \oint \lp \xi\lrcorner\bTheta-\delta \bj_\xi\rp\,.
\ee 
For the CDiff charge (\ref{charge}), the Dirichlet polarisation (\ref{dirichlet}) sets
\(\xi\lrcorner\bTheta=0\) on \(\mS\), for fixed corner vector field \(\xi\).
Thus the surface variation is exact, $\delta H_\xi = -\delta \oint_S \bj_\xi$,
and hence (\ref{charge}) is already the integrated Hamiltonian charge in this
Dirichlet sector\footnote{Integrability would have to be reconsidered for
field-dependent or varied symmetry parameters, moving corners, non-vanishing
symplectic flux, or mixed/Neumann boundary conditions \cite{Wald:1999wa,Chandrasekaran:2021vyu,Ciambelli:2024vhy}. 
A parameter-dependent boundary frame can lead to apparent violation of the first law \cite{Awad:2007me}. 
In the present case, although a canonical frame is determined by the bulk fields, its
pullback to \(\Gamma_{\mS}\) is part of the fixed Dirichlet boundary data.
The corner displacement is therefore kept fixed when the charge is varied.}. 
Thus, up to a fixed additive reference, $H_\xi= P[\xi]$, and we shall
identify the Hamiltonian surface energy $E \equiv H_{\bbie_0}
 = P_0$, while $P_i$ are the spatial momenta.

For the GL charge the same conclusion follows directly from the connection
part of the presymplectic potential. In the metric-affine formulation,
Eq.~(\ref{symplectic}) contains $\bTheta_A=\delta \bA_a{}^b\wedge \br^a{}_b$,
so \(\br^a{}_b\) is the momentum conjugate to the independent connection.
With \(\bA_a{}^b|_{\Gamma_\mS}\) fixed, and with a field-independent GL
parameter \(M_a{}^b\) retained at the corner, the surface variation is
\be
        \delta H_M
        =
        -\delta\oint_S M_a{}^b \br^a{}_b\,.
\ee
Therefore the integrated Hamiltonian corner generator $H_M = P[M]$ is precisely the GL Noether charge (\ref{GLcharge}).

\subsection{The Diff and GL sectors}
\label{expressions}

We now explain how the two surface potentials arise in the metric-affine
and absolute-parallel formulations, and why the (C)Diff charge and
the GL curvature-momentum charge have different status in the two
formulations. 

\subsubsection{The matter-completed (C)Diff charge}

The result (\ref{charge}) is most transparent in absolute teleparallelism. 
The covariant result (\ref{charge}) follows directly and unambiguously, as both the Diff (\ref{gaugeDiff}) and CDiff  (\ref{gaugeCDiff}) give the same
surface potential and the same Noether identity.
In the metric-affine formulation, the ordinary Diff gives the surface
potential (\ref{charge}) only up to the representation-dependent additional term
$(\xi\!\lrcorner \bA_a{}^b)\br^a{}_b$.
This term involves both the chosen representative of the flat
connection and the multiplier imposing flatness. The term is removed either
in the Weitzenböck gauge, \(\bA_a{}^b=0\), or more invariantly by using
the CDiff for which the tangent-space transport combined with
the spacetime drag eliminates the ambiguous contribution. Note that the Diff is 
the CDiff in the Weitzenb{\"o}ck gauge. 

The appearance of a matter surface term in (\ref{charge}) is natural from the
covariant phase-space viewpoint. In coupled Einstein--Maxwell systems
the Noether charge contains gravitational and electromagnetic
surface contributions \cite{Frodden:2017suc,Elgood:2020svt,Prabhu:2015vua}.
This completion has not, to our knowledge, been isolated in teleparallel applications.
Investigations of the Noether structure in metric teleparallelism
\cite{deAndrade:2000kr,Itin:2001tv,Obukhov:2002tm,Obukhov:2006sk,Capozziello:2018qcp,Hammad:2019oyb,Emtsova:2019moq}, 
symmetric teleparallelism \cite{BeltranJimenez:2021edg,Gomes:2022vrc,
Heisenberg:2022gqk,Emtsova:2023mwd,Emtsova:2025,Capozziello:2026rrc} and general teleparallelism
\cite{BeltranJimenez:2021edg,Gomes:2022vrc} have mostly been confined to set-ups for
which the gravitational superpotential is sufficient. Likewise,  
gravitational Hamiltonian prescriptions \cite{Nester:1989TeleHamiltonian,Maluf:2000ag,Maluf:2002zc}, 
or definitions read directly from the gravitational field equation \cite{deAndrade:2000kr,Itin:2001tv}, 
omit the symplectic matter contribution. Dirac--Bergmann Hamiltonian analyses of teleparallel theories have addressed
the complementary problem of constructing a bulk Hamiltonian,
constraints and degree-of-freedom count of the gravitational sector \cite{Maluf:1994ji,Blagojevic:2000qs,Maluf:2000ag,Okolow:2011nq,Blixt:2020ekl,DAmbrosio:2023asf,Tomonari:2023wcs,Bajardi:2024xxx,Chen:2025mhu,Capozziello:2026ccp}.

\subsubsection{The GL curvature-momentum charge}

The ordinary GL transformation displays the distinction between
metric-affine and absolute parallelism from another angle. In absolute
parallelism the GL transformation is a trivial Noether symmetry: the
associated surface charge vanishes. In the metric-affine formulation,
by contrast, the connection is an independent field and its conjugate
momentum \(\br^a{}_b\) appears as the GL Noether potential. The expression
(\ref{GLcharge}) is therefore the tangent-space curvature-momentum charge of the
metric-affine formulation.

This is the first-order counterpart of the curvature momentum that enters
the Iyer--Wald entropy formula. In tensor notation that object is
\(\partial L/\partial R_{\mu\nu\rho\sigma}\), contracted with the horizon
binormal. In the present differential-form notation the corresponding
object is \(\br^a{}_b\), and the GL parameter \(M_a{}^b\) selects the
tangent-space transformation whose momentum is being measured. The
normal-plane boost is the particular projection relevant for a horizon
cross-section. 

The charge formula (\ref{GLcharge}) itself is general,
but \(\br^a{}_b\) must be determined from the connection field
equation. 
The present parallel-geometric restriction
is a special case in which the curvature momentum is the multiplier
imposing \(\bR^a{}_b=0\). Below we evaluate it in G$_\parallel$R and show that its
normal-plane projection is the horizon area charge\footnote{In absolute parallelism this charge is absent. Quantum corrections may, however, generate curvature-dependent terms carrying a momentum of the same type as $\br^{a}{}_{b}$, suggesting a dynamical origin of the multiplier charge. This provides a possible link to effective and induced-gravity accounts of horizon entropy \cite{Jacobson:1994iw,Frolov:1997up,Solodukhin:2011gn}.}.

\subsubsection{Isolated \underline{GL} frame shifts}

Finally, consider the isolated frame GL shift. In metric-affine
parallelism the transformation (\ref{frameGL1})
is not an off-shell symmetry of the action. It is only the connection part
of the ordinary GL gauge transformation. Consequently, it gives neither
an independent Noether identity nor an independent surface charge. The
ordinary GL gauge symmetry is recovered by combining 
transformations (\ref{frameGL1}) with (\ref{frameGL2}).

The situation is slightly different in absolute parallelism. There the
flat connection is generated by the matrix \(\Lambda^a{}_b\), and the
corresponding frame variation is well-defined within the
space of parallel configurations, provided \ref{postulate1}. However, it still does not generate
an additional Noether charge. The variation changes the Lagrangian by a
boundary term which is equal to the symplectic contribution:
\be
    \bTheta_m =m_a{}^b\bmE^a{}_b\,, \quad 
    {\delta}_m \bL=\bdiff \bbell_m \quad \text{where} \quad
    \quad
    \bbell_m=m_a{}^b\bmE^a{}_b \qquad \Rightarrow \quad  \bJ_m=\bTheta_m-\bbell_m=0\,. 
\ee
Hence the associated Noether current vanishes identically, and in this Noether sense, the isolated frame \underline{GL} is a trivial symmetry.

The computation is nevertheless useful. It identifies $\bmE^a{}_b$
as the momentum conjugate to tangent-space frame transport. 
\(\bmE^a{}_b\) is not a surface charge by itself, but the (pre)symplectic
response of the action to a change of parallel frame. Its geometrical
meaning depends on the constitutive law, but becomes transparent
in the case of G$_\parallel$R.

\subsection{Entropy from the normal-plane boost}
\label{implications}

In G$_{\parallel}$R, the connection conjugate $(n-1)$-form admits the special
geometrical representation
\be \label{relation}
        \bmE^a{}_b
        \equiv 2\bq^a{}_b-\bbe^a\wedge \bbs_b
        =
        \frac{\mu_n}{2}\bDiff\area^a{}_b\,,
        \quad \text{where} \quad 
        \area^a{}_b=\ast(\bbe^a\wedge\bbe_b)\,.
\ee
$\area^a{}_b$ is the tangent-space codimension-two area form. (The notation is motivated by the analogous structure in $\bF\wedge\area$-theories \cite{Krasnov:2020lku}.) $\bmE^a{}_b$ measures the covariant change of this area $(n-2)$-form under parallel transport. This interpretation is special to the G$_{\parallel}$R constitutive law.

To verify the relation (\ref{relation}) explicitly, we note that since \(\mathring{\bDiff}\area^a{}_b=0\), one has
\bs
\be
\bDiff\area^a{}_b
 =
\bX^a{}_c\wedge \area^c{}_b
-
\bX^c{}_b\wedge \area^a{}_c\,.
\ee
Using the elementary identity $\bbe^d\wedge \area^a{}_b
=
\delta^d_b\ast\bbe^a-g^{ad}g_{bc}\ast\bbe^c$\,, 
this becomes
\be
\bDiff\area^a{}_b
=
\left(
X^a{}_{bc}
+X_{c}{}^a{}_b
-g_{bc}X^a{}_{d}{}^d
-\delta^a_c X^d{}_{db}
\right)\ast\bbe^c\,. \label{toinen}
\ee
\es
On the other hand, the G$_\parallel$R constitutive relations (\ref{potentials}) give
\bs
\ba
2\bq^a{}_b
& = &
\frac{\mu_n}{2}g^{ad}
\Big[
Q^c{}_{db}
-2Q_{(db)}{}^c
+\delta^c_{(d}(Q_{b)}+2S_{b)})
- g_{db}\V^c
-2S_{(db)}{}^c
\Big]\ast\bbe_c \nn \\
& = &
\mu_n
\left[
g^{ad}X^c{}_{(db)}
-\frac12\delta^a_b X^c
-\frac12 g^{ac}X^d{}_{db}
-\frac12\delta^c_b X^d{}_d{}^a
\right]\ast\bbe_c\,, \\ 
-\bbe^a\wedge\bbs_b
& = &
\mu_n\left(
Q^{[ac]}{}_{b}
+\frac12 S_b{}^{ac}
-S^{[ac]}{}_{b}
+\delta_b^{[a}\V^{c]}
\right)\ast \bbe_c
=
\mu_n\left(
X^{[a}{}_{b}{}^{c]}
+ \delta_b^{[a}\V^{c]}
\right)\ast \bbe_c\,. 
\ea
\es
By comparing the sum of these two ($n$-1)-forms,  
\be
 \bmE^a{}_b
        = 2\bq^a{}_b-\bbe^a\wedge \bbs_b
        = \frac{\mu_n}{2}\lp X^a{}_{bc}
+X_{c}{}^a{}_b
-g_{bc}X^a{}_{d}{}^d
-\delta^a_c X^d{}_{db}
\right)\ast\bbe^c\,, 
\ee
with the expression (\ref{toinen}), one may verify the relation (\ref{relation}). 

We may then explicitly compute the GL charge (\ref{GLcharge}). 
The curvature momentum \(\br^a{}_b\), which is the boundary momentum
conjugate to the independent metric-affine connection, is fixed
by the connection equation, $\hat{\bE}^a{}_b \approx 0$, as follows,
\be
\bDiff\br^a{}_b
=
-\frac{\mu_n}{2}\bDiff\area^a{}_b 
\qquad\Rightarrow\qquad
\br^a{}_b
=
-\frac{\mu_n}{2}\area^a{}_b+\bY^a{}_b\,,
\qquad
\bDiff\bY^a{}_b=0\,.
\ee
The covariantly constant homogeneous term corresponds to the
reference ambiguity in the connection momentum.  
The zero-reference choice $\bY^a{}_b=0$ is appropriate
to the applications below, as a non-zero \(\bY^a{}_b\) would assign an additional reference contribution to the geometric area term.

Consider the GL transformation which acts as a boost in the normal plane of a codimension-two surface.
For concreteness, consider a horizon cross-section \(\mathcal H\). The horizon binormal $\epsilon^\mu{}_\nu$ can be represented in a canonical frame as
$\epsilon^a{}_b = \e^a{}_\mu \epsilon^\mu{}_\nu\ie_b{}^\nu $. 
The normal-plane boost charge is therefore given by the formula (\ref{GLcharge}) with $M_a{}^b=\epsilon^b{}_a$ and the closed integration surface chosen as \(\mathcal H\).
We choose the orientation such that 
\be \label{binormal}
    \frac12\,\epsilon^b{}_a\,\iota_{\mathcal H}^*\area^a{}_b
    = \bdiff\bA_{\mathcal H}\,,
\ee
where \(\bdiff\bA_{\mathcal H}\) is the induced $(n-2)$-dimensional area element and
\(A_{\mathcal H}=\oint_{\mathcal H}\bdiff\bA_{\mathcal H}\).
This fixes the orientation convention for the unit boost generator.
(In the case that the canonical frame is adapted to the horizon we can simply write $\bdiff \bA_{\mathcal H} = \iota_{\mathcal H}^*\area^0{}_1$.) 
We obtain $P[M=\epsilon] \equiv P_{\rm boost}=-\oint\epsilon^b{}_a\br^a{}_b = \mu_nA_{\mathcal H}$.  
The Hamiltonian meaning of this projection can be seen directly from
the connection contribution to the presymplectic potential in
Eq.~(\ref{symplectic}), $\bTheta_A=\delta \bA_a{}^b\wedge \br^a{}_b$.

Upon Euclidean continuation of the normal two-plane, the unit
Lorentzian boost generator becomes the rotation generator of the
normal disk. Consider a small Euclidean tube $\Gamma_\epsilon \simeq \mS^1_\epsilon\times \mH$ 
surrounding the horizon cross-section. In the normal-plane sector we
write the connection as
\be
\left.\bA_a{}^b\right|_{\perp}
=
\epsilon^b{}_a\,\bba_\perp\,,
\ee
where the holonomy of \(\bba_\perp\) around the contractible normal
circle is the Euclidean opening angle,
\be
\vartheta
=
\oint_{\mS^1_\epsilon} \bba_\perp\,.
\ee
For the Euclidean entropy interpretation we pass from the Dirichlet polarisation to the normal-plane corner polarisation. 
Restricting the boundary variation to this normal-plane sector, with the unit internal binormal $\epsilon^a{}_b$ held fixed, gives
\be
\int_{\Gamma_\epsilon}\bTheta_A^{\perp}
=
\int_{\mS^1_\epsilon}\delta \bba_\perp
\oint_{\mH} \epsilon^b{}_a \br^a{}_b
=
- P_{\rm boost}\,\delta\vartheta .
\ee
Thus the boost charge $P_{\rm boost}=\mu_n A_{\mH}$ is the momentum conjugate to the
Euclidean opening angle. 
For a smooth Euclidean horizon the normal disk closes without a
conical defect, fixing $\vartheta=2\pi$. The corresponding entropy is therefore
\be
    S = 2\pi P_{\rm boost} =  
     2\pi \mu_n A_{\mathcal H}\,. \label{entropy}
\ee
Thus the usual Iyer--Wald area term is recovered here as a tangent-space GL charge. 
The conjugacy of the Euclidean opening angle and the horizon area is a standard off-shell black-hole relation \cite{Carlip:1993sa}.
A Lorentzian counterpart of the area--angle conjugacy
was derived in Ref.~\cite{Chandrasekaran:2023vzb}, where the
area element of two intersecting null horizons is canonically
conjugate to their relative boost angle and generates its shifts
under the Poisson bracket.

The reduction to the area term here follows from the pullback to the codimension-two surface and the projection onto its normal-plane binormal. It requires neither spherical symmetry nor a Killing realisation of the boost. Its thermodynamic interpretation, however, is more specific: at a smooth horizon corner, the contractible Euclidean normal circle fixes the opening angle conjugate to this boost charge. Although the full GL charge has $n^2$ components,
the pullback to the corner selects only the normal-plane boost projection\footnote{On a codimension-two corner
$\mathcal S$, the pullback of $\area^a{}_b$ is a simple binormal
area bivector. Hence all symmetric GL projections vanish, and,
in a frame adapted to $\mathcal S$, also the tangential and mixed
Lorentz projections vanish. The only non-vanishing local projection
is the boost in the normal two-plane.}. In 
spherical symmetry this is particularly
transparent, as the constant Cartesian boost components integrate to zero over the
sphere, while the local radial boost is proportional to the entropy.

\subsubsection*{Relation to previous constructions}

The relevant structural precedents for the present boost-charge construction are found in the curvature-based first-order and isolated-horizon frameworks discussed below\footnote{Among previous teleparallel treatments of black-hole entropy,
Hammad {\it et al.} derive a bare-Diff Noether charge and infer the entropy
from the first law \cite{Hammad:2019oyb}; Heisenberg {\it et al.} adapt the
Iyer--Wald construction to symmetric teleparallelism \cite{Heisenberg:2022gqk}; in the teleparallel
sector of Poincar\'e gauge theory, Blagojevi\'c and Cvetkovi\'c identify black-hole entropy with the horizon canonical charge required for differentiability of the generator \cite{Blagojevic:2019gsd,Blagojevic:2019bqg}; teleparallel Euclidean-action treatments include \cite{BeltranJimenez:2018vdo,Fiorini:2023axr,Kuntz:2024qgs,Stano:2025eje}. 
}.

Jacobson and Mohd have observed that, in an orthonormal frame
formalism, black-hole entropy is not naturally the Noether charge of a
bare Diff \cite{Jacobson:2015uqa}. A regular frame at the
bifurcation surface is not invariant under the horizon Killing flow
unless the Diff is supplemented by a compensating internal
Lorentz transformation. The entropy is then associated with this
Lorentz-Diff. In the present metric-affine parallel setting,
the spacetime drag and the tangent-space transport are separated from the outset, and the
curvature momentum \(\br^{a}{}_{b}\) is exposed directly as the Noether
potential of the GL sector. Whilst a Lorentz corner phase space
has been constructed for orthonormal tetrad gravity
\cite{Freidel:2020svx}, the charge (\ref{GLcharge}) and its normal-plane 
projection leading to (\ref{entropy}) provide the starting point for the corresponding
GL corner phase-space analysis in parallel geometry.

A closely related Lorentzian construction appears in the isolated-horizon phase space. 
Weak-isolated-horizon boundary conditions leave a horizon-preserving internal Lorentz boost at the boundary, 
and the Hamiltonian charge is proportional to the horizon area \cite{ChatterjeeGhosh2018}. With the boost parameter normalised to unity,
this yields the direct analogue of $P_{\rm boost}$ although it follows from Lorentzian horizon boundary conditions 
rather than from the Euclidean corner polarisation.

A different normalisation of this boost flow gives a local-observer energy.
In particular, stationary near-horizon observers 
at proper distance \(l_0\) define a local horizon energy $E_{\rm loc}$ via translations corresponding to boosts
at rate $1/l_0$ such that 
\(E_{\rm loc}=P_{\rm boost}/l_0=m_P^2 A_{\mH}/ l_0\), and the entropy at fixed $l_0$ is recovered by
using the corresponding local Unruh/Hawking temperature $T_{\mathrm{loc}}=(2\pi l_0)^{-1}$ via 
$\bdiff E_{\mathrm{loc}}/T_{\mathrm{loc}} =2\pi\bdiff P_{\mathrm{boost}} =\bdiff S$
\cite{FroddenGhoshPerez2013,ChatterjeeGhosh2018,Bianchi:2012vp}. 
An analogous distinction occurs in Diff-Noether constructions, wherein entropy uses the horizon generator normalised 
to unit surface gravity, whereas energy requires a choice of time-flow vector \cite{Wald:1993nt,Hayward:2000fi}.
In the present Euclidean corner approach, these roles are disentangled geometrically,
as smoothness fixes the opening angle to $2\pi$, yielding $S=2\pi P_{\mathrm{boost}}$ in a canonical frame, without an additional choice of a local 
observer or a proper-time normalisation.

\section{Applications}
\label{applications}

In the remainder of the article, we specialise the general \(n\)-dimensional
charge formulae to four spacetime dimensions, $n=4$. Thus, the horizon
cross-sections are two-dimensional, \(\mu_4=m_P^2\), and GL stands for 
GL(4).

\subsection{Spherical symmetry}

Consider the generic static spherically symmetric Ansatz for the metric 
in terms of the two functions $f(r)$ and $\sigma(r)$ of the radial coordinate $r$,
\bs
\label{metric}
\be \label{sssmetric}
\bdiff s^2 = -f^2\bdiff T^2 + \frac{\sigma^2\bdiff r^2}{f^2} + r^2\lp \bdiff\theta^2 + \sin^2\theta\bdiff\phi^2\rp\,. 
\ee
Transforming to Cartesian Kerr--Schild-type coordinates whose time coordinate $t$ is obtained from
$\bdiff t=\bdiff T+(\sigma/f^2-1/\sigma)\bdiff r$, this becomes
\be
\bdiff s^2
=
\eta_{\mu\nu}\bdiff x^\mu \bdiff x^\nu
+
\lp 1 -\sigma^2\rp\bdiff t^2 + \lp 1 - \frac{f^2}{\sigma^2}\rp\lp \sigma\bdiff t + \bdiff r\rp^2\,,
\ee 
\es
where the middle $\bdiff t^2$ term signals that the metric is not within the Kerr-Schild class when $\sigma\neq 1$.
A canonical representative satisfying postulate \ref{postulate2}  for the metric (\ref{metric}) is\footnote{In Hohmann's classification in Ref.~\cite{Hohmann:2019fvf}, the pair (\ref{metric},\ref{CANONICAL}) corresponds, within the general flat spherically symmetric
metric-affine class, to $F_1=\sigma$, $F_2=\lambda$, $F_5=r\lambda$, and $F_3=F_4=F_6=0$. \ref{postulate2} further determines $\lambda$.  The metric-compatible specialisation of the general flat spherically symmetric class was analysed in Ref.~\cite{Coley:2024tqe}, and the symmetric specialisation in Ref.~\cite{Hohmann:2024phz}.}
\bs
\label{CANONICAL}
\ba
\bbe^0
=& \sigma\bdiff t
=&
\sigma\bdiff T+
\lp\sigma^2/f^2-1\rp\bdiff r \,, \label{canonical0}
\\
\bbe^1
=& \lambda\bdiff x
=&
\lambda\left(
\sin\theta\cos\phi\bdiff r
+r\cos\theta\cos\phi\bdiff \theta
-r\sin\theta\sin\phi\bdiff\phi
\right) \,, 
\\
\bbe^2
=& \lambda\bdiff y
=&
\lambda\left(
\sin\theta\sin\phi\bdiff r
+r\cos\theta\sin\phi\bdiff \theta
+r\sin\theta\cos\phi\bdiff \phi
\right), 
\\
\bbe^3
=& \lambda\bdiff z
=&
\lambda\left(
\cos\theta\bdiff r-r\sin\theta\bdiff\theta
\right)\,, 
\ea
\es
where an integral of the metric functions gives the spatial scale factor $\lambda(r)$, 
\be \label{s}
\frac{\lambda'}{\lambda}
=
\frac{(f^2-\sigma^2)\sigma'}
     {(f^2+2rff'-\sigma^2)\sigma} \quad \Rightarrow \quad
\lambda = \lambda_0\exp\lb \int\frac{\lp f^2-\sigma^2\rp \bdiff\sigma}{\lp f^2+2ff'r-\sigma^2\rp\sigma}\rb\,. 
\ee
The solution determines $\lambda$ up to a nonzero multiplicative
constant $\lambda_0$.  
For the asymptotically flat solutions considered below we
fix this residual spatial scale by $\lambda(\infty)=1$. The temporal
leg of the frame (\ref{CANONICAL}) is chosen to be the future-directed, geometrically
defined Kodama vector $K$ \cite{Kodama:1979vn,Hayward1996,AbreuVisser2010}, 
$K = \bbie_0=\sigma^{-1}\partial_T$, so that
$K\rightarrow\partial_T$ as $r\rightarrow\infty$. These choices
specify the canonical representative\footnote{The canonical condition fixes a frame orbit, leaving a rigid affine freedom
corresponding to a global change of tangent-space basis. The overall normalisation of the time leg is a choice of time unit rather than a 
local physical ambiguity. Our global normalisation of the time leg $\bbie_0=K$ was fixed by the asymptotics, and any other choice would
renormalise the conventional charges. A solution-dependent replacement like \(\bbie_0\to\sigma_{+}K\)  
would require corrections to the Hamiltonian variations, due to $\delta\sigma_+  \neq 0$,
whereas a local rescaling like \(\bbie_0\to\sigma(r)K=\partial_T\) is not part of the residual freedom.}.

The temporal lapse 
factor $\sigma$ introduces anholonomy of this frame,
\bs
\label{canonicalS}
\ba
\bS^0 & = & \bdiff\sigma\wedge\bdiff t\,, \\
\bS^i & = &  \frac{\lp f^2-\sigma^2\rp \lambda}{\lp f^2+2ff'r-\sigma^2\rp\sigma}\bdiff\sigma\wedge\bdiff x^i\,. 
\ea
\es
The corresponding nonmetricity 1-forms are compactly expressed in terms of
unit vectors \(n_i=x_i/r\) and
the function \(h=2-f^2+(\sigma^2-1)/f^2\),
\bs
\label{canonicalQ}
\ba
\bQ_{00} & = &
-\bdiff \left(\frac{h}{\sigma^2}\right)\,, \\ 
\bQ_{0i}
& = &
\bdiff\left(\frac{1-h}{\sigma \lambda}\,n_i\right)\,, \\
\bQ_{ij}
& = &
\lp \delta_{ij}-n_in_j\rp\bdiff \left(\frac{1}{\lambda^2}\right)
+
n_i n_j\bdiff\left(\frac{h}{\lambda^2}\right)
+
\frac{h-1}{\lambda^2}\bdiff\lp n_i n_j\rp\,. 
\ea
\es
This demonstrates how Riemannian and teleparallel geometries are linked in G$_\parallel$R: the torsion (\ref{canonicalS}) and the nonmetricity (\ref{canonicalQ}) of the canonical parallel-geometric sector are determined by the Ansatz for the Riemannian sector (\ref{metric}). 

Besides using our results for the energy and normal-plane boost charges, we shall apply the Euclidean boundary prescription of Ref.~\cite{BeltranJimenez:2024ufa}. The relevance of the vector $\V^a$ introduced at (\ref{vectors}) is that it determines the horizon potential $\termopot$ as
\bs
\label{F}
\be \label{FreeEnergy}
\termopot = \frac{m_P^2}{2\beta}\int\ast\bX\,.
\ee
Here $\beta$ is the Euclidean period conjugate to the boundary
restriction of the chosen generator $K$. The
thermodynamic interpretation of $\Omega$ depends on the boundary
conditions and will be made explicit in the Reissner--Nordstr\"om example
below. The integral is evaluated on the Euclidean continuation of the timelike boundary $\mS^1_\beta\times \mS^2_r$, with the orientation inherited from the outward radial unit normal $\boldsymbol{\nu} = (\sigma/f)\bdiff r$ to the sphere \(r=\mathrm{const.}\) on a static spatial slice, on which the induced metric is ${\gamma}{}_{ab}$, 
\be
\termopot = -\frac{m_P^2}{2\beta}\int \sqrt{\gamma}X^a\nu_a\bdiff^3 x =  -\frac{m_P^2}{\beta}\int \frac{r\sin\theta}{\sigma}
\left(
rff'+f^2-\sigma^2
\right)\bdiff T\bdiff\theta\bdiff\phi = - \frac{4\pi m_P^2 r }{\sigma^2} \left(
rff'+f^2-\sigma^2
\right)\,. \label{freeformula}
\ee
\es
In the second step, we noted from Appendix \ref{localcurrents} that $X^i
=
\frac{2\lambda}{r\sigma^2}
(
rff'+f^2-\sigma^2)n^i$. 
In the third step, and in particular in the integration over the Euclidean $T$-coordinate, we took into account that its period is not $\beta$ but $\beta/\sigma$
on the constant-$r$ boundary.
At the horizon $r=r_+$ this becomes the normalisation factor \(\beta_T=\beta/\sigma_+\).
This rescaling, which stems from the relative lapse function associated with the $T$-coordinate with respect to the canonical frame time in (\ref{canonical0}), will be relevant for the dilatonic black hole which is characterised by $\sigma \neq 1$.  

The canonical frame for static spherical symmetry is summarised in Table \ref{kodamaframe}. 

\begin{table}[h]
\centering
\caption{The canonical Kodama frame representative. \label{kodamaframe}}
\renewcommand{\arraystretch}{1.25}
\begin{tabular}{p{0.28\linewidth} p{0.62\linewidth}}
\hline
Metric input &
\[
\bdiff s^2
=
-f^2\bdiff T^2+\frac{\sigma^2}{f^2}\bdiff r^2+r^2\bdiff\Omega^2
\]
\\

Canonical coframe &
\[
\bbe^0
=
\sigma \bdiff T+\frac{\sigma^2-f^2}{f^2}\bdiff r,
\qquad
\bbe^i_{\rm can}=\lambda\bdiff x^i
\]
\\

Canonical frame &
\[
\bbie_0
=
\frac{1}{\sigma}\,\partial_T,
\qquad
\bbie_i
=
\frac{1}{\lambda}\,\partial_i
-
\frac{\sigma^2-f^2}{\lambda\sigma f^2}\,n_i\,\partial_T
\]
\\

Frame algebra &
\[
[\bbie_0,\bbie_i]
=
\frac{\sigma'}{\lambda\sigma}\,n_i\,\bbie_0,
\qquad
[\bbie_i,\bbie_j]
=
\frac{\lambda'}{\lambda^2}
\left(n_j\bbie_i-n_i\bbie_j\right)
\]
\\

Horizon normalisation &
\[
\partial_T\big|_{r_+}
=
\sigma_+\,\bbie_0\big|_{r_+},
\qquad
\beta=\sigma_+\beta_T,
\qquad
\termopot=\frac{\termopot_T}{\sigma_+}
\]
\\
\hline
\end{tabular}
\end{table}

\subsection{The Reissner--Nordstr\"om black hole}
\label{RNexample}

Before turning on the dilaton, it is useful to consider the well-known Reissner--Nordstr\"om solution.  We take the matter Lagrangian to be
\begin{equation} \label{maxwellaction}
        \bL_M =-\frac{1}{8\pi}\bF\wedge \ast \bF\,,
        \qquad \bF=\bdiff \bA\,.
\end{equation}
We normalise the radial electric field by the charge $q$ such that 
\be \label{maxwell}
\bF = -\frac{q}{r^2}\bdiff t\wedge\bdiff r\,,
\ee
fix a sign convention by
\be \label{Echarge}
q \equiv -\frac{1}{4\pi}\oint\ast\bF = \frac{1}{4\pi}\oint  q \sin\theta\bdiff \theta\wedge\bdiff \phi\,,
\ee
and choose the electric potential
\begin{equation}
        \bA = \lp A_\infty-\frac{q}{r}\rp\bdiff t\,.
\end{equation}
The additive constant \(A_\infty\) is not fixed by the field strength.  This constant is irrelevant for the Maxwell equations, but it is relevant for the
symplectic potential and therefore has to be regarded as part of the boundary data.

The metric functions of the Reissner--Nordstr\"om solution are given by
\begin{equation}
        f^2=1-\frac{r_S}{r}+\frac{r_Q^2}{r^2}\,,
        \qquad
        \sigma^2=1\,, \quad \text{where}  \quad r_Q^2=\frac{q^2}{8\pi m_P^2}
        =\frac{q^2 r_S}{2m_S},
        \qquad
        m_S=4\pi m_P^2 r_S\,. 
\end{equation}
Since now \(\lambda'=0\), equation (\ref{s}) gives \(\lambda={\rm const.}\), and we set
\(\lambda=1\).  The canonical frame is therefore simply
\bs
\ba
        \bbe^0 & = &  \bdiff t
        = \bdiff T+\frac{1-f^2}{f^2}\bdiff r\,,
        \\
        \bbe^i & = & \bdiff x^i\,. 
\ea
\es
Thus the Reissner--Nordstr\"om geometry belongs to the Kerr--Schild class, and
the canonical frame is holonomic, $\bS^a=0$. 
The gravitational field is then carried by the nonmetricity. 
The nonmetricity 1-forms read, compactly in terms of $n^i=x^i/r$, 
\be
        \bQ_{00}
        =
        \bdiff f^2\,, 
        \quad
        \bQ_{0i}
        =
        \bdiff \left[
        \left(f^2-1\right)n_i
        \right]\,,
        \quad
        \bQ_{ij}
         =
        -n_i n_j\bdiff f^2
        +
        \left(1-f^2\right)\bdiff (n_i n_j)\,. 
\ee
The surface components of the gravitational potential are then obtained from the constitutive relation (\ref{spotential}),
\bs
\begin{align}
        \bbs_0
        &=
        -m_P^2 r\left(1-f^2\right)
        \sin\theta\bdiff \theta\wedge \bdiff\phi
        =
        m_P^2
        \left(
        r_S-\frac{r_Q^2}{r}
        \right)
        \bdiff\cos\theta\wedge \bdiff\phi ,
        \\
        \bbs_i
        &=
        m_P^2 r\left(1-f^2\right)
        n_i\bdiff\cos \theta\wedge \bdiff\phi+\cdots ,
\end{align}
\es
where the omitted terms vanish on a spherical surface \(\bdiff t=\bdiff r=0\).  
Hence
\bs
\begin{align}
       -\oint_{S^2_r}\bbs_0
        &=
        4\pi m_P^2
        \left(
        r_S-\frac{r_Q^2}{r}
        \right)
        =
        m_S-\frac{q^2}{2r}\,, \label{Fenergy}
        \\
       -\oint_{S^2_r}\bbs_i
        &=
        0\,.
\end{align}
\es
The result is the Misner--Sharp mass  \cite{MisnerSharp1964,Hayward1996} of the Reissner--Nordstr\"om geometry. 
It equals the asymptotic mass minus the Maxwell energy outside $S_r^2$.  

The Misner--Sharp mass, obtained here as the integral of the gravitational Freud superpotential $\bbs_0$, is not, in general,
the Hamiltonian energy of the coupled Einstein--Maxwell system.
Unless confined to a specific ensemble, the Maxwell symplectic potential supplies the boundary contribution
required by the variational problem, and hence also contributes to the Noether potential. 
In the present case, 
the relevant term is
\bs
\begin{equation} 
        -\oint\lp\bbie_a\lrcorner\bA\rp\bPi
        = 
        -\delta_a^0\oint\lp A_\infty - \frac{q}{r}\rp\frac{q}{4\pi} \sin\theta\bdiff \theta\wedge\bdiff \phi = \delta^0_a\lp\frac{q^2}{r} - qA_\infty\rp\,. \label{Menergy}
\end{equation}
The role of the additive constant is fixed by the boundary variation.
On the static worldtube
\(\Gamma_r\simeq \mI_\tau\times \mS_r^2\), generated by the canonical
time direction \(\xi\), the pullback of the Maxwell part of the
presymplectic potential gives
\be
\int_{\Gamma_r}\bTheta_M
=
\int_{I_\tau}
\left(
\oint_{\mS_r^2}\delta \bA\wedge\bPi
\right)
=
\int_{\mI_\tau}
q\,\delta\Phi(r)\bdiff\tau,
\qquad
\Phi(r)\equiv \bbie_0\lrcorner\bA\,.
\ee
\es
Thus \(\Phi(r)\) is the boundary potential conjugate to the
electric charge. In the fixed-potential polarisation its value is
held fixed, although the Hamiltonian surface charge retains the term
\(-q\Phi(r)\). With the canonical normalisation used here, $\Phi(r)=A_\infty-q/r$, i.e. asymptotically the fixed datum reduces 
to $\Phi(\infty)=A_\infty$.

Finally, the full Dirichlet Hamiltonian energy is  
\begin{equation} \label{RNenergy}
        E_{\rm RN} =   (\ref{Fenergy}) + (\ref{Menergy})    
        = m_S + \frac{q^2}{2r} - qA_\infty\,.
\end{equation} 
For the
asymptotically normalised gauge, $A_\infty=0$,
the electromagnetic contribution vanishes at infinity and the asymptotic
charge is $E_{\rm RN}(\infty;0)=m_S$.  In black-hole thermodynamics it is often more convenient to use instead the
horizon-regular gauge $A_t(r_+)=0$, which fixes $A_\infty=\frac{q}{r_+}$. In particular, the boundary condition $A_\infty=\frac{q}{r_+}$ makes the potential regular at the horizon \cite{Hawking:1995ap}. In this gauge the Maxwell boundary contribution vanishes at the horizon,
but the total canonical energy is shifted by the boundary datum relative
to the asymptotically normalised gauge.

\subsubsection{Horizon thermodynamics}
\label{horizonthermo}

We next examine the horizon variational problem and its Hamiltonian charges, whilst the asymptotic Hamiltonian remains the $r \rightarrow \infty$ limit of (\ref{RNenergy}).
Denote the two
Reissner--Nordstr\"om roots by\footnote{We assume $r_+>r_-$. The extremal geometry is obtained only as a limiting case, since the smooth Euclidean normal disk used below degenerates at extremality.}
\be
r_\pm=\frac12\left(r_S\pm\sqrt{r_S^2-4r_Q^2}\right)\,,
\qquad
r_S=r_++r_-\,,
\qquad
r_Q^2=r_+r_-=\frac{q^2}{8\pi m_P^2}\,.
\ee
The Euclidean boundary formula (\ref{freeformula}), specialised to \(\sigma=1\) and evaluated at $r_+$, gives
\be \label{kaavat1}
\termopot
=
-4\pi m_P^2 r
\left(rff'+f^2-1\right)\Big|_{r=r_+}
=
\frac{m_S}{2}\,.
\ee
At the outer horizon, the energy is given by the matter-completed charge (\ref{RNenergy}), $E(r_+) = E_+$. 
Euclidean regularity gives the temperature \cite{PhysRevD.15.2752}, while (\ref{entropy}) gives the entropy, 
\be \label{kaavat2}
E_+ 
= m_S+\frac{q^2}{2r_+}-qA_\infty\,, \qquad
\beta^{-1}
=
\frac{r_+-r_-}{4\pi r_+^2}\,,
\qquad
S=8\pi^2m_P^2r_+^2\,. 
\ee
Combining the results (\ref{kaavat2}) and (\ref{kaavat1}), one obtains
\be \label{statement}
E_+=\termopot+\beta^{-1}S-qA_+\,,
\ee
where $A_+=\Phi(r_+)$. 
Thus the horizon thermodynamic relation holds for arbitrary
\(A_\infty\). 
The thermodynamic interpretation nevertheless depends on which
electromagnetic variable is fixed at the boundary. We shall use the Maxwell work density and areal volume \cite{Hayward:1997jp}
\be
W_+\equiv\frac{q^2}{8\pi r_+^4}\,,
\qquad
V_+\equiv\frac{4\pi r_+^3}{3}\,.
\label{RNwork}
\ee
The horizon relations in (\ref{kaavat2}) then imply the identities
\be
\beta^{-1}\bdiff S+W_+\bdiff V_+
=4\pi m_P^2\bdiff r_+\,,
\qquad
2\beta^{-1}S+3W_+V_+
=4\pi m_P^2r_+\,.
\label{RNUSmarr}
\ee
They will be used below in the two boundary polarisations of the same
Reissner--Nordstr\"om geometry.
\newline\newline

\paragraph{\bf Fixed potential.} The Maxwell Lagrangian (\ref{maxwellaction}) generates the symplectic potential \(\bTheta_M=\delta\bA\wedge\bPi\), which is naturally adapted to the
Dirichlet polarisation. The horizon Hamiltonian is the
matter-completed charge in (\ref{kaavat2}), which we henceforth denote by $E_+ = E_+^{(A)}$ to emphasise the fixed-potential boundary condition. On the fixed-potential phase space, \(\delta A_+=0\), equations
(\ref{RNUSmarr}) give
\be 
\bdiff E_+^{(A)}
=\beta^{-1}\bdiff S+W_+\bdiff V_+-A_+\bdiff q,
\qquad
E_+^{(A)}
=2\beta^{-1}S+3W_+V_+-qA_+\,.
\label{RNgrandlaws}
\ee
Thus \(-A_+\) is the chemical potential conjugate to \(q\), and
(\ref{statement}) is the grand-canonical relation
\be
\termopot^{(A)}=E_+^{(A)}-\beta^{-1}S+qA_+
=4\pi m_P^2r_+-\beta^{-1}S
=\frac{m_S}{2}\,.
\label{RNgrandF}
\ee
The boundary-datum dependence of \(E_+^{(A)}\) is precisely cancelled
by the source term \(qA_+\). The first law and the Smarr relation (\ref{RNgrandlaws}) verify the consistent interpretation of $\termopot^{(A)}$ as the grand potential of the electric sector. 
\newline\newline

\paragraph{\bf Fixed charge.} Fixing the charge requires the Maxwell Legendre boundary term
\cite{Hawking:1995ap},
\be
\bL_M^{(q)}=\bL_M-\bdiff(\bA\wedge\bPi)\,,
\quad
\bTheta_M^{(q)}=-\bA\wedge\delta\bPi\,.
\label{RNLegendre}
\ee
The boundary condition is now \(\delta\bPi=0\), and hence
\(\delta q=0\). The same Legendre boundary term shifts the Noether
surface potential according to 
\be
\bj_{\xi}^{(q)}
= \bj_{\xi}^{(A)} - \xi\lrcorner(\bA\wedge\bPi)\,.
\ee
For the static radial electric configuration $\bPi=(q/4\pi)\sin\theta\bdiff\theta\wedge\bdiff\phi$
and the generator $\xi=\bbie_0$ without angular component,  
\(\xi\lrcorner\bPi=0\), and thus the induced change
of the Noether potential precisely cancels its
Dirichlet Maxwell term (\ref{Menergy}). The Hamiltonian energy of this polarisation is
therefore
\be
E_+^{(q)}
=-\oint_{S^2_+}\bbs_0
=m_S-\frac{q^2}{2r_+}
=E_+^{(A)}+qA_+
=4\pi m_P^2r_+\,.
\label{RNcanonicalenergy}
\ee
Its first law and Smarr relation are
\be
\bdiff E_+^{(q)}
=\beta^{-1}\bdiff S+W_+\bdiff V_+\,,
\qquad
E_+^{(q)}
=2\beta^{-1}S+3W_+V_+\,,
\qquad
\delta q=0\,,
\label{RNcanonicallaws}
\ee
and the corresponding Helmholtz free energy is
\be
\termopot^{(q)}=E_+^{(q)}-\beta^{-1}S
=\frac{m_S}{2}\,.
\label{RNHelmholtz}
\ee
The equality \(\termopot^{(q)}=\termopot^{(A)}\) is numerical: the actions, symplectic
potentials and allowed variations of the two ensembles remain
different. In either polarisation, differentiating the corresponding
Smarr relation and using its first law gives
\be
\beta^{-1}\bdiff S
+2S\,\bdiff(\beta^{-1})
+2W_+\bdiff V_+
+3V_+\bdiff W_+=0\,,
\label{RNGD}
\ee
with \(\delta A_+=0\) or \(\delta q=0\), respectively. 
At a general spherical corner the same Legendre relation gives
\begin{equation}
 E^{(q)}_{\rm RN}(r)
 =E^{(A)}_{\rm RN}(r;A_\infty)+q\Phi(r)
 =m_S-\frac{q^2}{2r}\,.
\end{equation}
It is useful to isolate the positive Coulomb energy outside the corner,
\begin{equation}
 U_{\rm ext}(r)\equiv\frac{q^2}{2r}
 =E^{(A)}_{\rm RN}(r;0)-m_S
 =m_S-E^{(q)}_{\rm RN}(r)>0\,.
\end{equation}
In the fixed-charge polarisation, $E^{(q)}_{\rm RN}=m_S-U_{\rm ext}$ is the Misner--Sharp
mass enclosed by the surface.  At the horizon, a displacement at fixed charge therefore obeys
\begin{equation}
 W_+\bdiff V_+=-\left.\bdiff U_{\rm ext}(r_+)\right|_q\,.
\end{equation}
In the fixed-potential polarisation, charge may instead be exchanged with
the boundary reservoir.  For $A_\infty=0$, the potential drop is
$\Delta\Phi=-A_+=q/r_+$, and hence
\begin{equation}
 U_{\rm ext}(r_+)=\frac12 q\Delta\Phi,\qquad
 -qA_+=q\Delta\Phi=2U_{\rm ext}(r_+),\qquad
 E^{(A)}_+=E^{(q)}_+-qA_+=m_S+U_{\rm ext}(r_+)\,.
\end{equation}
The factor of two has the familiar electrostatic origin\footnote{At fixed radius $r$ the Coulomb potential difference from infinity is
$\Delta\Phi(q)=q/r$. Since electrostatic potential difference is the work per unit charge, transferring charge $q$ across a
maintained $\Delta\Phi$ exchanges the work $q\Delta\Phi$ with the reservoir. By contrast, the Coulomb
field energy accumulated in charging the configuration from zero to $q$ is $U_{\rm ext}=\int_0^q\Delta\Phi(q')\bdiff q'
=\frac12 q\Delta\Phi$. Hence $q\Delta\Phi=2U_{\rm ext}$; see e.g.\ Chap.~9 of
Ref.~\cite{AnsermetBrechet}.}. 
\newline\newline

\paragraph{\bf Polarisation-matched horizon charge.} The two boundary polarisations define different variational problems and
different Hamiltonians. 
They can nevertheless be compared by matching the symmetry generator to the boundary data.
We therefore combine the CDiff (\ref{gaugeCDiff}) with the Maxwell U(1) transformation required
to match the generator to the boundary
data. For the Dirichlet Maxwell action, a time translation combined
with a U(1) gauge transformation \cite{Elgood:2020svt} acts as
\be
\delta_{\Xi,\alpha}\bA
=\delta_\Xi\bA + \bdiff\alpha\,,
\qquad
P_+^{(A)}[\Xi,\alpha]
=E_+^{(A)}-q\alpha\,.
\label{RNcombined}
\ee
The second relation follows because the combined matter surface
potential contains $(\Xi\lrcorner\bA+\alpha)\bPi$, with $\oint\bPi=q$.
In the fixed-potential sector, the horizon-adapted lift $\alpha_+=-A_+$ is field-independent on the
allowed phase space and defines an admissible reducibility parameter
\cite{Elgood:2020svt,Prabhu:2015vua}. 

The fixed-charge polarisation instead uses the
ordinary time translation,
$P_+^{(q)}[\Xi,0]=E_+^{(q)}$. We may therefore introduce the
polarisation-matched horizon charge
\be
\widehat E_+
\equiv
P_+^{(A)}[\Xi,-A_+]
=P_+^{(q)}[\Xi,0]
=E_+^{(A)}+qA_+
=E_+^{(q)}
=4\pi m_P^2r_+\,.
\label{RNcommonenergy}
\ee
This does not define a single generator on an unrestricted phase space, nor a replacement for the asymptotic charge from
(\ref{RNenergy}). It compares separately differentiable generators on the two restricted phase spaces, and 
the first law follows separately on those allowed phase spaces:
\begin{equation}
 \bdiff\widehat E_+
 =
 \begin{cases}
  \bdiff E_+^{(A)}+A_+\bdiff q
   =\beta^{-1}\bdiff S+W_+\bdiff V_+\,,
   & \bdiff A_+=0\,, \\[1mm]
  \bdiff E_+^{(q)}
   =\beta^{-1}\bdiff S+W_+\bdiff V_+\,,
   & \bdiff q=0\,.
 \end{cases}
\end{equation}
Thus both descriptions take the common algebraic form
\be
\widehat E_+
=2\beta^{-1}S+3W_+V_+\,,
\qquad
\termopot^{(A)}=\termopot^{(q)}
=\widehat E_+-\beta^{-1}S
=\frac{m_S}{2}\,.
\label{RNcommonlaws}
\ee
The construction is independent of the representative of the Maxwell
potential. Under $A_\infty\rightarrow A_\infty+c$, one has
$E_+^{(A)}\rightarrow E_+^{(A)}-qc$,
$A_+\rightarrow A_++c$, and
$\alpha_+\rightarrow\alpha_+-c$, so that
$P_+^{(A)}[\Xi,\alpha_+]$, and hence $\widehat E_+$, remains
unchanged. For the representative $A_+=0$, the lift
reduces to $\alpha_+=0$ and
$E_+^{(A)}=E_+^{(q)}=\widehat E_+$.

\subsection{Dilatonic charged black hole}
\label{dilatonExample}

The dilatonic solution provides a nontrivial test because the canonical Kodama generator has a radius-dependent normalisation relative to the static Killing generator.


We take the source Lagrangian to be the dilaton-Maxwell theory
\bs \label{solution}
\be
\bL_M = -\frac{\e^{-2\varphi/m_P}}{8\pi}\bF\wedge\ast\bF - \bdiff\varphi\wedge\ast\bdiff\varphi\,. 
\ee
The electric field, the electromagnetic momentum and charge are, respectively,
\be \label{maxwelldila}
\bF = -\frac{q\sigma}{r^2}\e^{2\varphi/m_P}\bdiff t\wedge\bdiff r\,, \quad
 \bPi = -\frac{1}{4\pi}e^{-2\varphi/m_P}\ast \bF\,,
 \quad
 q=\oint \bPi\,.
\ee
\es
The matter symplectic potential associated with (\ref{solution}) is $\bTheta_{M}^{(A)}
 =\delta \bA\wedge\bPi-2\delta\varphi\ast \bdiff\varphi$. 
The asymptotic scalar datum $\varphi_\infty$ is held
fixed. Although the scalar contributes to the symplectic potential,
it does not contribute directly to the diffeomorphism surface
potential, since $\bbie_a\lrcorner\varphi=0$ for the 0-form $\varphi$.
There is an exact solution with the dilaton profile \cite{GibbonsMaeda1988,GarfinkleHorowitzStrominger1991}
\be
e^{2\varphi/m_P} = e^{2\varphi_\infty /m_P}\lp 1 - \frac{2}{\sqrt{1+(r/r_q)^2}+1}\rp\,, \quad \text{where} \quad r_q =  e^{2\varphi_\infty/m_P}\frac{q^2}{2m_S}\,, \quad
m_S \equiv 4\pi m_P^2 r_S\,. 
\ee
The metric functions corresponding to this solution are
\bs \label{dilatonfunctions}
\ba
f^2 & = & 1- \frac{r_S}{r}\sqrt{1+(r_q/r)^2} + \frac{r_S r_q}{r^2}\,, \label{mf1} \\
\sigma^2 & = & \frac{r^2}{r^2 + r_q^2}\,.  \label{mf2} 
\ea
\es
Then, the scale factor $\lambda$ is deduced from (\ref{s}) as
\be
\lambda = \lp 1- \frac{r_q^2}{\lp r_S-r_q\rp R}\rp\sqrt{\frac{R-r_q}{R+r_q}}\,, \quad \text{where} \quad R \equiv \sqrt{r^2+r_q^2}\,.
\ee
We may now consider the gravitational fields,
\bs
\ba
\bbs_0
 & = &
\frac{m_P^2 r\lp\sigma^2-f^2\rp}{\sigma^2}\,
\bdiff\cos\theta\wedge \bdiff\phi\,, \\
\mathbf{s}_i & = & 
\frac{m_P^2 r(\sigma^2-f^2)}{\lambda\sigma^2}\,
n_i\bdiff\cos\theta\wedge\bdiff\phi\,\,\, +\,\, \dots\,,
\ea
\es
where the ellipsis denotes terms that vanish when setting $\bdiff t = \bdiff r=0$. We have omitted these terms since they do not contribute to the charge integrals on a spherical surface. Evaluating those integrals for the solution (\ref{dilatonfunctions}), we obtain
\bs
\ba \label{Fenergy2}
-\oint \bbs_0 & = & \frac{4\pi m_P^2}{r}\lp\frac{r_S R^2}{R+r_q}-r_q^2\rp\,,  \\
-\oint \bbs_i & = & 0\,. 
\ea
\es
For $r_q=0$, the gravitational charge reduces to the Schwarzschild mass. In general, the expression (\ref{Fenergy2}) is the Misner-Sharp mass of the dilatonic black hole. 
In the fixed-potential polarisation considered below, it must be supplemented by the matter surface term to obtain the full Hamiltonian energy.

 To compute the electromagnetic contribution to the surface charge in full generality, it is useful to keep both the asymptotic value of the dilaton and the additive constant in the electric potential arbitrary. In the following, \(\varphi_\infty\) is held fixed as boundary data\footnote{Allowing it to vary would introduce the corresponding scalar-charge term in the first law \cite{Ballesteros:2023}.}.  
We write
\be \label{Ainfty}
\bA = \lp A_\infty
        -
        q\int_r^\infty\frac{e^{2\varphi/m_P}\sigma\bdiff \rho }{\rho^2}\rp\bdiff t 
        = 
\left(
A_\infty
-
\frac{q e^{2\varphi_\infty/m_P}}
{R+r_q}
\right)\bdiff t\,. 
\ee
The solution has the correct asymptotic behaviour, $A_t = A_\infty - qe^{2\varphi_\infty/m_P}/r + O(r^{-2})$ as $r \rightarrow \infty$. We compute the electromagnetic contribution analogously to the simpler case (\ref{Menergy}),
\begin{equation} 
        -\oint\lp\bbie_a\lrcorner\bA\rp\bPi
        = 
        \frac{\delta^0_a}{\sigma}
\left[
\frac{q^2 e^{2\varphi_\infty/m_P}}
{\sqrt{r^2+r_q^2}+r_q}
-
qA_\infty
\right]\,.
       \label{Menergy2}
\end{equation}
The factor \(\sigma^{-1}\) appears because the surface charge is
associated with the canonical generator \(\bbie_0=\sigma^{-1}\partial_t\),
whereas \(A_\infty\) was introduced at (\ref{Ainfty}) as the additive constant in the coordinate
component \(A_t\).

The total energy is the sum of (\ref{Fenergy2}) and (\ref{Menergy2}),  
\be \label{vmaxwell}
E^{(A)}_{\text{dil}} \equiv P^{(A)}_0 
=  \frac{4\pi m_P^2}{r}\lp\frac{r_S R^2}{R+r_q}-r_q^2\rp + \frac{1}{\sigma}\lp\frac{2m_S r_q}{R+r_q}-qA_\infty\rp\,. 
\ee
For the representative with
\(A_\infty=0\), the electromagnetic contribution vanishes at infinity and
the asymptotic energy is \(E^{(A)}_{\text{dil}}(\infty)=m_S\). In the horizon-regular gauge
\(A_t(r_+)=0\), the charge is shifted by the corresponding electrostatic potential term. 
One can check that for the $A_\infty=0$ representative, the Hamiltonian excess $\Delta E^{(A)}_{\text{dil}}=E^{(A)}_{\text{dil}}-m_S>0$ is positive
throughout the nonextremal black-hole regime  $r_S>2r_q$. It can be decomposed as $\Delta E^{(A)}_{\text{dil}} = U_{\text{ext}} - U_\phi$, where $U_\phi>0$ is
the exterior dilaton energy.  The Maxwell surface term in Eq.~(\ref{Menergy2})
required by the fixed-potential variational problem reverses the
sign of the resulting finite-radius Hamiltonian correction with respect to the Misner--Sharp part in Eq.~(\ref{Fenergy2}). 
The matter term is absent in calculations based solely on a gravitational superpotential. 
The familiar negative finite-radius charge correction appears in several such calculations for the Reissner-Nordstr{\"o}m black hole
\cite{Aguirregabiria:1995qz,Herrera:2003pu,Xulu:2006zza,SaltiAydogdu2006RNAdS,Nashed2006RN,Vagenas:2006pj,Dadhich2020,Koivisto:2022nar,Gomes:2022vrc}, as well as for dilatonic charged black holes \cite{Chamorro:1994vw,Yang:1996iq,Nashed:2005kn,Nashed2012Dyadosphere,Radinschi:2010zza,MouradAbdelgaber2023StringyEntropy}. 
In the present Hamiltonian framework, the gravitational-superpotential result is naturally associated with the fixed-charge side of the variational problem.
 
The expression (\ref{vmaxwell}) now provides a stringent consistency check against the thermodynamic relation obtained independently from the Euclidean boundary action (\ref{F}).
The event horizon is located at the largest positive zero of \(f^2\), giving $r_+^2 = r_S\lp r_S-2r_q\rp$. 
The entropy (\ref{entropy}) is given by the area law, $S = 8\pi^2m_P^2r_+^2$.
The temperature follows from regularity of the Euclidean section. The standard derivation gives the temperature $\beta^{-1}_T$
conjugate to the static Killing
generator \(\partial_T\). The Hamiltonian charge (\ref{vmaxwell}), however, has been evaluated for the Kodama generator $K=\bbie_0$. 
At the horizon, $\partial_T = \sigma_+ K$, and the Euclidean quantities compared with this charge must therefore be 
referred to $K$ as well\footnote{The period conjugate to $K$ may also be obtained directly by
introducing, in the near-horizon expansion, $\tau=\sigma_+T$, so that $\partial_\tau=K|_{r_+}$.
The Euclidean near-horizon metric is
\[
\bdiff s^2_E=
\frac{f^2}{\sigma_+^2}\bdiff\tau_E^2
+
\frac{\sigma^2}{f^2}\bdiff r^2+\cdots = \bdiff\rho^2+
\left(\frac{(f^2)'_+}{2\sigma_+^2}\right)^2
\rho^2\bdiff\tau_E^2+\cdots\,,
\]
where the second equality follows by using $\rho=2\sigma_+\sqrt{(r-r_+)/(f^2)'_+}$, and \(f^2=(f^2)'_+(r-r_+)+\cdots\), \(\sigma=\sigma_++\cdots\). 
Regularity therefore gives $\beta=4\pi\sigma_+^2/(f^2)'_+$, which for the dilatonic charged black hole yields (\ref{dilbeta}).}. Equivalently, the Euclidean action $I_E$ is invariant, but its decomposition as
\(I_E=\beta \termopot = \beta_T \termopot_T\) depends on which Hamiltonian generator is used. The period and the thermodynamic potential conjugate to $K$ are obtained from the standard Killing-normalised quantities by the rescalings $\beta = \sigma_+ \beta_T$ and ${\termopot} = \frac{\termopot_T}{\sigma_+}$, where in particular
\begin{equation}
\beta^{-1}_T=\frac{(f^2)'(r_+)}{4\pi\sigma(r_+)} = \frac{1}{4\pi r_S}\,, \quad \text {whereas} \quad 
\beta^{-1}
= \frac{r_S-r_q}{4\pi r_S r_+}\,. \label{dilbeta}
\end{equation}
Using (\ref{freeformula}),
\bs
\begin{equation} \label{dilF}
 \termopot^{(A)}
=
\frac{2\pi m_P^2}{r_+}\lp r_S-2r_q\rp\lp r_S+r_q\rp\,.
\end{equation}
Now, using (\ref{vmaxwell}), (\ref{dilbeta}) and (\ref{dilF}), we confirm that
\be \label{dilresult}
\termopot^{(A)} = E^{(A)}_{\text{dil}+} - \beta^{-1}S + qA_+\,, \quad \text{where} \quad A_+ = \bbie_0\lrcorner\bA\mid_{r=r_+} = \sigma_+^{-1}\lp A_\infty -  q e^{2\varphi_\infty/m_P}/(R_++r_q)\rp\,. 
\ee
\es
For the horizon-regular U(1) representative $A_+=0$, we recover the standard free energy form. 

Equation~(\ref{dilresult}) is the fixed-potential thermodynamic relation. 
It naturally features the polarisation-matched horizon charge
\begin{equation}
\widehat E_+
\equiv
E_+^{(A)}+qA_+
=
4\pi m_P^2r_+\,.
\end{equation}
This is the Misner--Sharp mass evaluated at the horizon. 
Equation~(\ref{dilresult}) can therefore be written transparently as
\begin{equation}
\Omega^{(A)}
=
\widehat E_+-\beta^{-1}S
=
2\pi m_P^2r_+
\left(1+\frac{r_q}{r_S}\right)\,.
\end{equation}
In the fixed-charge polarisation, the matched charge becomes the
Hamiltonian energy itself. As in Section~\ref{horizonthermo}, following Ref.~\cite{Hawking:1995ap}, one adds the Maxwell
Legendre boundary term, $\bL_{M}^{(q)} = \bL_{M}^{(A)}-\bdiff(\bA\wedge\bPi)$, obtains $\bTheta_{M}^{(q)} =-\bA\wedge\delta\bPi-2\delta\varphi\ast\bdiff\varphi$, and imposes
$\delta\bPi=0$, hence $\delta q=0$. 
On a static spherical corner the Legendre term cancels the Maxwell contribution in
Eq.~(\ref{Menergy2}). 
Consequently, at any radius,
\begin{equation}
E_{\rm dil}^{(q)}(r)
=
E_{\rm dil}^{(A)}(r)+q(K\lrcorner \bA)\,, \quad \text{and in particular:} \quad E_+^{(q)}
=
\widehat E_+
=
4\pi m_P^2r_+\,. 
\end{equation}
The fixed-charge thermodynamic potential is therefore
\begin{equation}
\Omega^{(q)}
=
E_+^{(q)}-\beta^{-1}S
=
\widehat E_+-\beta^{-1}S
=
\Omega^{(A)}\,.
\end{equation}
Thus, the same polarisation structure is recovered as in the Reissner--Nordstr\"om example.
The Smarr relation and the first law take the same form. 
The conserved local energy--momentum currents for both black hole examples are recorded in Appendix~\ref{localcurrents}.

\section{Conclusions}
\label{conclusions}

\begin{itemize}
\item \emph{The canonical frame provides the local reference structure}.
\end{itemize}
A new result of this work is the canonical representative for generic static spherical symmetry, summarised in Table \ref{kodamaframe}.  
Its temporal leg is the geometrically determined Kodama vector $K$, related to the static Killing generator by a radius-dependent normalisation in the dilatonic example.
When both the Hamiltonian charge and the Euclidean thermodynamic quantities are referred to \(K\), they satisfy the relation derived from the Euclidean boundary action.


\begin{itemize}
\item \emph{Energy and momentum are the matter-completed (C)Diff charges (\ref{charge})}. 
\end{itemize}
This is the main result of the article. The matter contribution from the symplectic potential changes the interpretation of the surface charge. 
$P_a$ is not simply a gravitational superpotential evaluated on a $(n-2)$-surface, but the Hamiltonian generator appropriate to the chosen boundary polarisation of the full gravity--matter system.  
We highlight the following corollary:  

\begin{itemize}
\item \emph{The electric energy outside a charged black hole is positive}.
\end{itemize}
The exterior energy \(U_{\rm ext}(r)\), equal to \(q^2/2r\) for the
Reissner--Nordstr\"om black hole, is positive and independent of
\(A_\infty\). In the fixed-charge polarisation it is subtracted from
the total mass, giving the Hamiltonian charge enclosed within radius
\(r\). In the fixed-potential polarisation, the Maxwell surface term
combines with the Misner--Sharp mass, reversing
the \(U_{\rm ext}(r)\) correction, whilst retaining the
\(A_\infty\)-dependent electric source coupling.

  

\begin{itemize}
\item \emph{Horizon entropy is $2\pi$ times the unit normal-plane GL-boost charge}.
\end{itemize}
In the metric-affine formulation, the normal-plane projection of the
GL curvature momentum is conjugate to the Euclidean opening angle of
the horizon corner. For a smooth horizon the opening angle is \(2\pi\), and $S=2\pi P_{\rm boost}$ reproduces the Iyer--Wald area entropy. 
This picture has precursors in the isolated-horizon and local-observer treatments of boost charges, but here it arises in the same parallel-geometric framework as the energy--momentum charge. 
\newline\newline

The uniqueness and global existence of the canonical frame have not been
addressed here beyond static spherical symmetry, and the construction remains to be tested in qualitatively
different settings, including rotating geometries, non-stationary horizons,
moving corners, and more generic boundary conditions. 
A systematic comparison with standard Riemannian quasi-local constructions, such as the Brown--York boundary stress tensor, would further clarify how the parallel reference structure encodes familiar boundary observables.
These are natural directions for future work.

\acknowledgments{
This work was supported by the Estonian Research Council through grants  TARISTU24-TK10, TARISTU24-TK3, CoE TK202 ``Foundations of the Universe'', and PRG2608 ``Space--Time--Matter''. The author is grateful to Adel Awad for drawing his attention to the electromagnetic polarisation issue, which prompted the systematic treatment of the corresponding boundary conditions. 
}

\bibliography{absolute}

\begin{thebibliography}{126}%
\makeatletter
\providecommand \@ifxundefined [1]{%
 \@ifx{#1\undefined}
}%
\providecommand \@ifnum [1]{%
 \ifnum #1\expandafter \@firstoftwo
 \else \expandafter \@secondoftwo
 \fi
}%
\providecommand \@ifx [1]{%
 \ifx #1\expandafter \@firstoftwo
 \else \expandafter \@secondoftwo
 \fi
}%
\providecommand \natexlab [1]{#1}%
\providecommand \enquote  [1]{``#1''}%
\providecommand \bibnamefont  [1]{#1}%
\providecommand \bibfnamefont [1]{#1}%
\providecommand \citenamefont [1]{#1}%
\providecommand \href@noop [0]{\@secondoftwo}%
\providecommand \href [0]{\begingroup \@sanitize@url \@href}%
\providecommand \@href[1]{\@@startlink{#1}\@@href}%
\providecommand \@@href[1]{\endgroup#1\@@endlink}%
\providecommand \@sanitize@url [0]{\catcode `\\12\catcode `\$12\catcode
  `\&12\catcode `\#12\catcode `\^12\catcode `\_12\catcode `\%12\relax}%
\providecommand \@@startlink[1]{}%
\providecommand \@@endlink[0]{}%
\providecommand \url  [0]{\begingroup\@sanitize@url \@url }%
\providecommand \@url [1]{\endgroup\@href {#1}{\urlprefix }}%
\providecommand \urlprefix  [0]{URL }%
\providecommand \Eprint [0]{\href }%
\providecommand \doibase [0]{http://dx.doi.org/}%
\providecommand \selectlanguage [0]{\@gobble}%
\providecommand \bibinfo  [0]{\@secondoftwo}%
\providecommand \bibfield  [0]{\@secondoftwo}%
\providecommand \translation [1]{[#1]}%
\providecommand \BibitemOpen [0]{}%
\providecommand \bibitemStop [0]{}%
\providecommand \bibitemNoStop [0]{.\EOS\space}%
\providecommand \EOS [0]{\spacefactor3000\relax}%
\providecommand \BibitemShut  [1]{\csname bibitem#1\endcsname}%
\let\auto@bib@innerbib\@empty
\bibitem [{\citenamefont {Szabados}(2009)}]{Szabados:2009eka}%
  \BibitemOpen
  \bibfield  {author} {\bibinfo {author} {\bibfnamefont {L.~B.}\ \bibnamefont
  {Szabados}},\ }\href {\doibase 10.12942/lrr-2009-4} {\bibfield  {journal}
  {\bibinfo  {journal} {Living Reviews in Relativity}\ }\textbf {\bibinfo
  {volume} {12}},\ \bibinfo {pages} {4} (\bibinfo {year} {2009})}\BibitemShut
  {NoStop}%
\bibitem [{\citenamefont {Iyer}\ and\ \citenamefont
  {Wald}(1994)}]{Iyer:1994ys}%
  \BibitemOpen
  \bibfield  {author} {\bibinfo {author} {\bibfnamefont {V.}~\bibnamefont
  {Iyer}}\ and\ \bibinfo {author} {\bibfnamefont {R.~M.}\ \bibnamefont
  {Wald}},\ }\href {\doibase 10.1103/PhysRevD.50.846} {\bibfield  {journal}
  {\bibinfo  {journal} {Phys. Rev. D}\ }\textbf {\bibinfo {volume} {50}},\
  \bibinfo {pages} {846} (\bibinfo {year} {1994})},\ \Eprint
  {http://arxiv.org/abs/gr-qc/9403028} {arXiv:gr-qc/9403028} \BibitemShut
  {NoStop}%
\bibitem [{\citenamefont {Regge}\ and\ \citenamefont
  {Teitelboim}(1974)}]{REGGE1974286}%
  \BibitemOpen
  \bibfield  {author} {\bibinfo {author} {\bibfnamefont {T.}~\bibnamefont
  {Regge}}\ and\ \bibinfo {author} {\bibfnamefont {C.}~\bibnamefont
  {Teitelboim}},\ }\href {\doibase
  https://doi.org/10.1016/0003-4916(74)90404-7} {\bibfield  {journal} {\bibinfo
   {journal} {Annals of Physics}\ }\textbf {\bibinfo {volume} {88}},\ \bibinfo
  {pages} {286} (\bibinfo {year} {1974})}\BibitemShut {NoStop}%
\bibitem [{\citenamefont {Brown}\ and\ \citenamefont
  {York}(1993)}]{Brown:1992br}%
  \BibitemOpen
  \bibfield  {author} {\bibinfo {author} {\bibfnamefont {J.~D.}\ \bibnamefont
  {Brown}}\ and\ \bibinfo {author} {\bibfnamefont {J.~W.}\ \bibnamefont
  {York}},\ }\href {\doibase 10.1103/PhysRevD.47.1407} {\bibfield  {journal}
  {\bibinfo  {journal} {Phys. Rev. D}\ }\textbf {\bibinfo {volume} {47}},\
  \bibinfo {pages} {1407} (\bibinfo {year} {1993})},\ \Eprint
  {http://arxiv.org/abs/gr-qc/9209012} {arXiv:gr-qc/9209012} \BibitemShut
  {NoStop}%
\bibitem [{\citenamefont {Harlow}\ and\ \citenamefont
  {Wu}(2020)}]{Harlow:2019yfa}%
  \BibitemOpen
  \bibfield  {author} {\bibinfo {author} {\bibfnamefont {D.}~\bibnamefont
  {Harlow}}\ and\ \bibinfo {author} {\bibfnamefont {J.-Q.}\ \bibnamefont
  {Wu}},\ }\href {\doibase 10.1007/JHEP10(2020)146} {\bibfield  {journal}
  {\bibinfo  {journal} {JHEP}\ }\textbf {\bibinfo {volume} {10}},\ \bibinfo
  {pages} {146} (\bibinfo {year} {2020})},\ \Eprint
  {http://arxiv.org/abs/1906.08616} {arXiv:1906.08616 [hep-th]} \BibitemShut
  {NoStop}%
\bibitem [{\citenamefont {Chandrasekaran}\ \emph {et~al.}(2022)\citenamefont
  {Chandrasekaran}, \citenamefont {Flanagan}, \citenamefont {Shehzad},\ and\
  \citenamefont {Speranza}}]{Chandrasekaran:2021vyu}%
  \BibitemOpen
  \bibfield  {author} {\bibinfo {author} {\bibfnamefont {V.}~\bibnamefont
  {Chandrasekaran}}, \bibinfo {author} {\bibfnamefont {E.~E.}\ \bibnamefont
  {Flanagan}}, \bibinfo {author} {\bibfnamefont {I.}~\bibnamefont {Shehzad}}, \
  and\ \bibinfo {author} {\bibfnamefont {A.~J.}\ \bibnamefont {Speranza}},\
  }\href {\doibase 10.1142/S0217751X22501056} {\bibfield  {journal} {\bibinfo
  {journal} {Int. J. Mod. Phys. A}\ }\textbf {\bibinfo {volume} {37}},\
  \bibinfo {pages} {2250105} (\bibinfo {year} {2022})},\ \Eprint
  {http://arxiv.org/abs/2111.11974} {arXiv:2111.11974 [gr-qc]} \BibitemShut
  {NoStop}%
\bibitem [{\citenamefont {Donnelly}\ and\ \citenamefont
  {Freidel}(2016)}]{Donnelly:2016auv}%
  \BibitemOpen
  \bibfield  {author} {\bibinfo {author} {\bibfnamefont {W.}~\bibnamefont
  {Donnelly}}\ and\ \bibinfo {author} {\bibfnamefont {L.}~\bibnamefont
  {Freidel}},\ }\href {\doibase 10.1007/JHEP09(2016)102} {\bibfield  {journal}
  {\bibinfo  {journal} {JHEP}\ }\textbf {\bibinfo {volume} {09}},\ \bibinfo
  {pages} {102} (\bibinfo {year} {2016})},\ \Eprint
  {http://arxiv.org/abs/1601.04744} {arXiv:1601.04744 [hep-th]} \BibitemShut
  {NoStop}%
\bibitem [{\citenamefont {Speranza}(2018)}]{Speranza:2017gxd}%
  \BibitemOpen
  \bibfield  {author} {\bibinfo {author} {\bibfnamefont {A.~J.}\ \bibnamefont
  {Speranza}},\ }\href {\doibase 10.1007/JHEP02(2018)021} {\bibfield  {journal}
  {\bibinfo  {journal} {JHEP}\ }\textbf {\bibinfo {volume} {02}},\ \bibinfo
  {pages} {021} (\bibinfo {year} {2018})},\ \Eprint
  {http://arxiv.org/abs/1706.05061} {arXiv:1706.05061 [hep-th]} \BibitemShut
  {NoStop}%
\bibitem [{\citenamefont {Freidel}\ \emph
  {et~al.}(2020{\natexlab{a}})\citenamefont {Freidel}, \citenamefont
  {Geiller},\ and\ \citenamefont {Pranzetti}}]{Freidel:2020xyx}%
  \BibitemOpen
  \bibfield  {author} {\bibinfo {author} {\bibfnamefont {L.}~\bibnamefont
  {Freidel}}, \bibinfo {author} {\bibfnamefont {M.}~\bibnamefont {Geiller}}, \
  and\ \bibinfo {author} {\bibfnamefont {D.}~\bibnamefont {Pranzetti}},\ }\href
  {\doibase 10.1007/JHEP11(2020)026} {\bibfield  {journal} {\bibinfo  {journal}
  {JHEP}\ }\textbf {\bibinfo {volume} {11}},\ \bibinfo {pages} {026} (\bibinfo
  {year} {2020}{\natexlab{a}})},\ \Eprint {http://arxiv.org/abs/2006.12527}
  {arXiv:2006.12527 [hep-th]} \BibitemShut {NoStop}%
\bibitem [{\citenamefont {M{\o}ller}(1961)}]{Moller1961Conservation}%
  \BibitemOpen
  \bibfield  {author} {\bibinfo {author} {\bibfnamefont {C.}~\bibnamefont
  {M{\o}ller}},\ }\href@noop {} {\bibfield  {journal} {\bibinfo  {journal}
  {Matematisk-fysiske Skrifter, Kongelige Danske Videnskabernes Selskab}\
  }\textbf {\bibinfo {volume} {1}},\ \bibinfo {pages} {1} (\bibinfo {year}
  {1961})}\BibitemShut {NoStop}%
\bibitem [{\citenamefont {Chang}\ \emph {et~al.}(1999)\citenamefont {Chang},
  \citenamefont {Nester},\ and\ \citenamefont {Chen}}]{Chang:1998wj}%
  \BibitemOpen
  \bibfield  {author} {\bibinfo {author} {\bibfnamefont {C.-C.}\ \bibnamefont
  {Chang}}, \bibinfo {author} {\bibfnamefont {J.~M.}\ \bibnamefont {Nester}}, \
  and\ \bibinfo {author} {\bibfnamefont {C.-M.}\ \bibnamefont {Chen}},\ }\href
  {\doibase 10.1103/PhysRevLett.83.1897} {\bibfield  {journal} {\bibinfo
  {journal} {Phys. Rev. Lett.}\ }\textbf {\bibinfo {volume} {83}},\ \bibinfo
  {pages} {1897} (\bibinfo {year} {1999})},\ \Eprint
  {http://arxiv.org/abs/gr-qc/9809040} {gr-qc/9809040} \BibitemShut {NoStop}%
\bibitem [{\citenamefont {Einstein}(1916)}]{Einstein1916Hamilton}%
  \BibitemOpen
  \bibfield  {author} {\bibinfo {author} {\bibfnamefont {A.}~\bibnamefont
  {Einstein}},\ }\href@noop {} {\bibfield  {journal} {\bibinfo  {journal}
  {Sitzungsberichte der Preussischen Akademie der Wissenschaften}\ ,\ \bibinfo
  {pages} {1111}} (\bibinfo {year} {1916})}\BibitemShut {NoStop}%
\bibitem [{\citenamefont {de~Andrade}\ \emph {et~al.}(2000)\citenamefont
  {de~Andrade}, \citenamefont {Guillen},\ and\ \citenamefont
  {Pereira}}]{deAndrade:2000kr}%
  \BibitemOpen
  \bibfield  {author} {\bibinfo {author} {\bibfnamefont {V.~C.}\ \bibnamefont
  {de~Andrade}}, \bibinfo {author} {\bibfnamefont {L.~C.~T.}\ \bibnamefont
  {Guillen}}, \ and\ \bibinfo {author} {\bibfnamefont {J.~G.}\ \bibnamefont
  {Pereira}},\ }\href {\doibase 10.1103/PhysRevLett.84.4533} {\bibfield
  {journal} {\bibinfo  {journal} {Phys. Rev. Lett.}\ }\textbf {\bibinfo
  {volume} {84}},\ \bibinfo {pages} {4533} (\bibinfo {year} {2000})},\ \Eprint
  {http://arxiv.org/abs/gr-qc/0003100} {arXiv:gr-qc/0003100} \BibitemShut
  {NoStop}%
\bibitem [{\citenamefont {Maluf}\ \emph {et~al.}(2002)\citenamefont {Maluf},
  \citenamefont {da~Rocha-Neto}, \citenamefont {Toribio},\ and\ \citenamefont
  {Castello-Branco}}]{Maluf:2002zc}%
  \BibitemOpen
  \bibfield  {author} {\bibinfo {author} {\bibfnamefont {J.~W.}\ \bibnamefont
  {Maluf}}, \bibinfo {author} {\bibfnamefont {J.~F.}\ \bibnamefont
  {da~Rocha-Neto}}, \bibinfo {author} {\bibfnamefont {T.~M.~L.}\ \bibnamefont
  {Toribio}}, \ and\ \bibinfo {author} {\bibfnamefont {K.~H.}\ \bibnamefont
  {Castello-Branco}},\ }\href {\doibase 10.1103/PhysRevD.65.124001} {\bibfield
  {journal} {\bibinfo  {journal} {Phys. Rev. D}\ }\textbf {\bibinfo {volume}
  {65}},\ \bibinfo {pages} {124001} (\bibinfo {year} {2002})},\ \Eprint
  {http://arxiv.org/abs/gr-qc/0204035} {arXiv:gr-qc/0204035} \BibitemShut
  {NoStop}%
\bibitem [{\citenamefont {Nashed}(2006{\natexlab{a}})}]{Nashed:2005kn}%
  \BibitemOpen
  \bibfield  {author} {\bibinfo {author} {\bibfnamefont {G.~G.~L.}\
  \bibnamefont {Nashed}},\ }\href {\doibase 10.1142/S0217751X06031478}
  {\bibfield  {journal} {\bibinfo  {journal} {Int. J. Mod. Phys. A}\ }\textbf
  {\bibinfo {volume} {21}},\ \bibinfo {pages} {3181} (\bibinfo {year}
  {2006}{\natexlab{a}})},\ \Eprint {http://arxiv.org/abs/gr-qc/0501002}
  {arXiv:gr-qc/0501002} \BibitemShut {NoStop}%
\bibitem [{\citenamefont {Lucas}\ \emph {et~al.}(2009)\citenamefont {Lucas},
  \citenamefont {Obukhov},\ and\ \citenamefont {Pereira}}]{Lucas:2009nq}%
  \BibitemOpen
  \bibfield  {author} {\bibinfo {author} {\bibfnamefont {T.~G.}\ \bibnamefont
  {Lucas}}, \bibinfo {author} {\bibfnamefont {Y.~N.}\ \bibnamefont {Obukhov}},
  \ and\ \bibinfo {author} {\bibfnamefont {J.~G.}\ \bibnamefont {Pereira}},\
  }\href {\doibase 10.1103/PhysRevD.80.064043} {\bibfield  {journal} {\bibinfo
  {journal} {Phys. Rev. D}\ }\textbf {\bibinfo {volume} {80}},\ \bibinfo
  {pages} {064043} (\bibinfo {year} {2009})},\ \Eprint
  {http://arxiv.org/abs/0909.2418} {arXiv:0909.2418 [gr-qc]} \BibitemShut
  {NoStop}%
\bibitem [{\citenamefont {Krssak}\ \emph {et~al.}(2019)\citenamefont {Krssak},
  \citenamefont {van~den Hoogen}, \citenamefont {Pereira}, \citenamefont
  {B{\"o}hmer},\ and\ \citenamefont {Coley}}]{Krssak:2018ywd}%
  \BibitemOpen
  \bibfield  {author} {\bibinfo {author} {\bibfnamefont {M.}~\bibnamefont
  {Krssak}}, \bibinfo {author} {\bibfnamefont {R.~J.}\ \bibnamefont {van~den
  Hoogen}}, \bibinfo {author} {\bibfnamefont {J.~G.}\ \bibnamefont {Pereira}},
  \bibinfo {author} {\bibfnamefont {C.~G.}\ \bibnamefont {B{\"o}hmer}}, \ and\
  \bibinfo {author} {\bibfnamefont {A.~A.}\ \bibnamefont {Coley}},\ }\href
  {\doibase 10.1088/1361-6382/ab2e1f} {\bibfield  {journal} {\bibinfo
  {journal} {Class. Quant. Grav.}\ }\textbf {\bibinfo {volume} {36}},\ \bibinfo
  {pages} {183001} (\bibinfo {year} {2019})},\ \Eprint
  {http://arxiv.org/abs/1810.12932} {arXiv:1810.12932 [gr-qc]} \BibitemShut
  {NoStop}%
\bibitem [{\citenamefont {Maluf}\ \emph {et~al.}(2023)\citenamefont {Maluf},
  \citenamefont {Carneiro}, \citenamefont {Ulhoa},\ and\ \citenamefont
  {Rocha-Neto}}]{Maluf:2023rwe}%
  \BibitemOpen
  \bibfield  {author} {\bibinfo {author} {\bibfnamefont {J.~W.}\ \bibnamefont
  {Maluf}}, \bibinfo {author} {\bibfnamefont {F.~L.}\ \bibnamefont {Carneiro}},
  \bibinfo {author} {\bibfnamefont {S.~C.}\ \bibnamefont {Ulhoa}}, \ and\
  \bibinfo {author} {\bibfnamefont {J.~F.~d.}\ \bibnamefont {Rocha-Neto}},\
  }\href {\doibase 10.1002/andp.202300241} {\bibfield  {journal} {\bibinfo
  {journal} {Annalen Phys.}\ }\textbf {\bibinfo {volume} {535}},\ \bibinfo
  {pages} {2300241} (\bibinfo {year} {2023})},\ \Eprint
  {http://arxiv.org/abs/2306.03676} {arXiv:2306.03676 [gr-qc]} \BibitemShut
  {NoStop}%
\bibitem [{\citenamefont {Bahamonde}\ \emph {et~al.}(2023)\citenamefont
  {Bahamonde}, \citenamefont {Dialektopoulos}, \citenamefont
  {Escamilla-Rivera}, \citenamefont {Farrugia}, \citenamefont {Gakis},
  \citenamefont {Hendry}, \citenamefont {Hohmann}, \citenamefont {Levi~Said},
  \citenamefont {Mifsud},\ and\ \citenamefont
  {Di~Valentino}}]{Bahamonde:2021gfp}%
  \BibitemOpen
  \bibfield  {author} {\bibinfo {author} {\bibfnamefont {S.}~\bibnamefont
  {Bahamonde}}, \bibinfo {author} {\bibfnamefont {K.~F.}\ \bibnamefont
  {Dialektopoulos}}, \bibinfo {author} {\bibfnamefont {C.}~\bibnamefont
  {Escamilla-Rivera}}, \bibinfo {author} {\bibfnamefont {G.}~\bibnamefont
  {Farrugia}}, \bibinfo {author} {\bibfnamefont {V.}~\bibnamefont {Gakis}},
  \bibinfo {author} {\bibfnamefont {M.}~\bibnamefont {Hendry}}, \bibinfo
  {author} {\bibfnamefont {M.}~\bibnamefont {Hohmann}}, \bibinfo {author}
  {\bibfnamefont {J.}~\bibnamefont {Levi~Said}}, \bibinfo {author}
  {\bibfnamefont {J.}~\bibnamefont {Mifsud}}, \ and\ \bibinfo {author}
  {\bibfnamefont {E.}~\bibnamefont {Di~Valentino}},\ }\href {\doibase
  10.1088/1361-6633/ac9cef} {\bibfield  {journal} {\bibinfo  {journal} {Rept.
  Prog. Phys.}\ }\textbf {\bibinfo {volume} {86}},\ \bibinfo {pages} {026901}
  (\bibinfo {year} {2023})},\ \Eprint {http://arxiv.org/abs/2106.13793}
  {arXiv:2106.13793 [gr-qc]} \BibitemShut {NoStop}%
\bibitem [{\citenamefont {Gomes}\ \emph
  {et~al.}(2023{\natexlab{a}})\citenamefont {Gomes}, \citenamefont
  {Beltr{\'a}n~Jim{\'e}nez},\ and\ \citenamefont {Koivisto}}]{Gomes:2023hyk}%
  \BibitemOpen
  \bibfield  {author} {\bibinfo {author} {\bibfnamefont {D.~A.}\ \bibnamefont
  {Gomes}}, \bibinfo {author} {\bibfnamefont {J.}~\bibnamefont
  {Beltr{\'a}n~Jim{\'e}nez}}, \ and\ \bibinfo {author} {\bibfnamefont {T.~S.}\
  \bibnamefont {Koivisto}},\ }\href {\doibase 10.1088/1475-7516/2023/12/010}
  {\bibfield  {journal} {\bibinfo  {journal} {JCAP}\ }\textbf {\bibinfo
  {volume} {12}},\ \bibinfo {pages} {010} (\bibinfo {year}
  {2023}{\natexlab{a}})},\ \Eprint {http://arxiv.org/abs/2309.08554}
  {arXiv:2309.08554 [gr-qc]} \BibitemShut {NoStop}%
\bibitem [{\citenamefont {Koivisto}(2026)}]{Koivisto:2026yxt}%
  \BibitemOpen
  \bibfield  {author} {\bibinfo {author} {\bibfnamefont {T.~S.}\ \bibnamefont
  {Koivisto}},\ }\href {\doibase 10.1007/s10714-026-03514-5} {\bibfield
  {journal} {\bibinfo  {journal} {Gen. Rel. Grav.}\ }\textbf {\bibinfo {volume}
  {58}},\ \bibinfo {pages} {10} (\bibinfo {year} {2026})},\ \Eprint
  {http://arxiv.org/abs/2602.06509} {arXiv:2602.06509 [gr-qc]} \BibitemShut
  {NoStop}%
\bibitem [{\citenamefont {Koivisto}(2018)}]{Koivisto:2018aip}%
  \BibitemOpen
  \bibfield  {author} {\bibinfo {author} {\bibfnamefont {T.}~\bibnamefont
  {Koivisto}},\ }\href {\doibase 10.1142/S0219887818400066} {\bibfield
  {journal} {\bibinfo  {journal} {Int. J. Geom. Meth. Mod. Phys.}\ }\textbf
  {\bibinfo {volume} {15}},\ \bibinfo {pages} {1840006} (\bibinfo {year}
  {2018})},\ \Eprint {http://arxiv.org/abs/1802.00650} {arXiv:1802.00650
  [gr-qc]} \BibitemShut {NoStop}%
\bibitem [{\citenamefont {Beltr{\'a}n~Jim{\'e}nez}\ \emph
  {et~al.}(2019{\natexlab{a}})\citenamefont {Beltr{\'a}n~Jim{\'e}nez},
  \citenamefont {Heisenberg},\ and\ \citenamefont
  {Koivisto}}]{BeltranJimenez:2019bnx}%
  \BibitemOpen
  \bibfield  {author} {\bibinfo {author} {\bibfnamefont {J.}~\bibnamefont
  {Beltr{\'a}n~Jim{\'e}nez}}, \bibinfo {author} {\bibfnamefont
  {L.}~\bibnamefont {Heisenberg}}, \ and\ \bibinfo {author} {\bibfnamefont
  {T.~S.}\ \bibnamefont {Koivisto}},\ }\href {\doibase
  10.1142/S0218271819440127} {\bibfield  {journal} {\bibinfo  {journal} {Int.
  J. Mod. Phys. D}\ }\textbf {\bibinfo {volume} {28}},\ \bibinfo {pages}
  {1944012} (\bibinfo {year} {2019}{\natexlab{a}})},\ \Eprint
  {http://arxiv.org/abs/1903.12072} {arXiv:1903.12072 [gr-qc]} \BibitemShut
  {NoStop}%
\bibitem [{\citenamefont {Beltr{\'a}n~Jim{\'e}nez}\ \emph
  {et~al.}(2019{\natexlab{b}})\citenamefont {Beltr{\'a}n~Jim{\'e}nez},
  \citenamefont {Heisenberg},\ and\ \citenamefont
  {Koivisto}}]{BeltranJimenez:2019esp}%
  \BibitemOpen
  \bibfield  {author} {\bibinfo {author} {\bibfnamefont {J.}~\bibnamefont
  {Beltr{\'a}n~Jim{\'e}nez}}, \bibinfo {author} {\bibfnamefont
  {L.}~\bibnamefont {Heisenberg}}, \ and\ \bibinfo {author} {\bibfnamefont
  {T.~S.}\ \bibnamefont {Koivisto}},\ }\href {\doibase 10.3390/universe5070173}
  {\bibfield  {journal} {\bibinfo  {journal} {Universe}\ }\textbf {\bibinfo
  {volume} {5}},\ \bibinfo {pages} {173} (\bibinfo {year}
  {2019}{\natexlab{b}})},\ \Eprint {http://arxiv.org/abs/1903.06830}
  {arXiv:1903.06830 [hep-th]} \BibitemShut {NoStop}%
\bibitem [{\citenamefont {Hehl}\ \emph {et~al.}(1995)\citenamefont {Hehl},
  \citenamefont {McCrea}, \citenamefont {Mielke},\ and\ \citenamefont
  {Ne'eman}}]{HEHL19951}%
  \BibitemOpen
  \bibfield  {author} {\bibinfo {author} {\bibfnamefont {F.~W.}\ \bibnamefont
  {Hehl}}, \bibinfo {author} {\bibfnamefont {J.}~\bibnamefont {McCrea}},
  \bibinfo {author} {\bibfnamefont {E.~W.}\ \bibnamefont {Mielke}}, \ and\
  \bibinfo {author} {\bibfnamefont {Y.}~\bibnamefont {Ne'eman}},\ }\href
  {\doibase https://doi.org/10.1016/0370-1573(94)00111-F} {\bibfield  {journal}
  {\bibinfo  {journal} {Physics Reports}\ }\textbf {\bibinfo {volume} {258}},\
  \bibinfo {pages} {1} (\bibinfo {year} {1995})}\BibitemShut {NoStop}%
\bibitem [{\citenamefont {Jim{\'e}nez~Cano}(2021)}]{JimenezCano:2021rlu}%
  \BibitemOpen
  \bibfield  {author} {\bibinfo {author} {\bibfnamefont {A.}~\bibnamefont
  {Jim{\'e}nez~Cano}},\ }\emph {\bibinfo {title} {{Metric-affine Gauge theories
  of gravity. Foundations and new insights}}},\ \href@noop {} {Ph.D. thesis},\
  \bibinfo  {school} {Granada U., Theor. Phys. Astrophys.} (\bibinfo {year}
  {2021}),\ \Eprint {http://arxiv.org/abs/2201.12847} {arXiv:2201.12847
  [gr-qc]} \BibitemShut {NoStop}%
\bibitem [{\citenamefont {Obukhov}\ and\ \citenamefont
  {Pereira}(2003)}]{Obukhov:2002tm}%
  \BibitemOpen
  \bibfield  {author} {\bibinfo {author} {\bibfnamefont {Y.~N.}\ \bibnamefont
  {Obukhov}}\ and\ \bibinfo {author} {\bibfnamefont {J.~G.}\ \bibnamefont
  {Pereira}},\ }\href {\doibase 10.1103/PhysRevD.67.044016} {\bibfield
  {journal} {\bibinfo  {journal} {Phys. Rev. D}\ }\textbf {\bibinfo {volume}
  {67}},\ \bibinfo {pages} {044016} (\bibinfo {year} {2003})},\ \Eprint
  {http://arxiv.org/abs/gr-qc/0212080} {arXiv:gr-qc/0212080} \BibitemShut
  {NoStop}%
\bibitem [{\citenamefont {Golovnev}\ \emph {et~al.}(2017)\citenamefont
  {Golovnev}, \citenamefont {Koivisto},\ and\ \citenamefont
  {Sandstad}}]{Golovnev:2017dox}%
  \BibitemOpen
  \bibfield  {author} {\bibinfo {author} {\bibfnamefont {A.}~\bibnamefont
  {Golovnev}}, \bibinfo {author} {\bibfnamefont {T.}~\bibnamefont {Koivisto}},
  \ and\ \bibinfo {author} {\bibfnamefont {M.}~\bibnamefont {Sandstad}},\
  }\href {\doibase 10.1088/1361-6382/aa7830} {\bibfield  {journal} {\bibinfo
  {journal} {Class. Quant. Grav.}\ }\textbf {\bibinfo {volume} {34}},\ \bibinfo
  {pages} {145013} (\bibinfo {year} {2017})},\ \Eprint
  {http://arxiv.org/abs/1701.06271} {arXiv:1701.06271 [gr-qc]} \BibitemShut
  {NoStop}%
\bibitem [{\citenamefont {Frodden}\ and\ \citenamefont
  {Hidalgo}(2018)}]{Frodden:2017suc}%
  \BibitemOpen
  \bibfield  {author} {\bibinfo {author} {\bibfnamefont {E.}~\bibnamefont
  {Frodden}}\ and\ \bibinfo {author} {\bibfnamefont {D.}~\bibnamefont
  {Hidalgo}},\ }\href {\doibase 10.1088/1361-6382/aa9ba5} {\bibfield  {journal}
  {\bibinfo  {journal} {Class. Quant. Grav.}\ }\textbf {\bibinfo {volume}
  {35}},\ \bibinfo {pages} {035002} (\bibinfo {year} {2018})},\ \Eprint
  {http://arxiv.org/abs/1703.10120} {arXiv:1703.10120 [gr-qc]} \BibitemShut
  {NoStop}%
\bibitem [{\citenamefont {Braden}\ \emph {et~al.}(1990)\citenamefont {Braden},
  \citenamefont {Brown}, \citenamefont {Whiting},\ and\ \citenamefont
  {York}}]{Braden:1990hw}%
  \BibitemOpen
  \bibfield  {author} {\bibinfo {author} {\bibfnamefont {H.~W.}\ \bibnamefont
  {Braden}}, \bibinfo {author} {\bibfnamefont {J.~D.}\ \bibnamefont {Brown}},
  \bibinfo {author} {\bibfnamefont {B.~F.}\ \bibnamefont {Whiting}}, \ and\
  \bibinfo {author} {\bibfnamefont {J.~W.}\ \bibnamefont {York}, \bibfnamefont
  {Jr.}},\ }\href {\doibase 10.1103/PhysRevD.42.3376} {\bibfield  {journal}
  {\bibinfo  {journal} {Phys. Rev. D}\ }\textbf {\bibinfo {volume} {42}},\
  \bibinfo {pages} {3376} (\bibinfo {year} {1990})}\BibitemShut {NoStop}%
\bibitem [{\citenamefont {Louko}\ and\ \citenamefont
  {Winters-Hilt}(1996)}]{Louko:1996dw}%
  \BibitemOpen
  \bibfield  {author} {\bibinfo {author} {\bibfnamefont {J.}~\bibnamefont
  {Louko}}\ and\ \bibinfo {author} {\bibfnamefont {S.~N.}\ \bibnamefont
  {Winters-Hilt}},\ }\href {\doibase 10.1103/PhysRevD.54.2647} {\bibfield
  {journal} {\bibinfo  {journal} {Phys. Rev. D}\ }\textbf {\bibinfo {volume}
  {54}},\ \bibinfo {pages} {2647} (\bibinfo {year} {1996})},\ \Eprint
  {http://arxiv.org/abs/gr-qc/9602003} {arXiv:gr-qc/9602003} \BibitemShut
  {NoStop}%
\bibitem [{\citenamefont {Hawking}\ and\ \citenamefont
  {Ross}(1995)}]{Hawking:1995ap}%
  \BibitemOpen
  \bibfield  {author} {\bibinfo {author} {\bibfnamefont {S.~W.}\ \bibnamefont
  {Hawking}}\ and\ \bibinfo {author} {\bibfnamefont {S.~F.}\ \bibnamefont
  {Ross}},\ }\href {\doibase 10.1103/PhysRevD.52.5865} {\bibfield  {journal}
  {\bibinfo  {journal} {Phys. Rev. D}\ }\textbf {\bibinfo {volume} {52}},\
  \bibinfo {pages} {5865} (\bibinfo {year} {1995})},\ \Eprint
  {http://arxiv.org/abs/hep-th/9504019} {arXiv:hep-th/9504019} \BibitemShut
  {NoStop}%
\bibitem [{\citenamefont {Beltr{\'a}n~Jim{\'e}nez}\ and\ \citenamefont
  {Koivisto}(2026)}]{BeltranJimenez:2024ufa}%
  \BibitemOpen
  \bibfield  {author} {\bibinfo {author} {\bibfnamefont {J.}~\bibnamefont
  {Beltr{\'a}n~Jim{\'e}nez}}\ and\ \bibinfo {author} {\bibfnamefont {T.~S.}\
  \bibnamefont {Koivisto}},\ }\href {\doibase 10.1103/5wtj-7k2c} {\bibfield
  {journal} {\bibinfo  {journal} {Phys. Rev. D}\ }\textbf {\bibinfo {volume}
  {113}},\ \bibinfo {pages} {L021502} (\bibinfo {year} {2026})},\ \Eprint
  {http://arxiv.org/abs/2412.13946} {arXiv:2412.13946 [gr-qc]} \BibitemShut
  {NoStop}%
\bibitem [{\citenamefont {Koivisto}(2023)}]{Koivisto:2022oyt}%
  \BibitemOpen
  \bibfield  {author} {\bibinfo {author} {\bibfnamefont {T.~S.}\ \bibnamefont
  {Koivisto}},\ }\href {\doibase 10.1103/PhysRevD.107.024019} {\bibfield
  {journal} {\bibinfo  {journal} {Phys. Rev. D}\ }\textbf {\bibinfo {volume}
  {107}},\ \bibinfo {pages} {024019} (\bibinfo {year} {2023})},\ \Eprint
  {http://arxiv.org/abs/2209.14959} {arXiv:2209.14959 [gr-qc]} \BibitemShut
  {NoStop}%
\bibitem [{\citenamefont {Beltr{\'a}n~Jim{\'e}nez}\ \emph
  {et~al.}(2020{\natexlab{a}})\citenamefont {Beltr{\'a}n~Jim{\'e}nez},
  \citenamefont {Heisenberg},\ and\ \citenamefont
  {Koivisto}}]{BeltranJimenez:2020sih}%
  \BibitemOpen
  \bibfield  {author} {\bibinfo {author} {\bibfnamefont {J.}~\bibnamefont
  {Beltr{\'a}n~Jim{\'e}nez}}, \bibinfo {author} {\bibfnamefont
  {L.}~\bibnamefont {Heisenberg}}, \ and\ \bibinfo {author} {\bibfnamefont
  {T.}~\bibnamefont {Koivisto}},\ }\href {\doibase 10.1088/1361-6382/aba31b}
  {\bibfield  {journal} {\bibinfo  {journal} {Class. Quant. Grav.}\ }\textbf
  {\bibinfo {volume} {37}},\ \bibinfo {pages} {195013} (\bibinfo {year}
  {2020}{\natexlab{a}})},\ \Eprint {http://arxiv.org/abs/2004.04606}
  {arXiv:2004.04606 [hep-th]} \BibitemShut {NoStop}%
\bibitem [{\citenamefont {Li}\ \emph {et~al.}(2011)\citenamefont {Li},
  \citenamefont {Sotiriou},\ and\ \citenamefont {Barrow}}]{Li:2010cg}%
  \BibitemOpen
  \bibfield  {author} {\bibinfo {author} {\bibfnamefont {B.}~\bibnamefont
  {Li}}, \bibinfo {author} {\bibfnamefont {T.~P.}\ \bibnamefont {Sotiriou}}, \
  and\ \bibinfo {author} {\bibfnamefont {J.~D.}\ \bibnamefont {Barrow}},\
  }\href {\doibase 10.1103/PhysRevD.83.064035} {\bibfield  {journal} {\bibinfo
  {journal} {Phys. Rev. D}\ }\textbf {\bibinfo {volume} {83}},\ \bibinfo
  {pages} {064035} (\bibinfo {year} {2011})},\ \Eprint
  {http://arxiv.org/abs/1010.1041} {arXiv:1010.1041 [gr-qc]} \BibitemShut
  {NoStop}%
\bibitem [{\citenamefont {Gomes}\ \emph {et~al.}(2024)\citenamefont {Gomes},
  \citenamefont {Beltr{\'a}n~Jim{\'e}nez}, \citenamefont {Cano},\ and\
  \citenamefont {Koivisto}}]{Gomes:2023tur}%
  \BibitemOpen
  \bibfield  {author} {\bibinfo {author} {\bibfnamefont {D.~A.}\ \bibnamefont
  {Gomes}}, \bibinfo {author} {\bibfnamefont {J.}~\bibnamefont
  {Beltr{\'a}n~Jim{\'e}nez}}, \bibinfo {author} {\bibfnamefont {A.~J.}\
  \bibnamefont {Cano}}, \ and\ \bibinfo {author} {\bibfnamefont {T.~S.}\
  \bibnamefont {Koivisto}},\ }\href {\doibase 10.1103/PhysRevLett.132.141401}
  {\bibfield  {journal} {\bibinfo  {journal} {Phys. Rev. Lett.}\ }\textbf
  {\bibinfo {volume} {132}},\ \bibinfo {pages} {141401} (\bibinfo {year}
  {2024})},\ \Eprint {http://arxiv.org/abs/2311.04201} {arXiv:2311.04201
  [gr-qc]} \BibitemShut {NoStop}%
\bibitem [{\citenamefont {Bello-Morales}\ \emph {et~al.}(2024)\citenamefont
  {Bello-Morales}, \citenamefont {Beltr{\'a}n~Jim{\'e}nez}, \citenamefont
  {Jim{\'e}nez~Cano}, \citenamefont {Koivisto},\ and\ \citenamefont
  {Maroto}}]{Bello-Morales:2024vqk}%
  \BibitemOpen
  \bibfield  {author} {\bibinfo {author} {\bibfnamefont {A.~G.}\ \bibnamefont
  {Bello-Morales}}, \bibinfo {author} {\bibfnamefont {J.}~\bibnamefont
  {Beltr{\'a}n~Jim{\'e}nez}}, \bibinfo {author} {\bibfnamefont
  {A.}~\bibnamefont {Jim{\'e}nez~Cano}}, \bibinfo {author} {\bibfnamefont
  {T.~S.}\ \bibnamefont {Koivisto}}, \ and\ \bibinfo {author} {\bibfnamefont
  {A.~L.}\ \bibnamefont {Maroto}},\ }\href {\doibase 10.1007/JHEP12(2024)146}
  {\bibfield  {journal} {\bibinfo  {journal} {JHEP}\ }\textbf {\bibinfo
  {volume} {12}},\ \bibinfo {pages} {146} (\bibinfo {year} {2024})},\ \Eprint
  {http://arxiv.org/abs/2406.19355} {arXiv:2406.19355 [gr-qc]} \BibitemShut
  {NoStop}%
\bibitem [{\citenamefont {Aldrovandi}\ and\ \citenamefont
  {Pereira}(2013)}]{Aldrovandi:2013wha}%
  \BibitemOpen
  \bibfield  {author} {\bibinfo {author} {\bibfnamefont {R.}~\bibnamefont
  {Aldrovandi}}\ and\ \bibinfo {author} {\bibfnamefont {J.~G.}\ \bibnamefont
  {Pereira}},\ }\href {\doibase 10.1007/978-94-007-5143-9} {\emph {\bibinfo
  {title} {Teleparallel Gravity: An Introduction}}},\ \bibinfo {series}
  {Fundamental Theories of Physics}, Vol.\ \bibinfo {volume} {173}\ (\bibinfo
  {publisher} {Springer},\ \bibinfo {address} {Dordrecht},\ \bibinfo {year}
  {2013})\BibitemShut {NoStop}%
\bibitem [{\citenamefont {Nester}\ and\ \citenamefont
  {Yo}(1999)}]{Nester:1998mp}%
  \BibitemOpen
  \bibfield  {author} {\bibinfo {author} {\bibfnamefont {J.~M.}\ \bibnamefont
  {Nester}}\ and\ \bibinfo {author} {\bibfnamefont {H.-J.}\ \bibnamefont
  {Yo}},\ }\href@noop {} {\bibfield  {journal} {\bibinfo  {journal} {Chin. J.
  Phys.}\ }\textbf {\bibinfo {volume} {37}},\ \bibinfo {pages} {113} (\bibinfo
  {year} {1999})},\ \Eprint {http://arxiv.org/abs/gr-qc/9809049}
  {arXiv:gr-qc/9809049 [gr-qc]} \BibitemShut {NoStop}%
\bibitem [{\citenamefont {Adak}\ \emph {et~al.}(2006)\citenamefont {Adak},
  \citenamefont {Kalay},\ and\ \citenamefont {Sert}}]{Adak:2005cd}%
  \BibitemOpen
  \bibfield  {author} {\bibinfo {author} {\bibfnamefont {M.}~\bibnamefont
  {Adak}}, \bibinfo {author} {\bibfnamefont {M.}~\bibnamefont {Kalay}}, \ and\
  \bibinfo {author} {\bibfnamefont {O.}~\bibnamefont {Sert}},\ }\href {\doibase
  10.1142/S0218271806008474} {\bibfield  {journal} {\bibinfo  {journal} {Int.
  J. Mod. Phys. D}\ }\textbf {\bibinfo {volume} {15}},\ \bibinfo {pages} {619}
  (\bibinfo {year} {2006})},\ \Eprint {http://arxiv.org/abs/gr-qc/0505025}
  {arXiv:gr-qc/0505025 [gr-qc]} \BibitemShut {NoStop}%
\bibitem [{\citenamefont {Adak}\ \emph {et~al.}(2013)\citenamefont {Adak},
  \citenamefont {Sert}, \citenamefont {Kalay},\ and\ \citenamefont
  {Sari}}]{Adak:2008gd}%
  \BibitemOpen
  \bibfield  {author} {\bibinfo {author} {\bibfnamefont {M.}~\bibnamefont
  {Adak}}, \bibinfo {author} {\bibfnamefont {O.}~\bibnamefont {Sert}}, \bibinfo
  {author} {\bibfnamefont {M.}~\bibnamefont {Kalay}}, \ and\ \bibinfo {author}
  {\bibfnamefont {M.}~\bibnamefont {Sari}},\ }\href {\doibase
  10.1142/S0217751X13501674} {\bibfield  {journal} {\bibinfo  {journal} {Int.
  J. Mod. Phys. A}\ }\textbf {\bibinfo {volume} {28}},\ \bibinfo {pages}
  {1350167} (\bibinfo {year} {2013})},\ \Eprint
  {http://arxiv.org/abs/0810.2388} {arXiv:0810.2388 [gr-qc]} \BibitemShut
  {NoStop}%
\bibitem [{\citenamefont {Jimenez}\ \emph {et~al.}(2018)\citenamefont
  {Jimenez}, \citenamefont {Heisenberg},\ and\ \citenamefont
  {Koivisto}}]{Jimenez:2017tkx}%
  \BibitemOpen
  \bibfield  {author} {\bibinfo {author} {\bibfnamefont {J.~B.}\ \bibnamefont
  {Jimenez}}, \bibinfo {author} {\bibfnamefont {L.}~\bibnamefont {Heisenberg}},
  \ and\ \bibinfo {author} {\bibfnamefont {T.}~\bibnamefont {Koivisto}},\
  }\href {\doibase 10.1103/PhysRevD.98.044048} {\bibfield  {journal} {\bibinfo
  {journal} {Phys. Rev. D}\ }\textbf {\bibinfo {volume} {98}},\ \bibinfo
  {pages} {044048} (\bibinfo {year} {2018})},\ \Eprint
  {http://arxiv.org/abs/1710.03116} {arXiv:1710.03116 [gr-qc]} \BibitemShut
  {NoStop}%
\bibitem [{\citenamefont {J{\"a}rv}\ \emph {et~al.}(2018)\citenamefont
  {J{\"a}rv}, \citenamefont {R{\"u}nkla}, \citenamefont {Saal},\ and\
  \citenamefont {Vilson}}]{Jarv:2018bgs}%
  \BibitemOpen
  \bibfield  {author} {\bibinfo {author} {\bibfnamefont {L.}~\bibnamefont
  {J{\"a}rv}}, \bibinfo {author} {\bibfnamefont {M.}~\bibnamefont
  {R{\"u}nkla}}, \bibinfo {author} {\bibfnamefont {M.}~\bibnamefont {Saal}}, \
  and\ \bibinfo {author} {\bibfnamefont {O.}~\bibnamefont {Vilson}},\ }\href
  {\doibase 10.1103/PhysRevD.97.124025} {\bibfield  {journal} {\bibinfo
  {journal} {Phys. Rev. D}\ }\textbf {\bibinfo {volume} {97}},\ \bibinfo
  {pages} {124025} (\bibinfo {year} {2018})},\ \Eprint
  {http://arxiv.org/abs/1802.00492} {arXiv:1802.00492 [gr-qc]} \BibitemShut
  {NoStop}%
\bibitem [{\citenamefont {Heisenberg}(2024)}]{Heisenberg:2023lru}%
  \BibitemOpen
  \bibfield  {author} {\bibinfo {author} {\bibfnamefont {L.}~\bibnamefont
  {Heisenberg}},\ }\href {\doibase 10.1016/j.physrep.2024.02.001} {\bibfield
  {journal} {\bibinfo  {journal} {Phys. Rept.}\ }\textbf {\bibinfo {volume}
  {1066}},\ \bibinfo {pages} {1} (\bibinfo {year} {2024})},\ \Eprint
  {http://arxiv.org/abs/2309.15958} {arXiv:2309.15958 [gr-qc]} \BibitemShut
  {NoStop}%
\bibitem [{\citenamefont {Hohmann}(2021)}]{Hohmann:2021ast}%
  \BibitemOpen
  \bibfield  {author} {\bibinfo {author} {\bibfnamefont {M.}~\bibnamefont
  {Hohmann}},\ }\href {\doibase 10.3390/universe7050114} {\bibfield  {journal}
  {\bibinfo  {journal} {Universe}\ }\textbf {\bibinfo {volume} {7}},\ \bibinfo
  {pages} {114} (\bibinfo {year} {2021})},\ \Eprint
  {http://arxiv.org/abs/2104.00536} {arXiv:2104.00536 [gr-qc]} \BibitemShut
  {NoStop}%
\bibitem [{\citenamefont {Hohmann}(2024)}]{Hohmann:2023xxx}%
  \BibitemOpen
  \bibfield  {author} {\bibinfo {author} {\bibfnamefont {M.}~\bibnamefont
  {Hohmann}},\ }\href {\doibase 10.1103/PhysRevD.109.064003} {\bibfield
  {journal} {\bibinfo  {journal} {Phys. Rev. D}\ }\textbf {\bibinfo {volume}
  {109}},\ \bibinfo {pages} {064003} (\bibinfo {year} {2024})},\ \Eprint
  {http://arxiv.org/abs/2312.17609} {arXiv:2312.17609 [gr-qc]} \BibitemShut
  {NoStop}%
\bibitem [{\citenamefont {Heisenberg}\ \emph {et~al.}(2023)\citenamefont
  {Heisenberg}, \citenamefont {Hohmann},\ and\ \citenamefont
  {Kuhn}}]{Heisenberg:2022xxx}%
  \BibitemOpen
  \bibfield  {author} {\bibinfo {author} {\bibfnamefont {L.}~\bibnamefont
  {Heisenberg}}, \bibinfo {author} {\bibfnamefont {M.}~\bibnamefont {Hohmann}},
  \ and\ \bibinfo {author} {\bibfnamefont {S.}~\bibnamefont {Kuhn}},\ }\href
  {\doibase 10.1140/epjc/s10052-023-11462-6} {\bibfield  {journal} {\bibinfo
  {journal} {Eur. Phys. J. C}\ }\textbf {\bibinfo {volume} {83}},\ \bibinfo
  {pages} {315} (\bibinfo {year} {2023})},\ \Eprint
  {http://arxiv.org/abs/2212.14324} {arXiv:2212.14324 [gr-qc]} \BibitemShut
  {NoStop}%
\bibitem [{\citenamefont {Heisenberg}\ and\ \citenamefont
  {Hohmann}(2024)}]{Heisenberg:2023tho}%
  \BibitemOpen
  \bibfield  {author} {\bibinfo {author} {\bibfnamefont {L.}~\bibnamefont
  {Heisenberg}}\ and\ \bibinfo {author} {\bibfnamefont {M.}~\bibnamefont
  {Hohmann}},\ }\href {\doibase 10.1140/epjc/s10052-024-12810-w} {\bibfield
  {journal} {\bibinfo  {journal} {Eur. Phys. J. C}\ }\textbf {\bibinfo {volume}
  {84}},\ \bibinfo {pages} {462} (\bibinfo {year} {2024})},\ \Eprint
  {http://arxiv.org/abs/2311.05597} {arXiv:2311.05597 [gr-qc]} \BibitemShut
  {NoStop}%
\bibitem [{\citenamefont {Adak}\ \emph {et~al.}(2023)\citenamefont {Adak},
  \citenamefont {Dereli}, \citenamefont {Koivisto},\ and\ \citenamefont
  {Pala}}]{Adak:2023xxx}%
  \BibitemOpen
  \bibfield  {author} {\bibinfo {author} {\bibfnamefont {M.}~\bibnamefont
  {Adak}}, \bibinfo {author} {\bibfnamefont {T.}~\bibnamefont {Dereli}},
  \bibinfo {author} {\bibfnamefont {T.~S.}\ \bibnamefont {Koivisto}}, \ and\
  \bibinfo {author} {\bibfnamefont {C.}~\bibnamefont {Pala}},\ }\href {\doibase
  10.1142/S0219887823502158} {\bibfield  {journal} {\bibinfo  {journal} {Int.
  J. Geom. Meth. Mod. Phys.}\ }\textbf {\bibinfo {volume} {20}},\ \bibinfo
  {pages} {2350215} (\bibinfo {year} {2023})},\ \Eprint
  {http://arxiv.org/abs/2303.17812} {arXiv:2303.17812 [gr-qc]} \BibitemShut
  {NoStop}%
\bibitem [{\citenamefont {Bajardi}\ and\ \citenamefont
  {Blixt}(2024)}]{Bajardi:2024xxx}%
  \BibitemOpen
  \bibfield  {author} {\bibinfo {author} {\bibfnamefont {F.}~\bibnamefont
  {Bajardi}}\ and\ \bibinfo {author} {\bibfnamefont {D.}~\bibnamefont
  {Blixt}},\ }\href {\doibase 10.1103/PhysRevD.109.084078} {\bibfield
  {journal} {\bibinfo  {journal} {Phys. Rev. D}\ }\textbf {\bibinfo {volume}
  {109}},\ \bibinfo {pages} {084078} (\bibinfo {year} {2024})},\ \Eprint
  {http://arxiv.org/abs/2401.11591} {arXiv:2401.11591 [gr-qc]} \BibitemShut
  {NoStop}%
\bibitem [{\citenamefont {Beltr{\'a}n~Jim{\'e}nez}\ and\ \citenamefont
  {Koivisto}(2022{\natexlab{a}})}]{BeltranJimenez:2022azb}%
  \BibitemOpen
  \bibfield  {author} {\bibinfo {author} {\bibfnamefont {J.}~\bibnamefont
  {Beltr{\'a}n~Jim{\'e}nez}}\ and\ \bibinfo {author} {\bibfnamefont {T.~S.}\
  \bibnamefont {Koivisto}},\ }\href {\doibase 10.1142/S0219887822501080}
  {\bibfield  {journal} {\bibinfo  {journal} {Int. J. Geom. Meth. Mod. Phys.}\
  }\textbf {\bibinfo {volume} {19}},\ \bibinfo {pages} {2250108} (\bibinfo
  {year} {2022}{\natexlab{a}})},\ \Eprint {http://arxiv.org/abs/2202.01701}
  {arXiv:2202.01701 [gr-qc]} \BibitemShut {NoStop}%
\bibitem [{\citenamefont {Beltr{\'a}n~Jim{\'e}nez}\ \emph
  {et~al.}(2020{\natexlab{b}})\citenamefont {Beltr{\'a}n~Jim{\'e}nez},
  \citenamefont {Heisenberg}, \citenamefont {Iosifidis}, \citenamefont
  {Jim{\'e}nez-Cano},\ and\ \citenamefont {Koivisto}}]{BeltranJimenez:2019odq}%
  \BibitemOpen
  \bibfield  {author} {\bibinfo {author} {\bibfnamefont {J.}~\bibnamefont
  {Beltr{\'a}n~Jim{\'e}nez}}, \bibinfo {author} {\bibfnamefont
  {L.}~\bibnamefont {Heisenberg}}, \bibinfo {author} {\bibfnamefont
  {D.}~\bibnamefont {Iosifidis}}, \bibinfo {author} {\bibfnamefont
  {A.}~\bibnamefont {Jim{\'e}nez-Cano}}, \ and\ \bibinfo {author}
  {\bibfnamefont {T.~S.}\ \bibnamefont {Koivisto}},\ }\href {\doibase
  10.1016/j.physletb.2020.135422} {\bibfield  {journal} {\bibinfo  {journal}
  {Phys. Lett. B}\ }\textbf {\bibinfo {volume} {805}},\ \bibinfo {pages}
  {135422} (\bibinfo {year} {2020}{\natexlab{b}})},\ \Eprint
  {http://arxiv.org/abs/1909.09045} {arXiv:1909.09045 [gr-qc]} \BibitemShut
  {NoStop}%
\bibitem [{\citenamefont {Einstein}(1928)}]{Einstein:1928tp}%
  \BibitemOpen
  \bibfield  {author} {\bibinfo {author} {\bibfnamefont {A.}~\bibnamefont
  {Einstein}},\ }\href@noop {} {\bibfield  {journal} {\bibinfo  {journal}
  {Sitzungsberichte der Preussischen Akademie der Wissenschaften,
  Physikalisch-Mathematische Klasse}\ ,\ \bibinfo {pages} {217}} (\bibinfo
  {year} {1928})}\BibitemShut {NoStop}%
\bibitem [{\citenamefont {Lee}\ and\ \citenamefont {Wald}(1990)}]{Lee:1990nz}%
  \BibitemOpen
  \bibfield  {author} {\bibinfo {author} {\bibfnamefont {J.}~\bibnamefont
  {Lee}}\ and\ \bibinfo {author} {\bibfnamefont {R.~M.}\ \bibnamefont {Wald}},\
  }\href {\doibase 10.1063/1.528801} {\bibfield  {journal} {\bibinfo  {journal}
  {J. Math. Phys.}\ }\textbf {\bibinfo {volume} {31}},\ \bibinfo {pages} {725}
  (\bibinfo {year} {1990})}\BibitemShut {NoStop}%
\bibitem [{\citenamefont {Wald}\ and\ \citenamefont
  {Zoupas}(2000)}]{Wald:1999wa}%
  \BibitemOpen
  \bibfield  {author} {\bibinfo {author} {\bibfnamefont {R.~M.}\ \bibnamefont
  {Wald}}\ and\ \bibinfo {author} {\bibfnamefont {A.}~\bibnamefont {Zoupas}},\
  }\href {\doibase 10.1103/PhysRevD.61.084027} {\bibfield  {journal} {\bibinfo
  {journal} {Phys. Rev. D}\ }\textbf {\bibinfo {volume} {61}},\ \bibinfo
  {pages} {084027} (\bibinfo {year} {2000})},\ \Eprint
  {http://arxiv.org/abs/gr-qc/9911095} {arXiv:gr-qc/9911095} \BibitemShut
  {NoStop}%
\bibitem [{\citenamefont {Ciambelli}\ and\ \citenamefont
  {Geiller}(2025)}]{Ciambelli:2024vhy}%
  \BibitemOpen
  \bibfield  {author} {\bibinfo {author} {\bibfnamefont {L.}~\bibnamefont
  {Ciambelli}}\ and\ \bibinfo {author} {\bibfnamefont {M.}~\bibnamefont
  {Geiller}},\ }\href {\doibase 10.1007/JHEP05(2025)022} {\bibfield  {journal}
  {\bibinfo  {journal} {JHEP}\ }\textbf {\bibinfo {volume} {05}},\ \bibinfo
  {pages} {022} (\bibinfo {year} {2025})},\ \Eprint
  {http://arxiv.org/abs/2412.14992} {arXiv:2412.14992 [hep-th]} \BibitemShut
  {NoStop}%
\bibitem [{\citenamefont {Awad}(2009)}]{Awad:2007me}%
  \BibitemOpen
  \bibfield  {author} {\bibinfo {author} {\bibfnamefont {A.~M.}\ \bibnamefont
  {Awad}},\ }\href {\doibase 10.1142/S0218271809014522} {\bibfield  {journal}
  {\bibinfo  {journal} {Int. J. Mod. Phys. D}\ }\textbf {\bibinfo {volume}
  {18}},\ \bibinfo {pages} {405} (\bibinfo {year} {2009})},\ \Eprint
  {http://arxiv.org/abs/0708.3458} {arXiv:0708.3458 [hep-th]} \BibitemShut
  {NoStop}%
\bibitem [{\citenamefont {Elgood}\ \emph {et~al.}(2020)\citenamefont {Elgood},
  \citenamefont {Meessen},\ and\ \citenamefont {Ort{\'\i}n}}]{Elgood:2020svt}%
  \BibitemOpen
  \bibfield  {author} {\bibinfo {author} {\bibfnamefont {Z.}~\bibnamefont
  {Elgood}}, \bibinfo {author} {\bibfnamefont {P.}~\bibnamefont {Meessen}}, \
  and\ \bibinfo {author} {\bibfnamefont {T.}~\bibnamefont {Ort{\'\i}n}},\
  }\href {\doibase 10.1007/JHEP09(2020)026} {\bibfield  {journal} {\bibinfo
  {journal} {JHEP}\ }\textbf {\bibinfo {volume} {09}},\ \bibinfo {pages} {026}
  (\bibinfo {year} {2020})},\ \Eprint {http://arxiv.org/abs/2006.02792}
  {arXiv:2006.02792 [hep-th]} \BibitemShut {NoStop}%
\bibitem [{\citenamefont {Prabhu}(2017)}]{Prabhu:2015vua}%
  \BibitemOpen
  \bibfield  {author} {\bibinfo {author} {\bibfnamefont {K.}~\bibnamefont
  {Prabhu}},\ }\href {\doibase 10.1088/1361-6382/aa536b} {\bibfield  {journal}
  {\bibinfo  {journal} {Class. Quant. Grav.}\ }\textbf {\bibinfo {volume}
  {34}},\ \bibinfo {pages} {035011} (\bibinfo {year} {2017})},\ \Eprint
  {http://arxiv.org/abs/1511.00388} {arXiv:1511.00388 [gr-qc]} \BibitemShut
  {NoStop}%
\bibitem [{\citenamefont {Itin}(2002)}]{Itin:2001tv}%
  \BibitemOpen
  \bibfield  {author} {\bibinfo {author} {\bibfnamefont {Y.}~\bibnamefont
  {Itin}},\ }\href {\doibase 10.1142/S0217751X02011928} {\bibfield  {journal}
  {\bibinfo  {journal} {Int. J. Mod. Phys. A}\ }\textbf {\bibinfo {volume}
  {17}},\ \bibinfo {pages} {2765} (\bibinfo {year} {2002})},\ \Eprint
  {http://arxiv.org/abs/gr-qc/0103017} {arXiv:gr-qc/0103017} \BibitemShut
  {NoStop}%
\bibitem [{\citenamefont {Obukhov}\ \emph {et~al.}(2006)\citenamefont
  {Obukhov}, \citenamefont {Rubilar},\ and\ \citenamefont
  {Pereira}}]{Obukhov:2006sk}%
  \BibitemOpen
  \bibfield  {author} {\bibinfo {author} {\bibfnamefont {Y.~N.}\ \bibnamefont
  {Obukhov}}, \bibinfo {author} {\bibfnamefont {G.~F.}\ \bibnamefont
  {Rubilar}}, \ and\ \bibinfo {author} {\bibfnamefont {J.~G.}\ \bibnamefont
  {Pereira}},\ }\href {\doibase 10.1103/PhysRevD.74.104007} {\bibfield
  {journal} {\bibinfo  {journal} {Phys. Rev. D}\ }\textbf {\bibinfo {volume}
  {74}},\ \bibinfo {pages} {104007} (\bibinfo {year} {2006})},\ \Eprint
  {http://arxiv.org/abs/gr-qc/0610092} {arXiv:gr-qc/0610092} \BibitemShut
  {NoStop}%
\bibitem [{\citenamefont {Capozziello}\ \emph {et~al.}(2018)\citenamefont
  {Capozziello}, \citenamefont {Capriolo},\ and\ \citenamefont
  {Transirico}}]{Capozziello:2018qcp}%
  \BibitemOpen
  \bibfield  {author} {\bibinfo {author} {\bibfnamefont {S.}~\bibnamefont
  {Capozziello}}, \bibinfo {author} {\bibfnamefont {M.}~\bibnamefont
  {Capriolo}}, \ and\ \bibinfo {author} {\bibfnamefont {M.}~\bibnamefont
  {Transirico}},\ }\href {\doibase 10.1142/S0219887818501645} {\bibfield
  {journal} {\bibinfo  {journal} {Int. J. Geom. Meth. Mod. Phys.}\ }\textbf
  {\bibinfo {volume} {15}},\ \bibinfo {pages} {1850164} (\bibinfo {year}
  {2018})},\ \Eprint {http://arxiv.org/abs/1804.08530} {arXiv:1804.08530
  [gr-qc]} \BibitemShut {NoStop}%
\bibitem [{\citenamefont {Hammad}\ \emph {et~al.}(2019)\citenamefont {Hammad},
  \citenamefont {Dijamco}, \citenamefont {Torres-Rivas},\ and\ \citenamefont
  {B{\'e}rub{\'e}}}]{Hammad:2019oyb}%
  \BibitemOpen
  \bibfield  {author} {\bibinfo {author} {\bibfnamefont {F.}~\bibnamefont
  {Hammad}}, \bibinfo {author} {\bibfnamefont {D.}~\bibnamefont {Dijamco}},
  \bibinfo {author} {\bibfnamefont {A.}~\bibnamefont {Torres-Rivas}}, \ and\
  \bibinfo {author} {\bibfnamefont {D.}~\bibnamefont {B{\'e}rub{\'e}}},\ }\href
  {\doibase 10.1103/PhysRevD.100.124040} {\bibfield  {journal} {\bibinfo
  {journal} {Phys. Rev. D}\ }\textbf {\bibinfo {volume} {100}},\ \bibinfo
  {pages} {124040} (\bibinfo {year} {2019})},\ \Eprint
  {http://arxiv.org/abs/1912.08811} {arXiv:1912.08811 [gr-qc]} \BibitemShut
  {NoStop}%
\bibitem [{\citenamefont {Emtsova}\ \emph {et~al.}(2020)\citenamefont
  {Emtsova}, \citenamefont {Petrov},\ and\ \citenamefont
  {Toporensky}}]{Emtsova:2019moq}%
  \BibitemOpen
  \bibfield  {author} {\bibinfo {author} {\bibfnamefont {E.~D.}\ \bibnamefont
  {Emtsova}}, \bibinfo {author} {\bibfnamefont {A.~N.}\ \bibnamefont {Petrov}},
  \ and\ \bibinfo {author} {\bibfnamefont {A.~V.}\ \bibnamefont {Toporensky}},\
  }\href {\doibase 10.1088/1361-6382/ab7715} {\bibfield  {journal} {\bibinfo
  {journal} {Class. Quant. Grav.}\ }\textbf {\bibinfo {volume} {37}},\ \bibinfo
  {pages} {095006} (\bibinfo {year} {2020})},\ \Eprint
  {http://arxiv.org/abs/1910.08960} {arXiv:1910.08960 [gr-qc]} \BibitemShut
  {NoStop}%
\bibitem [{\citenamefont {Beltr{\'a}n~Jim{\'e}nez}\ and\ \citenamefont
  {Koivisto}(2022{\natexlab{b}})}]{BeltranJimenez:2021edg}%
  \BibitemOpen
  \bibfield  {author} {\bibinfo {author} {\bibfnamefont {J.}~\bibnamefont
  {Beltr{\'a}n~Jim{\'e}nez}}\ and\ \bibinfo {author} {\bibfnamefont {T.~S.}\
  \bibnamefont {Koivisto}},\ }\href {\doibase 10.1103/PhysRevD.105.L021502}
  {\bibfield  {journal} {\bibinfo  {journal} {Phys. Rev. D}\ }\textbf {\bibinfo
  {volume} {105}},\ \bibinfo {pages} {L021502} (\bibinfo {year}
  {2022}{\natexlab{b}})},\ \Eprint {http://arxiv.org/abs/2111.04716}
  {arXiv:2111.04716 [gr-qc]} \BibitemShut {NoStop}%
\bibitem [{\citenamefont {Gomes}\ \emph
  {et~al.}(2023{\natexlab{b}})\citenamefont {Gomes}, \citenamefont
  {Beltr{\'a}n~Jim{\'e}nez},\ and\ \citenamefont {Koivisto}}]{Gomes:2022vrc}%
  \BibitemOpen
  \bibfield  {author} {\bibinfo {author} {\bibfnamefont {D.~A.}\ \bibnamefont
  {Gomes}}, \bibinfo {author} {\bibfnamefont {J.}~\bibnamefont
  {Beltr{\'a}n~Jim{\'e}nez}}, \ and\ \bibinfo {author} {\bibfnamefont {T.~S.}\
  \bibnamefont {Koivisto}},\ }\href {\doibase 10.1103/PhysRevD.107.024044}
  {\bibfield  {journal} {\bibinfo  {journal} {Phys. Rev. D}\ }\textbf {\bibinfo
  {volume} {107}},\ \bibinfo {pages} {024044} (\bibinfo {year}
  {2023}{\natexlab{b}})},\ \Eprint {http://arxiv.org/abs/2205.09716}
  {arXiv:2205.09716 [gr-qc]} \BibitemShut {NoStop}%
\bibitem [{\citenamefont {Heisenberg}\ \emph {et~al.}(2022)\citenamefont
  {Heisenberg}, \citenamefont {Kuhn},\ and\ \citenamefont
  {Walleghem}}]{Heisenberg:2022gqk}%
  \BibitemOpen
  \bibfield  {author} {\bibinfo {author} {\bibfnamefont {L.}~\bibnamefont
  {Heisenberg}}, \bibinfo {author} {\bibfnamefont {S.}~\bibnamefont {Kuhn}}, \
  and\ \bibinfo {author} {\bibfnamefont {L.}~\bibnamefont {Walleghem}},\ }\href
  {\doibase 10.1088/1361-6382/ac987d} {\bibfield  {journal} {\bibinfo
  {journal} {Class. Quant. Grav.}\ }\textbf {\bibinfo {volume} {39}},\ \bibinfo
  {pages} {235002} (\bibinfo {year} {2022})},\ \Eprint
  {http://arxiv.org/abs/2203.13914} {arXiv:2203.13914 [gr-qc]} \BibitemShut
  {NoStop}%
\bibitem [{\citenamefont {Emtsova}\ \emph {et~al.}(2023)\citenamefont
  {Emtsova}, \citenamefont {Petrov},\ and\ \citenamefont
  {Toporensky}}]{Emtsova:2023mwd}%
  \BibitemOpen
  \bibfield  {author} {\bibinfo {author} {\bibfnamefont {E.~D.}\ \bibnamefont
  {Emtsova}}, \bibinfo {author} {\bibfnamefont {A.~N.}\ \bibnamefont {Petrov}},
  \ and\ \bibinfo {author} {\bibfnamefont {A.~V.}\ \bibnamefont {Toporensky}},\
  }\href {\doibase 10.1140/epjc/s10052-023-11460-8} {\bibfield  {journal}
  {\bibinfo  {journal} {Eur. Phys. J. C}\ }\textbf {\bibinfo {volume} {83}},\
  \bibinfo {pages} {366} (\bibinfo {year} {2023})},\ \Eprint
  {http://arxiv.org/abs/2212.03755} {arXiv:2212.03755 [gr-qc]} \BibitemShut
  {NoStop}%
\bibitem [{\citenamefont {Emtsova}\ \emph {et~al.}(2025)\citenamefont
  {Emtsova}, \citenamefont {Petrov},\ and\ \citenamefont
  {Toporensky}}]{Emtsova:2025}%
  \BibitemOpen
  \bibfield  {author} {\bibinfo {author} {\bibfnamefont {E.~D.}\ \bibnamefont
  {Emtsova}}, \bibinfo {author} {\bibfnamefont {A.~N.}\ \bibnamefont {Petrov}},
  \ and\ \bibinfo {author} {\bibfnamefont {A.~V.}\ \bibnamefont {Toporensky}},\
  }\href {\doibase 10.3367/UFNe.2025.05.039924} {\bibfield  {journal} {\bibinfo
   {journal} {Phys. Usp.}\ } (\bibinfo {year} {2025}),\
  10.3367/UFNe.2025.05.039924},\ \bibinfo {note} {accepted for publication},\
  \Eprint {http://arxiv.org/abs/2505.18084} {arXiv:2505.18084 [gr-qc]}
  \BibitemShut {NoStop}%
\bibitem [{\citenamefont {Capozziello}\ \emph {et~al.}(2026)\citenamefont
  {Capozziello}, \citenamefont {Capriolo},\ and\ \citenamefont
  {Lambiase}}]{Capozziello:2026rrc}%
  \BibitemOpen
  \bibfield  {author} {\bibinfo {author} {\bibfnamefont {S.}~\bibnamefont
  {Capozziello}}, \bibinfo {author} {\bibfnamefont {M.}~\bibnamefont
  {Capriolo}}, \ and\ \bibinfo {author} {\bibfnamefont {G.}~\bibnamefont
  {Lambiase}},\ }\href {\doibase 10.1103/1kvt-vvcf} {\bibfield  {journal}
  {\bibinfo  {journal} {Phys. Rev. D}\ }\textbf {\bibinfo {volume} {113}},\
  \bibinfo {pages} {044017} (\bibinfo {year} {2026})},\ \Eprint
  {http://arxiv.org/abs/2601.12088} {arXiv:2601.12088 [gr-qc]} \BibitemShut
  {NoStop}%
\bibitem [{\citenamefont {Nester}(1989)}]{Nester:1989TeleHamiltonian}%
  \BibitemOpen
  \bibfield  {author} {\bibinfo {author} {\bibfnamefont {J.~M.}\ \bibnamefont
  {Nester}},\ }\href {\doibase 10.1142/S0217751X89000704} {\bibfield  {journal}
  {\bibinfo  {journal} {Int. J. Mod. Phys. A}\ }\textbf {\bibinfo {volume}
  {4}},\ \bibinfo {pages} {1755} (\bibinfo {year} {1989})}\BibitemShut
  {NoStop}%
\bibitem [{\citenamefont {Maluf}\ and\ \citenamefont
  {da~Rocha-Neto}(2000)}]{Maluf:2000ag}%
  \BibitemOpen
  \bibfield  {author} {\bibinfo {author} {\bibfnamefont {J.~W.}\ \bibnamefont
  {Maluf}}\ and\ \bibinfo {author} {\bibfnamefont {J.~F.}\ \bibnamefont
  {da~Rocha-Neto}},\ }\href@noop {} {\  (\bibinfo {year} {2000})},\ \Eprint
  {http://arxiv.org/abs/gr-qc/0002059} {arXiv:gr-qc/0002059} \BibitemShut
  {NoStop}%
\bibitem [{\citenamefont {Maluf}(1994)}]{Maluf:1994ji}%
  \BibitemOpen
  \bibfield  {author} {\bibinfo {author} {\bibfnamefont {J.~W.}\ \bibnamefont
  {Maluf}},\ }\href {\doibase 10.1063/1.530774} {\bibfield  {journal} {\bibinfo
   {journal} {J. Math. Phys.}\ }\textbf {\bibinfo {volume} {35}},\ \bibinfo
  {pages} {335} (\bibinfo {year} {1994})}\BibitemShut {NoStop}%
\bibitem [{\citenamefont {Blagojevic}\ and\ \citenamefont
  {Nikolic}(2000)}]{Blagojevic:2000qs}%
  \BibitemOpen
  \bibfield  {author} {\bibinfo {author} {\bibfnamefont {M.}~\bibnamefont
  {Blagojevic}}\ and\ \bibinfo {author} {\bibfnamefont {I.~A.}\ \bibnamefont
  {Nikolic}},\ }\href {\doibase 10.1103/PhysRevD.62.024021} {\bibfield
  {journal} {\bibinfo  {journal} {Phys. Rev. D}\ }\textbf {\bibinfo {volume}
  {62}},\ \bibinfo {pages} {024021} (\bibinfo {year} {2000})},\ \Eprint
  {http://arxiv.org/abs/hep-th/0002022} {arXiv:hep-th/0002022} \BibitemShut
  {NoStop}%
\bibitem [{\citenamefont {Okol{\'o}w}(2013)}]{Okolow:2011nq}%
  \BibitemOpen
  \bibfield  {author} {\bibinfo {author} {\bibfnamefont {A.}~\bibnamefont
  {Okol{\'o}w}},\ }\href {\doibase 10.1007/s10714-013-1605-y} {\bibfield
  {journal} {\bibinfo  {journal} {Gen. Rel. Grav.}\ }\textbf {\bibinfo {volume}
  {45}},\ \bibinfo {pages} {2569} (\bibinfo {year} {2013})},\ \Eprint
  {http://arxiv.org/abs/1111.5498} {arXiv:1111.5498 [gr-qc]} \BibitemShut
  {NoStop}%
\bibitem [{\citenamefont {Blixt}\ \emph {et~al.}(2021)\citenamefont {Blixt},
  \citenamefont {Guzm{\'a}n}, \citenamefont {Hohmann},\ and\ \citenamefont
  {Pfeifer}}]{Blixt:2020ekl}%
  \BibitemOpen
  \bibfield  {author} {\bibinfo {author} {\bibfnamefont {D.}~\bibnamefont
  {Blixt}}, \bibinfo {author} {\bibfnamefont {M.-J.}\ \bibnamefont
  {Guzm{\'a}n}}, \bibinfo {author} {\bibfnamefont {M.}~\bibnamefont {Hohmann}},
  \ and\ \bibinfo {author} {\bibfnamefont {C.}~\bibnamefont {Pfeifer}},\ }\href
  {\doibase 10.1142/S0219887821300051} {\bibfield  {journal} {\bibinfo
  {journal} {Int. J. Geom. Meth. Mod. Phys.}\ }\textbf {\bibinfo {volume}
  {18}},\ \bibinfo {pages} {2130005} (\bibinfo {year} {2021})},\ \Eprint
  {http://arxiv.org/abs/2012.09180} {arXiv:2012.09180 [gr-qc]} \BibitemShut
  {NoStop}%
\bibitem [{\citenamefont {D'Ambrosio}\ \emph {et~al.}(2023)\citenamefont
  {D'Ambrosio}, \citenamefont {Heisenberg},\ and\ \citenamefont
  {Zentarra}}]{DAmbrosio:2023asf}%
  \BibitemOpen
  \bibfield  {author} {\bibinfo {author} {\bibfnamefont {F.}~\bibnamefont
  {D'Ambrosio}}, \bibinfo {author} {\bibfnamefont {L.}~\bibnamefont
  {Heisenberg}}, \ and\ \bibinfo {author} {\bibfnamefont {S.}~\bibnamefont
  {Zentarra}},\ }\href {\doibase 10.1002/prop.202300185} {\bibfield  {journal}
  {\bibinfo  {journal} {Fortsch. Phys.}\ }\textbf {\bibinfo {volume} {71}},\
  \bibinfo {pages} {2300185} (\bibinfo {year} {2023})},\ \Eprint
  {http://arxiv.org/abs/2308.02250} {arXiv:2308.02250 [gr-qc]} \BibitemShut
  {NoStop}%
\bibitem [{\citenamefont {Tomonari}\ and\ \citenamefont
  {Bahamonde}(2024)}]{Tomonari:2023wcs}%
  \BibitemOpen
  \bibfield  {author} {\bibinfo {author} {\bibfnamefont {K.}~\bibnamefont
  {Tomonari}}\ and\ \bibinfo {author} {\bibfnamefont {S.}~\bibnamefont
  {Bahamonde}},\ }\href {\doibase 10.1140/epjc/s10052-024-12677-x} {\bibfield
  {journal} {\bibinfo  {journal} {Eur. Phys. J. C}\ }\textbf {\bibinfo {volume}
  {84}},\ \bibinfo {pages} {349} (\bibinfo {year} {2024})},\ \bibinfo {note}
  {[Erratum: Eur.Phys.J.C 84, 508 (2024)]},\ \Eprint
  {http://arxiv.org/abs/2308.06469} {arXiv:2308.06469 [gr-qc]} \BibitemShut
  {NoStop}%
\bibitem [{\citenamefont {Chen}\ \emph {et~al.}(2026)\citenamefont {Chen},
  \citenamefont {Chen},\ and\ \citenamefont {Gao}}]{Chen:2025mhu}%
  \BibitemOpen
  \bibfield  {author} {\bibinfo {author} {\bibfnamefont {J.-J.}\ \bibnamefont
  {Chen}}, \bibinfo {author} {\bibfnamefont {Z.}~\bibnamefont {Chen}}, \ and\
  \bibinfo {author} {\bibfnamefont {X.}~\bibnamefont {Gao}},\ }\href {\doibase
  10.1103/jv1c-417c} {\bibfield  {journal} {\bibinfo  {journal} {Phys. Rev. D}\
  }\textbf {\bibinfo {volume} {113}},\ \bibinfo {pages} {10} (\bibinfo {year}
  {2026})},\ \Eprint {http://arxiv.org/abs/2512.24298} {arXiv:2512.24298
  [gr-qc]} \BibitemShut {NoStop}%
\bibitem [{\citenamefont {Capozziello}\ and\ \citenamefont
  {Sauro}(2026)}]{Capozziello:2026ccp}%
  \BibitemOpen
  \bibfield  {author} {\bibinfo {author} {\bibfnamefont {S.}~\bibnamefont
  {Capozziello}}\ and\ \bibinfo {author} {\bibfnamefont {D.}~\bibnamefont
  {Sauro}},\ }\href@noop {} {\  (\bibinfo {year} {2026})},\ \Eprint
  {http://arxiv.org/abs/2604.19310} {arXiv:2604.19310 [gr-qc]} \BibitemShut
  {NoStop}%
\bibitem [{\citenamefont {Jacobson}(1994)}]{Jacobson:1994iw}%
  \BibitemOpen
  \bibfield  {author} {\bibinfo {author} {\bibfnamefont {T.}~\bibnamefont
  {Jacobson}},\ }\href@noop {} {\  (\bibinfo {year} {1994})},\ \Eprint
  {http://arxiv.org/abs/gr-qc/9404039} {arXiv:gr-qc/9404039} \BibitemShut
  {NoStop}%
\bibitem [{\citenamefont {Frolov}\ and\ \citenamefont
  {Fursaev}(1997)}]{Frolov:1997up}%
  \BibitemOpen
  \bibfield  {author} {\bibinfo {author} {\bibfnamefont {V.~P.}\ \bibnamefont
  {Frolov}}\ and\ \bibinfo {author} {\bibfnamefont {D.~V.}\ \bibnamefont
  {Fursaev}},\ }\href {\doibase 10.1103/PhysRevD.56.2212} {\bibfield  {journal}
  {\bibinfo  {journal} {Phys. Rev. D}\ }\textbf {\bibinfo {volume} {56}},\
  \bibinfo {pages} {2212} (\bibinfo {year} {1997})},\ \Eprint
  {http://arxiv.org/abs/hep-th/9703178} {arXiv:hep-th/9703178} \BibitemShut
  {NoStop}%
\bibitem [{\citenamefont {Solodukhin}(2011)}]{Solodukhin:2011gn}%
  \BibitemOpen
  \bibfield  {author} {\bibinfo {author} {\bibfnamefont {S.~N.}\ \bibnamefont
  {Solodukhin}},\ }\href {\doibase 10.12942/lrr-2011-8} {\bibfield  {journal}
  {\bibinfo  {journal} {Living Rev. Rel.}\ }\textbf {\bibinfo {volume} {14}},\
  \bibinfo {pages} {8} (\bibinfo {year} {2011})},\ \Eprint
  {http://arxiv.org/abs/1104.3712} {arXiv:1104.3712 [hep-th]} \BibitemShut
  {NoStop}%
\bibitem [{\citenamefont {Krasnov}(2020)}]{Krasnov:2020lku}%
  \BibitemOpen
  \bibfield  {author} {\bibinfo {author} {\bibfnamefont {K.}~\bibnamefont
  {Krasnov}},\ }\href {\doibase 10.1017/9781108674652} {\emph {\bibinfo {title}
  {{Formulations of General Relativity}}}},\ Cambridge Monographs on
  Mathematical Physics\ (\bibinfo  {publisher} {Cambridge University Press},\
  \bibinfo {year} {2020})\BibitemShut {NoStop}%
\bibitem [{\citenamefont {Carlip}\ and\ \citenamefont
  {Teitelboim}(1995)}]{Carlip:1993sa}%
  \BibitemOpen
  \bibfield  {author} {\bibinfo {author} {\bibfnamefont {S.}~\bibnamefont
  {Carlip}}\ and\ \bibinfo {author} {\bibfnamefont {C.}~\bibnamefont
  {Teitelboim}},\ }\href {\doibase 10.1088/0264-9381/12/7/011} {\bibfield
  {journal} {\bibinfo  {journal} {Class. Quant. Grav.}\ }\textbf {\bibinfo
  {volume} {12}},\ \bibinfo {pages} {1699} (\bibinfo {year} {1995})},\ \Eprint
  {http://arxiv.org/abs/gr-qc/9312002} {arXiv:gr-qc/9312002} \BibitemShut
  {NoStop}%
\bibitem [{\citenamefont {Chandrasekaran}\ and\ \citenamefont
  {Flanagan}(2024)}]{Chandrasekaran:2023vzb}%
  \BibitemOpen
  \bibfield  {author} {\bibinfo {author} {\bibfnamefont {V.}~\bibnamefont
  {Chandrasekaran}}\ and\ \bibinfo {author} {\bibfnamefont {E.~E.}\
  \bibnamefont {Flanagan}},\ }\href {\doibase 10.1007/JHEP07(2024)017}
  {\bibfield  {journal} {\bibinfo  {journal} {JHEP}\ }\textbf {\bibinfo
  {volume} {07}},\ \bibinfo {pages} {017} (\bibinfo {year} {2024})},\ \Eprint
  {http://arxiv.org/abs/2309.03871} {arXiv:2309.03871 [hep-th]} \BibitemShut
  {NoStop}%
\bibitem [{\citenamefont {Blagojevi{\'c}}\ and\ \citenamefont
  {Cvetkovi{\'c}}(2019{\natexlab{a}})}]{Blagojevic:2019gsd}%
  \BibitemOpen
  \bibfield  {author} {\bibinfo {author} {\bibfnamefont {M.}~\bibnamefont
  {Blagojevi{\'c}}}\ and\ \bibinfo {author} {\bibfnamefont {B.}~\bibnamefont
  {Cvetkovi{\'c}}},\ }\href {\doibase 10.1103/PhysRevD.99.104058} {\bibfield
  {journal} {\bibinfo  {journal} {Phys. Rev. D}\ }\textbf {\bibinfo {volume}
  {99}},\ \bibinfo {pages} {104058} (\bibinfo {year} {2019}{\natexlab{a}})},\
  \Eprint {http://arxiv.org/abs/1903.02263} {arXiv:1903.02263 [gr-qc]}
  \BibitemShut {NoStop}%
\bibitem [{\citenamefont {Blagojevi{\'c}}\ and\ \citenamefont
  {Cvetkovi{\'c}}(2019{\natexlab{b}})}]{Blagojevic:2019bqg}%
  \BibitemOpen
  \bibfield  {author} {\bibinfo {author} {\bibfnamefont {M.}~\bibnamefont
  {Blagojevi{\'c}}}\ and\ \bibinfo {author} {\bibfnamefont {B.}~\bibnamefont
  {Cvetkovi{\'c}}},\ }\href {\doibase 10.1103/PhysRevD.100.044029} {\bibfield
  {journal} {\bibinfo  {journal} {Phys. Rev. D}\ }\textbf {\bibinfo {volume}
  {100}},\ \bibinfo {pages} {044029} (\bibinfo {year} {2019}{\natexlab{b}})},\
  \Eprint {http://arxiv.org/abs/1905.04928} {arXiv:1905.04928 [gr-qc]}
  \BibitemShut {NoStop}%
\bibitem [{\citenamefont {Beltr{\'a}n~Jim{\'e}nez}\ \emph
  {et~al.}(2018)\citenamefont {Beltr{\'a}n~Jim{\'e}nez}, \citenamefont
  {Heisenberg},\ and\ \citenamefont {Koivisto}}]{BeltranJimenez:2018vdo}%
  \BibitemOpen
  \bibfield  {author} {\bibinfo {author} {\bibfnamefont {J.}~\bibnamefont
  {Beltr{\'a}n~Jim{\'e}nez}}, \bibinfo {author} {\bibfnamefont
  {L.}~\bibnamefont {Heisenberg}}, \ and\ \bibinfo {author} {\bibfnamefont
  {T.~S.}\ \bibnamefont {Koivisto}},\ }\href {\doibase
  10.1088/1475-7516/2018/08/039} {\bibfield  {journal} {\bibinfo  {journal}
  {JCAP}\ }\textbf {\bibinfo {volume} {08}},\ \bibinfo {pages} {039} (\bibinfo
  {year} {2018})},\ \Eprint {http://arxiv.org/abs/1803.10185} {arXiv:1803.10185
  [gr-qc]} \BibitemShut {NoStop}%
\bibitem [{\citenamefont {Fiorini}\ \emph {et~al.}(2023)\citenamefont
  {Fiorini}, \citenamefont {Gonz{\'a}lez},\ and\ \citenamefont
  {V{\'a}squez}}]{Fiorini:2023axr}%
  \BibitemOpen
  \bibfield  {author} {\bibinfo {author} {\bibfnamefont {F.}~\bibnamefont
  {Fiorini}}, \bibinfo {author} {\bibfnamefont {P.~A.}\ \bibnamefont
  {Gonz{\'a}lez}}, \ and\ \bibinfo {author} {\bibfnamefont {Y.}~\bibnamefont
  {V{\'a}squez}},\ }\href {\doibase 10.1088/1475-7516/2023/12/033} {\bibfield
  {journal} {\bibinfo  {journal} {JCAP}\ }\textbf {\bibinfo {volume} {12}},\
  \bibinfo {pages} {033} (\bibinfo {year} {2023})},\ \Eprint
  {http://arxiv.org/abs/2309.06293} {arXiv:2309.06293 [gr-qc]} \BibitemShut
  {NoStop}%
\bibitem [{\citenamefont {Kuntz}\ \emph {et~al.}(2025)\citenamefont {Kuntz},
  \citenamefont {Paci},\ and\ \citenamefont {Zanusso}}]{Kuntz:2024qgs}%
  \BibitemOpen
  \bibfield  {author} {\bibinfo {author} {\bibfnamefont {I.}~\bibnamefont
  {Kuntz}}, \bibinfo {author} {\bibfnamefont {G.}~\bibnamefont {Paci}}, \ and\
  \bibinfo {author} {\bibfnamefont {O.}~\bibnamefont {Zanusso}},\ }\href
  {\doibase 10.1088/1361-6382/ada867} {\bibfield  {journal} {\bibinfo
  {journal} {Class. Quant. Grav.}\ }\textbf {\bibinfo {volume} {42}},\ \bibinfo
  {pages} {045011} (\bibinfo {year} {2025})},\ \Eprint
  {http://arxiv.org/abs/2410.04510} {arXiv:2410.04510 [gr-qc]} \BibitemShut
  {NoStop}%
\bibitem [{\citenamefont {Stano}(2025)}]{Stano:2025eje}%
  \BibitemOpen
  \bibfield  {author} {\bibinfo {author} {\bibfnamefont {M.}~\bibnamefont
  {Stano}},\ }\emph {\bibinfo {title} {{Black Hole Action in Einstein's other
  Gravity}}},\ \href@noop {} {\bibinfo {type} {Other thesis}} (\bibinfo {year}
  {2025}),\ \Eprint {http://arxiv.org/abs/2503.11746} {arXiv:2503.11746
  [gr-qc]} \BibitemShut {NoStop}%
\bibitem [{\citenamefont {Jacobson}\ and\ \citenamefont
  {Mohd}(2015)}]{Jacobson:2015uqa}%
  \BibitemOpen
  \bibfield  {author} {\bibinfo {author} {\bibfnamefont {T.}~\bibnamefont
  {Jacobson}}\ and\ \bibinfo {author} {\bibfnamefont {A.}~\bibnamefont
  {Mohd}},\ }\href {\doibase 10.1103/PhysRevD.92.124010} {\bibfield  {journal}
  {\bibinfo  {journal} {Phys. Rev. D}\ }\textbf {\bibinfo {volume} {92}},\
  \bibinfo {pages} {124010} (\bibinfo {year} {2015})},\ \Eprint
  {http://arxiv.org/abs/1507.01054} {arXiv:1507.01054 [gr-qc]} \BibitemShut
  {NoStop}%
\bibitem [{\citenamefont {Freidel}\ \emph
  {et~al.}(2020{\natexlab{b}})\citenamefont {Freidel}, \citenamefont
  {Geiller},\ and\ \citenamefont {Pranzetti}}]{Freidel:2020svx}%
  \BibitemOpen
  \bibfield  {author} {\bibinfo {author} {\bibfnamefont {L.}~\bibnamefont
  {Freidel}}, \bibinfo {author} {\bibfnamefont {M.}~\bibnamefont {Geiller}}, \
  and\ \bibinfo {author} {\bibfnamefont {D.}~\bibnamefont {Pranzetti}},\ }\href
  {\doibase 10.1007/JHEP11(2020)027} {\bibfield  {journal} {\bibinfo  {journal}
  {JHEP}\ }\textbf {\bibinfo {volume} {11}},\ \bibinfo {pages} {027} (\bibinfo
  {year} {2020}{\natexlab{b}})},\ \Eprint {http://arxiv.org/abs/2007.03563}
  {arXiv:2007.03563 [hep-th]} \BibitemShut {NoStop}%
\bibitem [{\citenamefont {Chatterjee}\ and\ \citenamefont
  {Ghosh}(2018)}]{ChatterjeeGhosh2018}%
  \BibitemOpen
  \bibfield  {author} {\bibinfo {author} {\bibfnamefont {A.}~\bibnamefont
  {Chatterjee}}\ and\ \bibinfo {author} {\bibfnamefont {A.}~\bibnamefont
  {Ghosh}},\ }\href {\doibase 10.1140/epjc/s10052-018-6021-8} {\bibfield
  {journal} {\bibinfo  {journal} {The European Physical Journal C}\ }\textbf
  {\bibinfo {volume} {78}},\ \bibinfo {pages} {550} (\bibinfo {year} {2018})},\
  \Eprint {http://arxiv.org/abs/1511.01709} {arXiv:1511.01709 [gr-qc]}
  \BibitemShut {NoStop}%
\bibitem [{\citenamefont {Frodden}\ \emph {et~al.}(2013)\citenamefont
  {Frodden}, \citenamefont {Ghosh},\ and\ \citenamefont
  {Perez}}]{FroddenGhoshPerez2013}%
  \BibitemOpen
  \bibfield  {author} {\bibinfo {author} {\bibfnamefont {E.}~\bibnamefont
  {Frodden}}, \bibinfo {author} {\bibfnamefont {A.}~\bibnamefont {Ghosh}}, \
  and\ \bibinfo {author} {\bibfnamefont {A.}~\bibnamefont {Perez}},\ }\href
  {\doibase 10.1103/PhysRevD.87.121503} {\bibfield  {journal} {\bibinfo
  {journal} {Physical Review D}\ }\textbf {\bibinfo {volume} {87}},\ \bibinfo
  {pages} {121503} (\bibinfo {year} {2013})},\ \Eprint
  {http://arxiv.org/abs/1110.4055} {arXiv:1110.4055 [gr-qc]} \BibitemShut
  {NoStop}%
\bibitem [{\citenamefont {Bianchi}\ and\ \citenamefont
  {Wieland}(2012)}]{Bianchi:2012vp}%
  \BibitemOpen
  \bibfield  {author} {\bibinfo {author} {\bibfnamefont {E.}~\bibnamefont
  {Bianchi}}\ and\ \bibinfo {author} {\bibfnamefont {W.}~\bibnamefont
  {Wieland}},\ }\href@noop {} {\  (\bibinfo {year} {2012})},\ \Eprint
  {http://arxiv.org/abs/1205.5325} {arXiv:1205.5325 [gr-qc]} \BibitemShut
  {NoStop}%
\bibitem [{\citenamefont {Wald}(1993)}]{Wald:1993nt}%
  \BibitemOpen
  \bibfield  {author} {\bibinfo {author} {\bibfnamefont {R.~M.}\ \bibnamefont
  {Wald}},\ }\href {\doibase 10.1103/PhysRevD.48.R3427} {\bibfield  {journal}
  {\bibinfo  {journal} {Phys. Rev. D}\ }\textbf {\bibinfo {volume} {48}},\
  \bibinfo {pages} {R3427} (\bibinfo {year} {1993})},\ \Eprint
  {http://arxiv.org/abs/gr-qc/9307038} {arXiv:gr-qc/9307038} \BibitemShut
  {NoStop}%
\bibitem [{\citenamefont {Hayward}(2000)}]{Hayward:2000fi}%
  \BibitemOpen
  \bibfield  {author} {\bibinfo {author} {\bibfnamefont {S.~A.}\ \bibnamefont
  {Hayward}},\ }\href@noop {} {\  (\bibinfo {year} {2000})},\ \Eprint
  {http://arxiv.org/abs/gr-qc/0004042} {arXiv:gr-qc/0004042} \BibitemShut
  {NoStop}%
\bibitem [{\citenamefont {Hohmann}(2020)}]{Hohmann:2019fvf}%
  \BibitemOpen
  \bibfield  {author} {\bibinfo {author} {\bibfnamefont {M.}~\bibnamefont
  {Hohmann}},\ }\href {\doibase 10.3390/sym12030453} {\bibfield  {journal}
  {\bibinfo  {journal} {Symmetry}\ }\textbf {\bibinfo {volume} {12}},\ \bibinfo
  {pages} {453} (\bibinfo {year} {2020})},\ \Eprint
  {http://arxiv.org/abs/1912.12906} {arXiv:1912.12906 [math-ph]} \BibitemShut
  {NoStop}%
\bibitem [{\citenamefont {Coley}\ \emph {et~al.}(2024)\citenamefont {Coley},
  \citenamefont {Landry}, \citenamefont {van~den Hoogen},\ and\ \citenamefont
  {McNutt}}]{Coley:2024tqe}%
  \BibitemOpen
  \bibfield  {author} {\bibinfo {author} {\bibfnamefont {A.~A.}\ \bibnamefont
  {Coley}}, \bibinfo {author} {\bibfnamefont {A.}~\bibnamefont {Landry}},
  \bibinfo {author} {\bibfnamefont {R.~J.}\ \bibnamefont {van~den Hoogen}}, \
  and\ \bibinfo {author} {\bibfnamefont {D.~D.}\ \bibnamefont {McNutt}},\
  }\href {\doibase 10.1140/epjc/s10052-024-12629-5} {\bibfield  {journal}
  {\bibinfo  {journal} {Eur. Phys. J. C}\ }\textbf {\bibinfo {volume} {84}},\
  \bibinfo {pages} {334} (\bibinfo {year} {2024})},\ \Eprint
  {http://arxiv.org/abs/2402.07238} {arXiv:2402.07238 [gr-qc]} \BibitemShut
  {NoStop}%
\bibitem [{\citenamefont {Hohmann}\ and\ \citenamefont
  {Karanasou}(2025)}]{Hohmann:2024phz}%
  \BibitemOpen
  \bibfield  {author} {\bibinfo {author} {\bibfnamefont {M.}~\bibnamefont
  {Hohmann}}\ and\ \bibinfo {author} {\bibfnamefont {V.}~\bibnamefont
  {Karanasou}},\ }\href {\doibase 10.1103/PhysRevD.111.064057} {\bibfield
  {journal} {\bibinfo  {journal} {Phys. Rev. D}\ }\textbf {\bibinfo {volume}
  {111}},\ \bibinfo {pages} {064057} (\bibinfo {year} {2025})},\ \Eprint
  {http://arxiv.org/abs/2412.11730} {arXiv:2412.11730 [gr-qc]} \BibitemShut
  {NoStop}%
\bibitem [{\citenamefont {Kodama}(1980)}]{Kodama:1979vn}%
  \BibitemOpen
  \bibfield  {author} {\bibinfo {author} {\bibfnamefont {H.}~\bibnamefont
  {Kodama}},\ }\href {\doibase 10.1143/PTP.63.1217} {\bibfield  {journal}
  {\bibinfo  {journal} {Prog. Theor. Phys.}\ }\textbf {\bibinfo {volume}
  {63}},\ \bibinfo {pages} {1217} (\bibinfo {year} {1980})}\BibitemShut
  {NoStop}%
\bibitem [{\citenamefont {Hayward}(1996)}]{Hayward1996}%
  \BibitemOpen
  \bibfield  {author} {\bibinfo {author} {\bibfnamefont {S.~A.}\ \bibnamefont
  {Hayward}},\ }\href {\doibase 10.1103/PhysRevD.53.1938} {\bibfield  {journal}
  {\bibinfo  {journal} {Phys. Rev. D}\ }\textbf {\bibinfo {volume} {53}},\
  \bibinfo {pages} {1938} (\bibinfo {year} {1996})},\ \Eprint
  {http://arxiv.org/abs/gr-qc/9408002} {arXiv:gr-qc/9408002} \BibitemShut
  {NoStop}%
\bibitem [{\citenamefont {Abreu}\ and\ \citenamefont
  {Visser}(2010)}]{AbreuVisser2010}%
  \BibitemOpen
  \bibfield  {author} {\bibinfo {author} {\bibfnamefont {G.}~\bibnamefont
  {Abreu}}\ and\ \bibinfo {author} {\bibfnamefont {M.}~\bibnamefont {Visser}},\
  }\href {\doibase 10.1103/PhysRevD.82.044027} {\bibfield  {journal} {\bibinfo
  {journal} {Phys. Rev. D}\ }\textbf {\bibinfo {volume} {82}},\ \bibinfo
  {pages} {044027} (\bibinfo {year} {2010})},\ \Eprint
  {http://arxiv.org/abs/1004.1456} {arXiv:1004.1456} \BibitemShut {NoStop}%
\bibitem [{\citenamefont {Misner}\ and\ \citenamefont
  {Sharp}(1964)}]{MisnerSharp1964}%
  \BibitemOpen
  \bibfield  {author} {\bibinfo {author} {\bibfnamefont {C.~W.}\ \bibnamefont
  {Misner}}\ and\ \bibinfo {author} {\bibfnamefont {D.~H.}\ \bibnamefont
  {Sharp}},\ }\href {\doibase 10.1103/PhysRev.136.B571} {\bibfield  {journal}
  {\bibinfo  {journal} {Phys. Rev.}\ }\textbf {\bibinfo {volume} {136}},\
  \bibinfo {pages} {B571} (\bibinfo {year} {1964})}\BibitemShut {NoStop}%
\bibitem [{\citenamefont {Gibbons}\ and\ \citenamefont
  {Hawking}(1977)}]{PhysRevD.15.2752}%
  \BibitemOpen
  \bibfield  {author} {\bibinfo {author} {\bibfnamefont {G.~W.}\ \bibnamefont
  {Gibbons}}\ and\ \bibinfo {author} {\bibfnamefont {S.~W.}\ \bibnamefont
  {Hawking}},\ }\href {\doibase 10.1103/PhysRevD.15.2752} {\bibfield  {journal}
  {\bibinfo  {journal} {Phys. Rev. D}\ }\textbf {\bibinfo {volume} {15}},\
  \bibinfo {pages} {2752} (\bibinfo {year} {1977})}\BibitemShut {NoStop}%
\bibitem [{\citenamefont {Hayward}(1998)}]{Hayward:1997jp}%
  \BibitemOpen
  \bibfield  {author} {\bibinfo {author} {\bibfnamefont {S.~A.}\ \bibnamefont
  {Hayward}},\ }\href {\doibase 10.1088/0264-9381/15/10/017} {\bibfield
  {journal} {\bibinfo  {journal} {Class. Quant. Grav.}\ }\textbf {\bibinfo
  {volume} {15}},\ \bibinfo {pages} {3147} (\bibinfo {year} {1998})},\ \Eprint
  {http://arxiv.org/abs/gr-qc/9710089} {arXiv:gr-qc/9710089} \BibitemShut
  {NoStop}%
\bibitem [{\citenamefont {Ansermet}\ and\ \citenamefont
  {Br{\'e}chet}(2019)}]{AnsermetBrechet}%
  \BibitemOpen
  \bibfield  {author} {\bibinfo {author} {\bibfnamefont {J.-P.}\ \bibnamefont
  {Ansermet}}\ and\ \bibinfo {author} {\bibfnamefont {S.~D.}\ \bibnamefont
  {Br{\'e}chet}},\ }\href {\doibase 10.1017/9781108620932} {\emph {\bibinfo
  {title} {Principles of Thermodynamics}}}\ (\bibinfo  {publisher} {Cambridge
  University Press},\ \bibinfo {address} {Cambridge},\ \bibinfo {year}
  {2019})\BibitemShut {NoStop}%
\bibitem [{\citenamefont {Gibbons}\ and\ \citenamefont
  {Maeda}(1988)}]{GibbonsMaeda1988}%
  \BibitemOpen
  \bibfield  {author} {\bibinfo {author} {\bibfnamefont {G.~W.}\ \bibnamefont
  {Gibbons}}\ and\ \bibinfo {author} {\bibfnamefont {K.-i.}\ \bibnamefont
  {Maeda}},\ }\href {\doibase 10.1016/0550-3213(88)90006-5} {\bibfield
  {journal} {\bibinfo  {journal} {Nucl. Phys. B}\ }\textbf {\bibinfo {volume}
  {298}},\ \bibinfo {pages} {741} (\bibinfo {year} {1988})}\BibitemShut
  {NoStop}%
\bibitem [{\citenamefont {Garfinkle}\ \emph {et~al.}(1991)\citenamefont
  {Garfinkle}, \citenamefont {Horowitz},\ and\ \citenamefont
  {Strominger}}]{GarfinkleHorowitzStrominger1991}%
  \BibitemOpen
  \bibfield  {author} {\bibinfo {author} {\bibfnamefont {D.}~\bibnamefont
  {Garfinkle}}, \bibinfo {author} {\bibfnamefont {G.~T.}\ \bibnamefont
  {Horowitz}}, \ and\ \bibinfo {author} {\bibfnamefont {A.}~\bibnamefont
  {Strominger}},\ }\href {\doibase 10.1103/PhysRevD.43.3140} {\bibfield
  {journal} {\bibinfo  {journal} {Phys. Rev. D}\ }\textbf {\bibinfo {volume}
  {43}},\ \bibinfo {pages} {3140} (\bibinfo {year} {1991})},\ \bibinfo {note}
  {erratum: Phys. Rev. D 45, 3888 (1992)}\BibitemShut {NoStop}%
\bibitem [{\citenamefont {Ballesteros}\ \emph {et~al.}(2023)\citenamefont
  {Ballesteros}, \citenamefont {G{\'o}mez-Fayr{\'e}n}, \citenamefont
  {Ort{\'\i}n},\ and\ \citenamefont {Zatti}}]{Ballesteros:2023}%
  \BibitemOpen
  \bibfield  {author} {\bibinfo {author} {\bibfnamefont {R.}~\bibnamefont
  {Ballesteros}}, \bibinfo {author} {\bibfnamefont {C.}~\bibnamefont
  {G{\'o}mez-Fayr{\'e}n}}, \bibinfo {author} {\bibfnamefont {T.}~\bibnamefont
  {Ort{\'\i}n}}, \ and\ \bibinfo {author} {\bibfnamefont {M.}~\bibnamefont
  {Zatti}},\ }\href {\doibase 10.1007/JHEP05(2023)158} {\bibfield  {journal}
  {\bibinfo  {journal} {JHEP}\ }\textbf {\bibinfo {volume} {05}},\ \bibinfo
  {pages} {158} (\bibinfo {year} {2023})},\ \Eprint
  {http://arxiv.org/abs/2302.11630} {arXiv:2302.11630 [hep-th]} \BibitemShut
  {NoStop}%
\bibitem [{\citenamefont {Aguirregabiria}\ \emph {et~al.}(1996)\citenamefont
  {Aguirregabiria}, \citenamefont {Chamorro},\ and\ \citenamefont
  {Virbhadra}}]{Aguirregabiria:1995qz}%
  \BibitemOpen
  \bibfield  {author} {\bibinfo {author} {\bibfnamefont {J.~M.}\ \bibnamefont
  {Aguirregabiria}}, \bibinfo {author} {\bibfnamefont {A.}~\bibnamefont
  {Chamorro}}, \ and\ \bibinfo {author} {\bibfnamefont {K.~S.}\ \bibnamefont
  {Virbhadra}},\ }\href {\doibase 10.1007/BF02109529} {\bibfield  {journal}
  {\bibinfo  {journal} {Gen. Rel. Grav.}\ }\textbf {\bibinfo {volume} {28}},\
  \bibinfo {pages} {1393} (\bibinfo {year} {1996})},\ \Eprint
  {http://arxiv.org/abs/gr-qc/9501002} {arXiv:gr-qc/9501002} \BibitemShut
  {NoStop}%
\bibitem [{\citenamefont {Herrera}\ \emph {et~al.}(2003)\citenamefont
  {Herrera}, \citenamefont {Santos},\ and\ \citenamefont
  {Skea}}]{Herrera:2003pu}%
  \BibitemOpen
  \bibfield  {author} {\bibinfo {author} {\bibfnamefont {L.}~\bibnamefont
  {Herrera}}, \bibinfo {author} {\bibfnamefont {N.~O.}\ \bibnamefont {Santos}},
  \ and\ \bibinfo {author} {\bibfnamefont {J.~E.~F.}\ \bibnamefont {Skea}},\
  }\href {\doibase 10.1023/A:1026255228035} {\bibfield  {journal} {\bibinfo
  {journal} {Gen. Rel. Grav.}\ }\textbf {\bibinfo {volume} {35}},\ \bibinfo
  {pages} {2057} (\bibinfo {year} {2003})},\ \Eprint
  {http://arxiv.org/abs/gr-qc/0306005} {arXiv:gr-qc/0306005} \BibitemShut
  {NoStop}%
\bibitem [{\citenamefont {Xulu}(2006)}]{Xulu:2006zza}%
  \BibitemOpen
  \bibfield  {author} {\bibinfo {author} {\bibfnamefont {S.~S.}\ \bibnamefont
  {Xulu}},\ }\href {\doibase 10.1007/s10702-006-1013-6} {\bibfield  {journal}
  {\bibinfo  {journal} {Found. Phys. Lett.}\ }\textbf {\bibinfo {volume}
  {19}},\ \bibinfo {pages} {603} (\bibinfo {year} {2006})}\BibitemShut
  {NoStop}%
\bibitem [{\citenamefont {Salti}\ and\ \citenamefont {Aydo{\u
  g}du}(2006)}]{SaltiAydogdu2006RNAdS}%
  \BibitemOpen
  \bibfield  {author} {\bibinfo {author} {\bibfnamefont {M.}~\bibnamefont
  {Salti}}\ and\ \bibinfo {author} {\bibfnamefont {O.}~\bibnamefont {Aydo{\u
  g}du}},\ }\href {\doibase 10.1140/epjc/s2006-02550-6} {\bibfield  {journal}
  {\bibinfo  {journal} {European Physical Journal C}\ }\textbf {\bibinfo
  {volume} {47}},\ \bibinfo {pages} {247} (\bibinfo {year} {2006})}\BibitemShut
  {NoStop}%
\bibitem [{\citenamefont {Nashed}(2006{\natexlab{b}})}]{Nashed2006RN}%
  \BibitemOpen
  \bibfield  {author} {\bibinfo {author} {\bibfnamefont {G.~G.~L.}\
  \bibnamefont {Nashed}},\ }\href {\doibase 10.1142/S0217732306020445}
  {\bibfield  {journal} {\bibinfo  {journal} {Modern Physics Letters A}\
  }\textbf {\bibinfo {volume} {21}},\ \bibinfo {pages} {2241} (\bibinfo {year}
  {2006}{\natexlab{b}})},\ \Eprint {http://arxiv.org/abs/gr-qc/0401041}
  {arXiv:gr-qc/0401041} \BibitemShut {NoStop}%
\bibitem [{\citenamefont {Vagenas}(2006)}]{Vagenas:2006pj}%
  \BibitemOpen
  \bibfield  {author} {\bibinfo {author} {\bibfnamefont {E.~C.}\ \bibnamefont
  {Vagenas}},\ }\href {\doibase 10.1142/S0217732306020482} {\bibfield
  {journal} {\bibinfo  {journal} {Mod. Phys. Lett. A}\ }\textbf {\bibinfo
  {volume} {21}},\ \bibinfo {pages} {1947} (\bibinfo {year} {2006})},\ \Eprint
  {http://arxiv.org/abs/gr-qc/0602107} {arXiv:gr-qc/0602107} \BibitemShut
  {NoStop}%
\bibitem [{\citenamefont {Dadhich}(2020)}]{Dadhich2020}%
  \BibitemOpen
  \bibfield  {author} {\bibinfo {author} {\bibfnamefont {N.}~\bibnamefont
  {Dadhich}},\ }\href {\doibase 10.1088/1475-7516/2020/04/035} {\bibfield
  {journal} {\bibinfo  {journal} {JCAP}\ }\textbf {\bibinfo {volume} {04}},\
  \bibinfo {pages} {035} (\bibinfo {year} {2020})},\ \Eprint
  {http://arxiv.org/abs/1903.03436} {arXiv:1903.03436 [gr-qc]} \BibitemShut
  {NoStop}%
\bibitem [{\citenamefont {Koivisto}(2022)}]{Koivisto:2022nar}%
  \BibitemOpen
  \bibfield  {author} {\bibinfo {author} {\bibfnamefont {T.}~\bibnamefont
  {Koivisto}},\ }\href {\doibase 10.1142/S0219887822400072} {\bibfield
  {journal} {\bibinfo  {journal} {Int. J. Geom. Meth. Mod. Phys.}\ }\textbf
  {\bibinfo {volume} {19}},\ \bibinfo {pages} {2240007} (\bibinfo {year}
  {2022})},\ \Eprint {http://arxiv.org/abs/2202.11522} {arXiv:2202.11522
  [gr-qc]} \BibitemShut {NoStop}%
\bibitem [{\citenamefont {Chamorro}\ and\ \citenamefont
  {Virbhadra}(1996)}]{Chamorro:1994vw}%
  \BibitemOpen
  \bibfield  {author} {\bibinfo {author} {\bibfnamefont {A.}~\bibnamefont
  {Chamorro}}\ and\ \bibinfo {author} {\bibfnamefont {K.~S.}\ \bibnamefont
  {Virbhadra}},\ }\href {\doibase 10.1142/S0218271896000175} {\bibfield
  {journal} {\bibinfo  {journal} {Int. J. Mod. Phys. D}\ }\textbf {\bibinfo
  {volume} {5}},\ \bibinfo {pages} {251} (\bibinfo {year} {1996})},\ \Eprint
  {http://arxiv.org/abs/hep-th/9407187} {arXiv:hep-th/9407187} \BibitemShut
  {NoStop}%
\bibitem [{\citenamefont {Yang}\ \emph {et~al.}(1997)\citenamefont {Yang},
  \citenamefont {Yeh}, \citenamefont {Hsu},\ and\ \citenamefont
  {Lee}}]{Yang:1996iq}%
  \BibitemOpen
  \bibfield  {author} {\bibinfo {author} {\bibfnamefont {I.-C.}\ \bibnamefont
  {Yang}}, \bibinfo {author} {\bibfnamefont {C.-T.}\ \bibnamefont {Yeh}},
  \bibinfo {author} {\bibfnamefont {R.-R.}\ \bibnamefont {Hsu}}, \ and\
  \bibinfo {author} {\bibfnamefont {C.-R.}\ \bibnamefont {Lee}},\ }\href
  {\doibase 10.1142/S0218271897000194} {\bibfield  {journal} {\bibinfo
  {journal} {Int. J. Mod. Phys. D}\ }\textbf {\bibinfo {volume} {6}},\ \bibinfo
  {pages} {349} (\bibinfo {year} {1997})},\ \Eprint
  {http://arxiv.org/abs/gr-qc/9609038} {arXiv:gr-qc/9609038} \BibitemShut
  {NoStop}%
\bibitem [{\citenamefont {Nashed}(2012)}]{Nashed2012Dyadosphere}%
  \BibitemOpen
  \bibfield  {author} {\bibinfo {author} {\bibfnamefont {G.~G.~L.}\
  \bibnamefont {Nashed}},\ }\href {\doibase 10.1088/1674-1056/21/3/030401}
  {\bibfield  {journal} {\bibinfo  {journal} {Chinese Physics B}\ }\textbf
  {\bibinfo {volume} {21}},\ \bibinfo {pages} {030401} (\bibinfo {year}
  {2012})}\BibitemShut {NoStop}%
\bibitem [{\citenamefont {Radinschi}\ \emph {et~al.}(2010)\citenamefont
  {Radinschi}, \citenamefont {Rahaman},\ and\ \citenamefont
  {Ghosh}}]{Radinschi:2010zza}%
  \BibitemOpen
  \bibfield  {author} {\bibinfo {author} {\bibfnamefont {I.}~\bibnamefont
  {Radinschi}}, \bibinfo {author} {\bibfnamefont {F.}~\bibnamefont {Rahaman}},
  \ and\ \bibinfo {author} {\bibfnamefont {A.}~\bibnamefont {Ghosh}},\ }\href
  {\doibase 10.1007/s10773-010-0273-x} {\bibfield  {journal} {\bibinfo
  {journal} {Int. J. Theor. Phys.}\ }\textbf {\bibinfo {volume} {49}},\
  \bibinfo {pages} {943} (\bibinfo {year} {2010})},\ \Eprint
  {http://arxiv.org/abs/1003.2926} {arXiv:1003.2926 [gr-qc]} \BibitemShut
  {NoStop}%
\bibitem [{\citenamefont {Mourad}\ and\ \citenamefont
  {Abdelgaber}(2023)}]{MouradAbdelgaber2023StringyEntropy}%
  \BibitemOpen
  \bibfield  {author} {\bibinfo {author} {\bibfnamefont {M.~F.}\ \bibnamefont
  {Mourad}}\ and\ \bibinfo {author} {\bibfnamefont {M.}~\bibnamefont
  {Abdelgaber}},\ }\href {\doibase 10.1007/s12648-023-02783-z} {\bibfield
  {journal} {\bibinfo  {journal} {Indian Journal of Physics}\ }\textbf
  {\bibinfo {volume} {97}},\ \bibinfo {pages} {4503} (\bibinfo {year}
  {2023})}\BibitemShut {NoStop}%
\end{thebibliography}%

\appendix

\section{Derivations of the Noether currents and identities}
\label{derivations}

Here we consider the symmetries of the actions in both the metric-affine and absolute parallel geometries, working in the formalism set up in Section \ref{gravitytheory}. 

\subsection{Metric-affine parallelism}

\subsubsection{GL}
\label{MAGgl}

Consider the GL transformation (\ref{gaugeGL}). It is an exact symmetry, which means that $\delta_M \bL = 0$. The symplectic potential becomes 
\be
\bTheta = -2M^{ab}\bq_{ab} + M_a{}^b\bbe^a\wedge\bbs_b + \bDiff M_a{}^b\wedge\br^a{}_b = \bdiff\lp M_a{}^b\br^a{}_b\rp - M_a{}^b\hat{\bE}^a{}_b\,, 
\ee
where we did a partial integration and identified the equation of motion for the connection (\ref{EoMc}). The Noether current is thus given as
$\bJ_M \approx \bdiff\bj_M$, wherein the Noether-Wald $(n-2)$-form $\bj_M =  M_a{}^b\br^a{}_b$. We then derive the Noether identity associated with the GL symmetry, by considering the variations (\ref{variations}),
\ba
\delta\bL & = &  M^{ab}\lp \bbe_a\wedge\bE_b + 2\bE_{ab} \rp + \bDiff M_a{}^b\wedge\hat{\bE}^a{}_b + \lp M_c{}^a\br^c{}_b
- M_b{}^c\br^a{}_c\rp\wedge\bR_a{}^b + \bdiff\lb M^{ab}\lp  \bbe_a\wedge\bbs_b-2\bq_{ab} \rp + \bDiff M_a{}^b\wedge\br^a{}_b\rb  \nn \\
& = &  M_a{}^b\lp \bbe^a\wedge\bE_b + 2\bE^a{}_b - \bDiff\hat{\bE}^a{}_b + \br^a{}_c\wedge\bR_b{}^c - \br^c{}_b\wedge\bR_c{}^a\rp \nn \\
& + & \bdiff\lb   M^{ab}\lp \bbe_a\wedge\bbs_b-2\bq_{ab} + \hat{\bE}_{ab}\rp + \bDiff M_a{}^b\wedge\br^a{}_b\rb\,. 
\ea
The last line is a total differential which only contributes a surface term to the action. When considering transformations $M_a{}^b$ which vanish at the surface but are otherwise arbitrary, we obtain the Noether identity
\be
 \bbe^a\wedge\bE_b + 2\bE^a{}_b - \bDiff\hat{\bE}^a{}_b + \br^a{}_c\wedge\bR_b{}^c - \br^c{}_b\wedge\bR_c{}^a = 0\,,
\ee
which holds without imposing the equations of motion. 

\subsubsection{Diff}

Consider the Diff (\ref{gaugeDiff}). It is not an exact symmetry, but $\delta_\xi \bL = \bdiff\lp \xi\lrcorner\bL\rp$. We can expand the inner product as
\be
\xi\lrcorner\bL = -\xi\lrcorner\bbe^a\lp \bt_a + \bT_a\rp+ \xi\lrcorner\bS^a\wedge\bbs_a + \lp\xi\lrcorner\bQ^{ab}\rp\bq_{ab} 
+ \xi\lrcorner\lp \br^a{}_b\wedge\bR_a{}^b\rp + \xi\lrcorner\bpsi\wedge\frac{\partial \bL_M}{\partial \bpsi} 
 + \xi\lrcorner\bdiff\bpsi\wedge\bPi\,.   \label{dotL}
\ee
On the other hand, the symplectic potential may be written as
\ba
\bTheta_\xi & = & \bdiff\lb\lp\xi\lrcorner\bbe^a\rp\bbs_a + \lp\xi\lrcorner\bA_a{}^b\rp\br^a{}_b\rb + 
\lp\xi\lrcorner\bdiff\bbe^a\rp\wedge\bbs_a - \lp\xi\lrcorner\bbe^a\rp\bdiff\bbs_a - \lp\xi\lrcorner\bdiff g^{ab}\rp\bq_{ab} \nn \\
& + & \xi\lrcorner\bR_a{}^b\wedge\br^a{}_b - \lp\xi\lrcorner\bA_a{}^b\rp\bDiff\br^a{}_b + \delta_\xi\bpsi\wedge\bPi \nn \\
& = & \bdiff\bj_\xi + \xi\lrcorner\bS^a\wedge\bbs_a 
-\lp\xi\lrcorner\bbe^a\rp\bDiff\bbs_a + \lp\xi\lrcorner\bQ^{ab}\rp\bq_{ab} \nn \\
& + & \xi\lrcorner\bdiff\bpsi\wedge\bPi+ (-1)^p\xi\lrcorner\bpsi\wedge\bdiff \bPi - \lp \xi\lrcorner\bA_a{}^b\rp\hat{\bE}^a{}_b + \xi\lrcorner\bR_a{}^b\wedge\br^a{}_b\,,
\ea 
where
\be
\bj_\xi = \lp\xi\lrcorner\bbe^a\rp\bbs_a + \lp\xi\lrcorner\bA_a{}^b\rp\br^a{}_b + \xi\lrcorner\bpsi\wedge\bPi\,. 
\ee
This is the Noether-Wald potential for the on-shell conserved current
\be
\bJ_\xi = \bTheta_\xi - \xi\lrcorner\bL = \bdiff\bj_\xi  - \lp \xi\lrcorner\bA_a{}^b\rp\hat{\bE}^a{}_b + \xi\lrcorner\bR_a{}^b\wedge\br^a{}_b - \xi\lrcorner\bPsi\wedge\bE_\psi\,. 
\ee
We find the Noether identity
\be
-\lp\xi\lrcorner\bQ^{ab}\rp\bE_{ab}  +  \xi\lrcorner \bdiff\lp\bbe^a\wedge\bE_a + \bA_a{}^b\wedge\hat{\bE}{}^a{}_b + \bPsi\wedge\bE_\psi\rp
 +  \lp \xi\lrcorner\bdiff\br^a{}_b + \bA_c{}^a\xi\lrcorner\br^c{}_b
- \bA_b{}^c\xi\lrcorner\br^a{}_c\rp\wedge\bR_a{}^b  = 0\,.
\ee

\subsubsection{CDiff}

Consider the covariant Diff (\ref{gaugeCDiff}). It is not an exact symmetry, but we have $\delta_\Xi \bL = \bdiff\lp \Xi\lrcorner\bL\rp$ as given in (\ref{dotL}). Both the calculation and the result are now neater than in the previous, non-covariant case. We obtain
\be
\bJ_\Xi = \bTheta_\Xi - \Xi\lrcorner\bL = \bdiff\bj_\Xi - \lp\Xi\lrcorner\bbe^a\rp\bE_a - \Xi\lrcorner\bPsi\wedge\bE_\psi+ \Xi\lrcorner\br^a{}_b\wedge\bR_a{}^b\,,
\ee
where the Noether-Wald potential is now the manifestly covariant $(n-2)$-form,  
\be
\bj_\Xi = \lp\Xi\lrcorner\bbe^a\rp\bbs_a + \Xi\lrcorner\bPsi\wedge\bPi\,. 
\ee
The Noether identity stemming from the invariance under (\ref{gaugeCDiff}) is
\be
\lp\Xi\lrcorner\bQ^{ab}\rp\bE_{ab} + \lp\Xi\lrcorner\bbe^a\rp\bE_a - \Xi\lrcorner\bS^a\wedge\bE_a - \Xi\lrcorner\bR_a{}^b\wedge\hat{\bE}^a{}_b - \Xi\lrcorner\bDiff\br^a{}_b\wedge\bR_a{}^b 
= 0\,. 
\ee

\subsubsection{The \underline{GL} frame shift}

Consider the frame transformation $\underline{\delta}_m \bA_a{}^b = \bDiff m_a{}^b$. This transformation is not an off-shell gauge symmetry of the metric-affine action. Consequently it does not define an independent Noether charge.

To deduce the variation of the Lagrangian, we use the off-shell generalisation of the formula (\ref{formula})
$X = \mathring{R} - R + \mathring{\nabla}_\mu \V^\mu$,  and note the frame-independence of the metric objects $\mathring{R}$ and $\nabla_\mu$. The curvature transforms covariantly,
$\underline{\delta}_m \bR_a{}^b = m_c{}^b\bR_a{}^c - m_a{}^c \bR_c{}^b$, since it only depends on the connection. Then, the scalar curvature will not be invariant, but
\be
\underline{\delta}_m R = \underline{\delta}_m\lp g^{ac}\bbie_b\lrcorner\bbie_a\lrcorner \bR_c{}^b\rp = m^{ab}\lp R_{ab} - \tilde{R}_{ba} \rp\,. \label{deltaR1}
\ee
Due to this, the \underline{GL} is not strictly a symmetry of the Lagrangian. Indeed, the EoM (\ref{EoMc}) ${\bmE}^a{}_b \approx -\bDiff\br^a{}_b$ is not frame-covariant, 
\be
\underline{\delta}_m{\bmE}^a{}_b = \lp\bDiff m_b{}^c\rp\br^a{}_c - \lp\bDiff m_c{}^a\rp\br^c{}_b\,. 
\ee
The covariance is restored for covariantly constant transformation $\bDiff m_a{}^b$=0. 
The frame transformations of the non-metricity and the torsion are
\bs
\label{mtransA}
\ba
\underline{\delta}_m Q^\mu & = & \underline{\delta}_m \lp -g_{ab}g^{\mu\nu}\Diff_\nu g^{ab}\rp = -2g^{\mu\nu}g_{ab}g^{c(a}\Diff_\nu m_c{}^{b)} = -2g^{\mu\nu}\Diff_\nu m\,, \\
\underline{\delta}_m \tilde{Q}^\mu & = & \underline{\delta}_m \lp -\ie_a^\mu\ie_b^\nu\Diff_\nu g^{ab}\rp = 2\ie_a^\mu\ie_b^\nu g^{c(a}\Diff_\nu m_c{}^{b)}\,, \\
\underline{\delta}_m S^\mu & = & \underline{\delta}_m\lp 2\ie_a^\nu g^{\alpha\mu}\Diff_{[\alpha}\e^a_{\nu]}\rp = -2\ie_a^\nu g^{\mu\alpha}\e^c_{[\nu}\Diff_{\alpha]}m_c{}^a
= - g^{\mu\nu}\Diff_\nu m + \ie_a^\nu\e^c_\alpha g^{\mu\alpha}\Diff_\nu m_c{}^a\,,
\ea
and thus
\be
\underline{\delta}_m \V^\mu = 2 g^{\alpha[\mu}\ie_b^{\nu]}\e^a_\alpha\Diff_\nu m_a{}^b\,. 
\ee
\es
The symplectic potential (\ref{symplectic}) simplifies to
\be
\bTheta_m  = \bDiff m_a{}^b\wedge\br^a{}_b \approx \bdiff\lp   m_a{}^b\wedge\br^a{}_b\rp + m_a{}^b\bmE^a{}_b\,,
\ee
by using (\ref{important}). Combining these results one obtains the ``current'' (in quotations since it is not a Noether current) 
\be
``\bJ_m'' = \bdiff''\bj_m'' + m^{ab}\lp \bmE_{ab} + R_{ab}\ast\boldsymbol{1} - \tilde{R}_{ba} \ast\boldsymbol{1} \rp\,,  
\ee
where
\be
``\bj_m'' = m_a{}^b\br^a{}_b + \lp \nabla_\nu m_\mu{}^\nu - g_{\alpha\mu} \nabla^\nu m_\nu{}^\alpha\rp \ast \bdiff x^\mu\,. 
\ee
As expected, we do not obtain a proper Noether charge (though in restricted cases, the expression can reduce to the GL Noether charge considered in \ref{MAGgl}). 
Neither is there any usual Noether identity associated with the \underline{GL}. 

\subsubsection{Alternative \underline{GL}}

Consider the frame transformation $\underline{\delta}^{(e)}_m \bbe^a = m_b{}^a\bbe^b$. To deduce the variation of the Lagrangian, we use the off-shell generalisation of the formula (\ref{formula})
$X = \mathring{R} - R + \mathring{\nabla}_\mu \V^\mu$,  and note the frame-independence of the metric scalar curvature $\mathring{R}$. The general scalar curvature transforms as
\be
\underline{\delta}_m^{(e)} R = \delta_m^{(e)} \lp \bbie_b\lrcorner\bbie_a\lrcorner \bR^{ab} \rp = -m_a{}^c  \bbie_b\lrcorner\bbie_c\lrcorner \bR^{ab} -  m_b{}^c  \bbie_c\lrcorner\bbie_a\lrcorner \bR^{ab}
= -m^{ab}\lp R_{ab} + \tilde{R}_{ab}\rp\,.  \label{deltaR2}
\ee
The vectors transform as
\bs
\label{mtranse}
\ba
\underline{\delta}_m^{(e)} Q^\mu & = & \underline{\delta}_m^{(e)} \lp -g_{ab}g^{cd}\ie^\mu_c\ie^\nu_d\Diff_\nu g^{ab}\rp = -2m^{(\mu\nu)}Q_\nu\,, \\
\underline{\delta}_m^{(e)} \tilde{Q}^\mu & = & \underline{\delta}_m^{(e)} \lp \bbie_a\lrcorner\bQ^{ab}\ie_b^\mu\rp = -m^{ab}\lp Q_{ba}{}^\mu + \tilde{Q}_a\ie_b^\mu\rp\,, \\
\underline{\delta}_m^{(e)} S^\mu & = & \underline{\delta}_m^{(e)}\lp 2\ie_a^\nu\ie_b^\mu\ie_c^\alpha g^{bc}\Diff_{[\alpha}\e^a_{\nu]}\rp = \Diff^\mu m - \e^{a\mu}\Diff_b m_a{}^b - 2m^{(\mu\nu)}S_\nu\,, 
\ea
from which it follows that
\be
\underline{\delta}_m^{(e)} \V^\mu =  2\Diff^\mu m - 2\ie^\mu_a\ie^\nu_b\Diff_\nu m^{ab} -2m^{(\mu\nu)}\V_\nu - m^{\mu\nu}\tilde{Q}_\nu + m^{ab}\lp 2Q^\mu{}_{ab} - Q_{ba}{}^\mu\rp\,.
\ee
\es
Then consider independent GL transformations of the metric. We obtain
\be
\underline{\delta}_m^{(g)} R = \delta_m^{(g)}\lp R_a{}^b{}_{cb}g^{ac}\rp = 2m^{ab}\tilde{R}_{(ab)}\,.
\ee
Combined with (\ref{deltaR1}) and (\ref{deltaR2}), this confirms that the scalar $R$ is GL-invariant. Similarly, applying this transformation to the vectors
\bs
\label{transg}
\ba
\underline{\delta}_m^{(g)} Q^\mu & = & \underline{\delta}_m^{(g)} \lp -g_{ab}g^{cd}\ie^\mu_c\ie^\nu_d\Diff_\nu g^{ab}\rp = 2m^{(\mu\nu)}Q_\nu -2m^{ab}Q^\mu{}_{ab}
- 2g_{ab}\Diff^\mu m^{ab}\,,  \\
\underline{\delta}_m^{(g)} \tilde{Q}^\mu & = & \underline{\delta}_m^{(g)} \lp -\ie_a^\mu\ie_b^\nu \Diff_\nu g^{ab}\rp = -2\ie^\mu_a\ie^\nu_b\Diff_\nu m^{(ab)}\,, \\
\underline{\delta}_m^{(g)} S^\mu & = & \underline{\delta}_m^{(g)}\lp 2\ie_a^\nu \ie_b^\mu\ie^\alpha_c g^{bc}\Diff_{[\alpha}\e^a_{\nu]}\rp = 2m^{(\mu\nu)}S_\nu\,, \\
\underline{\delta}_m^{(g)} \V^\mu & = & -2\Diff^\mu m + 2\ie^\mu_a\ie^\nu_b\Diff_\nu m^{(ab)} + 2m^{(\mu\nu)}\lp \V_\nu + \tilde{Q}_\nu\rp\,,   
\ea
\es
we verify, by summing up (\ref{mtransA}), (\ref{mtranse}) and (\ref{transg}) that $\delta_m + \delta^{(e)}_m + \delta^{(g)}_m = 0$ for each vector separately. In varying the action with respect to either the frame or the metric, one should note that
\bs
\ba
\underline{\delta}_m^{(e)}\lp \mathring{\nabla}_\mu \V^\mu\rp & = &  \mathring{\nabla}_\mu \underline{\delta}_m^{(e)} \V^\mu + \V^\mu\Diff_\mu m\,,  \\
\underline{\delta}_m^{(g)}\lp \mathring{\nabla}_\mu \V^\mu\rp & = &  \mathring{\nabla}_\mu \underline{\delta}_m^{(g)} \V^\mu - \V^\mu\Diff_\mu m\,, 
\ea
\es
and thus neither transformation alone is a frame transformation, but combined together they constitute a frame transformation equivalent to $\underline{\delta}_m$. All these considerations are consistent with the identity $\underline{\delta}_m + \underline{\delta}_m^{(e)} + \underline{\delta}_m^{(g)} = \delta_{M=m}$. 

\subsection{Absolute teleparallelism}

Here we shall go through the derivations in the framework of the action (\ref{Aaction}). The EoMs (\ref{EoMg}) and (\ref{EoMe}) retain their form, but the EoM for the connection (\ref{EoMc}) is now replaced by the EoM for the matrix element $\aL^a_b$,  
\ba
\delta\bA_a{}^b\wedge{\bmE}^a{}_b & = & \delta\lp \iL_a^c\bdiff\aL^b_c\rp\wedge{\bmE}^a{}_b = \lp \delta \iL_a^c\bdiff\aL^b_c + \iL_a^c\bdiff\delta\aL^b_c\rp\wedge{\bmE}^a{}_b = 
 \delta\iL^c_a\bdiff\aL^b_c\wedge{\bmE}^a{}_b + \bdiff\lp \delta\aL_c^b\iL^c_a{\bmE}^a{}_b\rp - \delta\aL^b_c\bdiff\lp \iL^c_a{\bmE}^a{}_b\rp 
 \nn \\
 & = & \bdiff\lp\delta\aL^b_c\iL^c_a{\bmE}^a{}_b\rp - \delta\aL^b_c\iL^c_a\bDiff{\bmE}^a{}_b\,, \label{Avariation}
 \ea
where ${\bmE}^a{}_b = 2\bq^a{}_b - \bbe^a\wedge\bbs_b$. 
The G$_\parallel$R theory is again specified by the same constitutive law (\ref{potentials}) which, according to (\ref{important}), guarantees that 
(\ref{Avariation}) is a pure boundary term. It therefore contributes to the symplectic potential,
\be \label{symplectic2}
\bTheta = -\delta g^{ab}\bq_{ab} + \delta\bbe^a\wedge\bbs_a + \delta\aL^b_c\iL^c_a{\bmE}^a{}_b  + \delta\bpsi\wedge\bPi\,,
\ee
and the total variation of the Lagrangian is simplified into the form 
\be
\label{variations2}
\delta\bL = \delta g^{ab}\bE_{ab} + \delta\bbe^a\wedge\bE_a + \delta\bPsi\wedge\bE_\psi - \delta\aL^b_c\iL^c_a\bDiff{\bmE}^a{}_b + \bdiff\bTheta\,.
\ee
In a generic absolute parallel theory, one has to take into account also the nontrivial connection EoM read off directly from here: $\bDiff\bmE^a{}_b \approx 0$.

\subsubsection{GL}
\label{GL2}

The consistent GL transformations of the matrix elements are
\be
\delta_M \aL^a_b = M_c^a\aL^c_b\,, \quad \delta_M \iL^a_b = -M_b{}^c\iL^a_c \quad \Rightarrow \quad
\delta_M \bA_a{}^b = \delta_M\lp \iL_a^c\bdiff\aL^b_c\rp = \delta_M \lp -\aL^b_c\bdiff\iL_a^c\rp = \bDiff M_a{}^b\,, \label{matrixtransform}
\ee 
The GL symmetry of the Lagrangian is exact, $\delta_M \bL=0$. Thus, the symplectic potential (\ref{symplectic2}) gives directly the Noether current
\be
\bJ_M = \bTheta_M = - 2M^{ab}\bq_{ab} + M_b{}^a\bbe^b\wedge\bbs_a + M_d{}^b\aL^d_cV^c_a{\bmE}^a{}_b 
= M_a{}^b\lp -2\bq^a{}_b + \bbe^a\wedge\bbs_b + {\bmE}^a{}_b\rp = 0\,,
\ee
which vanishes identically. The total variation of the Lagrangian is thus very simple and results immediately in the Noether identity
\bs
\label{GLidentity}
\be
\bbe_a\wedge\bE_b = -2\bE_{ab}\,.
\ee
It is also simple to verify this explicitly,
\ba
\bbe_a\wedge\bE_b & = &  \bbe_a\wedge\lp\bDiff \bbs_b - \bt_b - \bT_b\rp = -2g_{ac}\bDiff\bq^c{}_b + \bS_a\wedge\bbs_b
- \bbe_a\wedge\lp \bt_b - \bbie_c\lrcorner\bT^c{}_b\rp \nn \\
& = &  -2g_{ac}\bDiff\bq^c{}_b + \bG_{ab} + \bT_{ab} = -2\bE_{ab}\,.
\ea
\es
In the \1 equality we used (\ref{EoMe}), in \2 we used (\ref{important}), in the \3 (\ref{Gt}) and in the \4 (\ref{EoMg}).

\subsubsection{Diff}

The matrix elements $\aL^a_b$ and $\iL^a_b$ are scalars under the Diff (\ref{gaugeDiff}). The $n$-form $\bL$ transforms as $\delta_\xi \bL = \bdiff\lp \xi\lrcorner\bL\rp$, where
\be \label{dotL4}
\xi\lrcorner\bL = -\xi\lrcorner\bbe^a\lp \bt_a - \bbie_b\lrcorner\bT^b{}_a\rp+ \xi\lrcorner\bS^a\wedge\bbs_a + \lp\xi\lrcorner\bQ^{ab}\rp\bq_{ab} 
 + \xi\lrcorner\bpsi\wedge\frac{\partial \bL_M}{\partial \bpsi} 
 + \xi\lrcorner\bdiff\bpsi\wedge\bPi\,.   
\ee
Combining this with the symplectic potential
\be
\bTheta_\xi = \bdiff\lp \xi\lrcorner\bbe^a\bbs_a\rp - \lp\xi\lrcorner\bbe^a\rp\bDiff\bbs_a + \xi\lrcorner\bS^a\wedge\bbs_a + \lp\xi\lrcorner\bQ^{ab}\rp\bq_{ab} + \delta_\xi\bpsi\wedge\bPi\,, \label{symp4}
\ee
we obtain the conserved current
\be
\bJ_\xi = \bdiff\bj_\xi - \lp\xi\lrcorner\bbe^a\rp\bE_a - \xi\lrcorner\bPsi\wedge\bE_\psi\,, \quad \text{where} \quad \bj_\xi = \lp\xi\lrcorner\bbe^a\rp\bbs_a + \xi\lrcorner\bpsi\wedge\bPi\,.  \label{curr4}
\ee
The Noether identity stemming from the Diff invariance is
\bs
\ba
0 & = & \lp\xi\lrcorner\bdiff g^{ab}\rp\bE_{ab} + \xi\lrcorner\bS^a\wedge\bE_a + \lp\xi\lrcorner\bA_a{}^b\rp\wedge\bbe^a\wedge\bE_b - \lp\xi\lrcorner\bbe^a\rp\bDiff\bE_a + \xi\lrcorner\bdiff\lp\bPsi\wedge\bE_\psi\rp \nn \\
& = & 
-\lp\xi\lrcorner\bQ^{ab}\rp\bE_{ab} + \xi\lrcorner\bdiff\lp\bbe^a\wedge\bE_a  
+ \bPsi\wedge\bE_\psi\rp\,.  \label{Diffidentity}
\ea
To arrive at the manifestly covariant form in the \2 line, we used the GL Noether identity (\ref{GLidentity}). 

It is useful to record the corresponding identity for the matter sector
alone. We may consider \(\bL_M\) alone, as it is independently Diff-invariant. One obtains
\be
0
=
\frac12(\xi\lrcorner \bQ^{ab})\bT_{ab}
+\xi\lrcorner \bdiff
\left(
        -\bbe^a\wedge\bT_a
        +\bpsi\wedge \bE_\psi
\right)\,.
\ee
\es
In the special case of G$_\parallel$R, $\bT_a$ is the local canonical energy current. We then see that   
\be \label{inertialforce}
\text{in G$_\parallel$R}: \quad  \bdiff \bT_a
        \approx
        (\bbie_a\lrcorner \bS^b)\wedge\bT_b
        -
        \frac12(\bbie_a\lrcorner \bQ^{bc})\bT_{bc}\,.
\ee        
The RHS of this equation represents the effect of inertial forces exerted by torsion and nonmetricity. In a canonical frame the force terms vanish and the
current is conserved, $\bdiff\bT_a \approx 0$.    

\subsubsection{CDiff}

The covariant Lie derivative results in the same transformation of the 4-form (\ref{dotL4}), and also in the same symplectic potential (\ref{symp4}). Thus we obtain the same manifestly covariant expression
for the current (\ref{curr4}). One can arrive at the same result by noticing that $\delta_\Xi \aL^a_b=0$, i.e. the matrix elements are, by construction, constant when performing a proper gauge-covariant CDiff along some vector $\Xi$. In particular, this is the consequence of (\ref{matrixtransform}). The Noether identity is also the same as (\ref{Diffidentity}).

\subsubsection{\underline{GL}}

We had deduced above that the variation of the connection in terms of the variation of the more elementary connection-generating matrix leaves the Lagrangian
invariant in (\ref{variations2}) but contributes to the symplectic potential in (\ref{symplectic2}),
\be
\bTheta_m = m_d{}^b\aL^d_c\iL^c_a{\bmE}^a{}_b =m_a{}^b{\bmE}^a{}_b\,.
\ee
The Lagrangian changes by the corresponding boundary term,
\be
    \underline{\delta}_m \bL=\bdiff \bK_m\,,\qquad \text{where} \quad \bK_m= m_a{}^b\bmE^a{}_b\,,
\ee
and hence the Noether current vanishes identically, $\bJ_m=\bTheta_m-\bK_m=0$.
The significance of the calculation is therefore not a new conservation law,
but the identification of \(\bmE^a{}_b\) as the momentum conjugate to
tangent-space frame transport.

\subsubsection{Alternative \underline{GL}}

The preceding transformation is only one piece of the ordinary
GL gauge transformation checked in \ref{GL2}.  In particular, it corresponds to the part
implemented by
\bs
\be \label{frame1}
  \underline{\delta}_m^{(\Lambda)}\Lambda^a{}_b=m^a{}_c\Lambda^c{}_b\,,
  \qquad
  \underline{\delta}_m^{(\Lambda)}\bA^a{}_b=\bDiff m^a{}_b\,,
\ee
with $g^{ab}$ and $e^a$ fixed, whereas the GL gauge transformation involves also the part 
\be \label{frame2}
  \underline{\delta}_m^{(e)} \bbe^a=m^a{}_b \bbe^b\,,
  \qquad
  \underline{\delta}_m^{(g)}g^{ab}=2m^{(ab)}\,.
\ee
\es
The three pieces add up to the GL gauge transformation,
\be
  \underline{\delta}_m^{(\Lambda)}+\underline{\delta}_m^{(e)}+\underline{\delta}_m^{(g)}
  =\delta_{M=m}.
\ee
At the level of the symplectic potential the corresponding pieces are
\be
  \bTheta_{\Lambda}=m_b{}^a\bmE^b{}_a,\qquad
  \bTheta_{e}=m_b{}^a \bbe^b\wedge \bbs_a,\qquad
  \bTheta_g=-2m^{ab}\bq_{ab}\,.
\ee
Equivalently, after aligning the indices,
\be
  \bTheta_{\Lambda}+\bTheta_e+\bTheta_g
  =
  m^a{}_b
  \left(
    \bmE^b{}_a+\bbe^b\wedge \bbs_a-2\bq^b{}_a
  \right)=0\,,
\ee
by the definition of ${\bmE}^a{}_b$. 

\section{Local currents}
\label{localcurrents}

In this appendix we work in the spherically symmetric geometry summarised in Table \ref{kodamaframe}. To evaluate the $\bX$ entering in $\termopot = \frac{m_P^2}{2\beta}\int\ast\bX$ (\ref{FreeEnergy}), we compute the traces defined in Eq.~(\ref{vectors}):
\bs
\ba
Q^0
& = &
\frac{6(f^2-1)}{\sigma}\frac{\lambda'}{\lambda}\,,
\qquad
Q^i
=
-\frac{6 f^2 \lambda'}{\sigma^2}\,n^i\,,
\\
\tilde{Q}^0
& = &
\frac{1}{\sigma}
\left[
2ff'
+(f^2-1)
\left(
\frac{\lambda'}{\lambda}
-\frac{\sigma'}{\sigma}
+\frac{2}{r}
\right)
\right]\,,
\quad
\tilde{Q}^i
=
\frac{\lambda}{\sigma^2}
\left[
2f^2
\left(
\frac{\sigma'}{\sigma}
-\frac{\lambda'}{\lambda}
\right)
-2ff'
+\frac{2(\sigma^2-f^2)}{r}
\right]n^i\,,
\\
S^0
& = &
\frac{1-f^2}{\sigma}
\left(
\frac{\sigma'}{\sigma}
+2\frac{\lambda'}{\lambda}
\right),
\qquad
S^i
=
\frac{\lambda f^2}{\sigma^2}
\left(
\frac{\sigma'}{\sigma}
+2\frac{\lambda'}{\lambda}
\right)n^i\,,
\\
\Rightarrow \quad
X^0
& = &
-\frac{1}{\sigma}
\left[
2ff'
+(f^2-1)
\left(
\frac{\sigma'}{\sigma}
-\frac{\lambda'}{\lambda}
+\frac{2}{r}
\right)
\right]\,,
\quad
X^i
=
\frac{2\lambda}{r\sigma^2}
\left(
rff'+f^2-\sigma^2
\right)n^i\,.
\ea
\es
This leads to the result (\ref{freeformula}). 

Whilst the energy-momentum of the frame field vanishes, there can nevertheless exist non-vanishing energy-momentum associated with the metric field. 
In the spherically symmetric case under consideration, this is expected to occur in geometries characterised by $\sigma' \neq 0$. Explicitly, we obtain
\bs
\ba
\bG^0{}_0
&=&
\ast m_P^2\,
\frac{\sigma'(\sigma^2-f^2)}{\sigma^3 r}\,,
\\
\bG^0{}_i
&=&
\ast m_P^2\,
\frac{\sigma'(1-f^2)}{\sigma^3 r}\,
n_i\,,
\\
\bG^i{}_0
&=&
0\,,
\\
\bG^i{}_j
&=&
\ast m_P^2\,
\frac{\sigma'(f^2-\sigma^2)}{2\sigma^3 r}\,
\left(\delta^i{}_j-n^i n_j\right)\,. 
\ea
\es
The metrical energy tensor does not, in general, represent a conserved current.

In contrast, the material energy tensors $\bT_{ab}$ are directly related to the conserved material energy currents $\bT_a$ via (\ref{canonicalmatter}). 
Let us demonstrate this by deducing the conserved matter currents for the Reissner--Nordstr\"om black hole of Section \ref{RNexample}. The components of the energy-momentum tensor $T^a{}_b=\ast \bT^a{}_b$ are particularly simple for this example, 
\be \label{RNdensity}
T^0{}_{0}=-\rho,\quad
T^0{}_{i}=T^i{}_{0}=0,\quad
T^i{}_{j}=\rho\delta^i{}_j-2\rho n^i n_j\,, \quad \text{where} \quad \rho=\frac{q^2}{8\pi r^4}\,. 
\ee
It is straightforward to translate this into the energy-momentum current 3-forms
\bs
\ba
\bT_0 & = & -\rho r^2 \bdiff r\wedge \bdiff\Omega\ \\  
\bT_i & = & 
\frac{\rho(1-f^2)r^2}{f^2} n_i \bdiff r\wedge \bdiff\Omega
+\rho r^2 n_i \bdiff T\wedge \bdiff\Omega
-\rho r \epsilon_{ijk}n^j\bdiff T\wedge \bdiff r\wedge \bdiff n^k\,. 
\ea
\es
These are the conserved currents. One may verify the conservation directly. The case $\bdiff \bT_0=0$ is obvious. 
The first term in \(\bT_i\) is
closed, since its radial derivative gives \(\bdiff r\wedge \bdiff r=0\), while
\(\bdiff n_i\wedge \bdiff\Omega=0\). Further, noting that $\epsilon_{ijk}\bdiff n^j\wedge \bdiff n^k=2n_i\bdiff \Omega$, the remaining terms give
\be
\bdiff \bT_i = -\Big[(\rho r^2)'+2\rho r\Big]n_i \bdiff T\wedge \bdiff r\wedge \bdiff \Omega = 0\,,
\ee
which follows directly from the $\rho \sim r^{-4}$ scaling of the energy density (\ref{RNdensity}). 

For completeness, we record the conserved material energy currents also for the dilatonic black hole of Section \ref{dilatonExample}. We define the total energy density $\rho$ and the total radial pressure $p$ in terms of the electric energy density $\rho_F$ and the dilaton energy density $\rho_\varphi$,
\begin{equation}
\rho=\rho_F+\rho_\varphi\,,
\qquad
p=-\rho_F+\rho_\varphi\,,
\qquad
\rho_F=
\frac{q^2 e^{2\varphi_\infty/m_P}}
     {8\pi r^2\left(\sqrt{r^2+r_q^2}+r_q\right)^2}\,,
\qquad
\rho_\varphi=
\frac{m_P^2 r_q^2 f^2}{r^4}\,.
\label{energies}
\end{equation}
In the canonical frame,
\bs
\label{canonicalcurrent}
\ba
\bT_0 & = & -\rho r^2\bdiff r\wedge\bdiff\Omega\,, \\
\bT_i & = & \frac{\rho\lp \sigma^2 - f^2\rp r^2}{\lambda f^2} n_i\bdiff r \wedge\bdiff\Omega
- \frac{p\sigma r^2}{\lambda}  n_i\bdiff T\wedge\bdiff\Omega 
+ \frac{p\sigma r}{\lambda} \epsilon_{ijk} n^j \bdiff T\wedge\bdiff r\wedge\bdiff n^k\,. \label{canonicalc2}
\ea
\es
The conservation of these currents follows from the matter Diff identity (\ref{inertialforce}),
\be
\bdiff \bT_a \approx (\bbie_a\lrcorner \bS^b)\wedge \bT_b
-\frac12(\bbie_a\lrcorner \bQ_{bc})\bT^{bc} = \frac{\sigma r^2}{\lambda}\,n_a
\left[
\left(
\frac{\sigma'}{\sigma}-\frac{\lambda'}{\lambda}
\right)p
-\frac{f'}{f}(\rho+p)
\right] \bdiff T\wedge \bdiff r\wedge \bdiff \Omega\,.
\ee
Since $n_0=0$, the zeroth component vanishes identically. Further,
using the canonical-frame condition for \(\lambda'/\lambda\) from  (\ref{s}), together with the field equations encoded in the explicit solution (\ref{energies}), the bracket vanishes identically. Thus,
the momentum currents are conserved as well and we conclude that $\bdiff \bT_a \approx 0$. Alternatively, the conclusions can be confirmed by applying $\bdiff$ directly to (\ref{canonicalcurrent}), as we did above for the Reissner--Nordstr\"om example.  

The interpretation of the local matter energy integral requires some care.
For the Reissner--Nordstr\"om solution, extending
$-\int \bT_0$ to $r=0$ reproduces the familiar divergent
Coulomb self-energy.  In the dilatonic solution the Maxwell sector is softened. Since
\be
  \rho_F r^2
  =
  \frac{q^2 e^{2\varphi_\infty/m_P}}
       {8\pi\bigl(\sqrt{r^2+r_q^2}+r_q\bigr)^2}\,, \quad   \rho_\varphi r^2
  =
  \frac{m_P^2 r_q^2 f^2}{r^2}\,, 
\ee
the Maxwell contribution is integrable at $r=0$.  Though the full continuation to the centre is singular, due to the dilaton
gradient contribution, this is not relevant for the exterior problem. On the static exterior region,
\be
   -\int_{\text{exterior}} \bT_0
  =
  4\pi\int_r^\infty
  \bigl(\rho_F+\rho_\varphi\bigr) r'^2\bdiff r'\,,
  \qquad r\geq r_+\,,
\ee
is finite and positive. 
The local exterior matter energy differs from the Hamiltonian charge, which also includes the electromagnetic surface contribution. We therefore decompose the Hamiltonian excess as 
\be
\Delta E_{\rm dil}^{(A)}(r;0) = E_{\rm dil}^{(A)}(r;0)-m_S
=
U_{\rm ext}(r)-U_\varphi(r)\,,
\ee
and the explicit computation gives  
\bs
\ba
U_\varphi(r)
& = &
\frac{m_S(1-x)}{12kx}
\left[
3(1+x)-k(1-x)(x^2+2x+3)
\right] >0\,,
\\
U_{\rm ext}(r)
& = &
\frac{m_S(1-x)}{6x}
\left(2x^3+2x^2-x+3\right) >0\,,
\ea
\es
with $x=r/(R+r_q)$ and $k=r_S/2r_q$. Thus, the Hamiltonian excess $\Delta E_{\rm dil}^{(A)}
=\frac{m_S(1-x)}{4kx}\Big[k(x^3+x^2-x+3)-(1+x)\Big]>0$ is positive, 
since $0<x<1$, $k>1$, and $x^3+x^2-x+3-(1+x)=x^2(1+x)+2(1-x)>0$.

Although the spatial currents $\bT_i$ need not vanish locally,
they do not represent a net momentum of the static solution.  Indeed,
the canonical spatial generators contain a temporal component relative
to the static slicing,
\be
  \bbie_i
  =
  \lambda^{-1}\partial_i
  -
  \frac{\sigma^2-f^2}{\lambda\sigma f^2}
  n_i\partial_T\,.
\ee
Accordingly, on a $T=\mathrm{const.}$ slice one finds a local term
proportional to $\rho n_i$.  On the canonical spatial leaf $\bbe^0=0$,
equation~\eqref{canonicalc2} instead reduces to
\be
  \left.\bT_i\right|_{\bbe^0=0}
  =
  \frac{(\rho+p)(\sigma^2-f^2)r^2}{\lambda f^2}
  n_i\bdiff r\wedge \bdiff\Omega
  =
  \frac{2\rho_\varphi(\sigma^2-f^2)r^2}{\lambda f^2}
  n_i\bdiff r\wedge \bdiff \Omega\,.
\ee
Hence the Reissner--Nordstr\"om spatial current vanishes pointwise on
the canonical leaf, while the dilatonic solution retains a local
contribution from the scalar radial stress.  Nevertheless, in both
cases the integrated Cartesian momentum charges vanish by spherical
symmetry, $-\int \bT_i = 0$,  noting that $\int n_i\bdiff\Omega=0$.

\end{document}